%% file: arxiv.tex
\documentclass[twocolumn, a4paper]{article}

\usepackage[a4paper, total={16.5cm, 24.5cm}]{geometry}
\usepackage{graphicx}         
\usepackage{parskip}
\usepackage{authblk}
\usepackage{amsmath}
\usepackage{amssymb}
\usepackage{amsthm}
\usepackage[utopia]{mathdesign}
\usepackage{mathtools}
\usepackage{color,soul}
\usepackage{nicefrac}
\usepackage{hyphenat}
\usepackage{isomath}
\usepackage{algpseudocode}
\usepackage{algorithm}
\usepackage{xparse}
\usepackage{cases}
\usepackage{url}

\include{tikzsettings}
\usepackage{grffile}
\usetikzlibrary{external}

\newtheorem{thm}{Theorem}

\newtheorem{lem}{Lemma}

\newtheorem{assum}{Assumption}

\newtheorem{defn}{Definition}

\newtheorem{rem}{Remark}
\newtheorem{prob}{Problem}

\let\OLDthebibliography\thebibliography
\renewcommand\thebibliography[1]{
	\OLDthebibliography{#1}
	\setlength{\parskip}{0pt}
	\setlength{\itemsep}{0pt plus 0.3ex}
}

\AtBeginDocument{%
	\abovedisplayskip=9pt plus 3pt minus 6pt
	\abovedisplayshortskip=0pt plus 3pt
	\belowdisplayskip=9pt plus 3pt minus 6pt
	\belowdisplayshortskip=7pt plus 3pt minus 4pt
}

\newlength\mytemplena
\newlength\mytemplenb
\DeclareDocumentCommand\myalignalign{sm}
{
	\settowidth{\mytemplena}{$\displaystyle #2$}%
	\setlength\mytemplenb{\widthof{$\displaystyle=$}/2}%
	\hskip-\mytemplena%
	\hskip\IfBooleanTF#1{-\mytemplenb}{+\mytemplenb}%
}

\newtheorem{sprob}{\changed{Subproblem}}

\newcommand*{\tran}{^{\mkern-1.5mu\mathsf{T}}\!}  
\newcommand*{\traninv}{^{-{\mkern-1.5mu\mathsf{T}}}\!}  

\DeclareMathOperator{\Tr}{Tr}
\DeclareMathOperator{\rank}{rank}
\DeclareMathOperator{\dom}{dom}
\def\d{\ensuremath{\mathrm{d}}}
\DeclareMathOperator{\reach}{reach}
\def\norm[#1]{\left|#1\right|}
\def\shortnorm[#1]{|#1|}
\DeclarePairedDelimiter{\floor}{\lfloor}{\rfloor}

\def\plant{{\mathrm{p}}}
\def\controller{{\mathrm{c}}}

\def\disturbance{{\mathrm{w}}}
\def\noise{{\mathrm{v}}}

\def\output{{\mathrm{y}}}

\def\Xs{\mathcal{X}}

\def\Us{\mathcal{U}}
\def\Ws{\mathcal{W}}
\def\Vs{\mathcal{V}}

\def\No{\mathbb{N}_{0}}
\def\N{\mathbb{N}}
\def\R{\mathbb{R}}
\def\S{\mathbb{S}}

\def\Linf{\mathcal{L}_{\infty}}
\def\Fsu{\mathcal{F}_{\Us}}
\def\Fsw{\mathcal{F}_{\Ws}}
\def\Ns{\mathcal{N}}
\def\Is{\mathcal{I}}
\def\Js{\mathcal{J}}

\def\Es{\mathcal{E}}
\def\Ks{\mathcal{C}}
\def\Bs{\mathcal{B}}
\def\Xstilde{\tilde{\Xs}}
\def\Xsy{\Xs_\output}
\def\Xw{\Xs_\disturbance}
\def\Xwtilde{\tilde{\Xs}_\disturbance}
\def\Ls{\mathcal{L}}
\def\Ts{\mathcal{T}}
\def\Cs{\mathcal{C}}
\def\Ds{\mathcal{D}}

\def\np{{n_\plant}}
\def\nup{{n_{\mathrm{u}}}}
\def\ny{{n_{\mathrm{y}}}}
\def\nw{{n_{\mathrm{w}}}}
\def\nv{{n_{\mathrm{v}}}}
\def\nd{{n_{\mathrm{d}}}}
\def\nc{{n_\controller}}

\def\xv{\boldsymbol{x}}
\def\yv{\boldsymbol{y}}
\def\uv{\boldsymbol{u}}
\def\wv{\boldsymbol{w}}
\def\vv{\boldsymbol{v}}
\def\xiv{\boldsymbol{\xi}}
\def\upsilonv{\boldsymbol{\upsilon}}
\def\psiv{\boldsymbol{\psi}}
\def\omegav{\boldsymbol{\omega}}
\def\nuv{\boldsymbol{\nu}}
\def\zetav{\boldsymbol{\zeta}}
\def\deltav{\boldsymbol{\delta}}
\def\xixuw{\xiv_{\xp\hat{\upsilonv}\omegav}}
\def\chiv{\boldsymbol{\chi}}

\def\dv{\boldsymbol{d}}
\def\sv{\boldsymbol{s}}
\def\cv{\boldsymbol{c}}
\def\qv{\boldsymbol{q}}
\def\fv{\boldsymbol{f}}
\def\gv{\boldsymbol{g}}
\def\hv{\boldsymbol{h}}
\def\mv{\boldsymbol{m}}
\def\lv{\boldsymbol{l}}
\def\ev{\boldsymbol{e}}
\def\bv{\boldsymbol{b}}
\def\xip{\xiv_\plant}
\def\xic{\xiv_\controller}
\def\xc{\xv_\controller}
\def\xp{\xv_\plant}

\def\zv{\boldsymbol{z}}
\def\pv{\boldsymbol{p}}
\def\xptilde{\tilde{\xv}_\plant}
\def\xiptilde{\tilde{\xiv}_\plant}

\def\ptilde{\tilde{\pv}}

\def\Am{\boldsymbol{A}}
\def\Bm{\boldsymbol{B}}
\def\Cm{\boldsymbol{C}}
\def\Dm{\boldsymbol{D}}
\def\Em{\boldsymbol{E}}

\def\I{\mathbf{I}}
\def\O{\mathbf{0}}
\def\Fm{\boldsymbol{F}}
\def\Gm{\boldsymbol{G}}
\def\Hm{\boldsymbol{H}}
\def\Jm{\boldsymbol{J}}
\def\Mm{\boldsymbol{M}}
\def\Nm{\boldsymbol{N}}
\def\CE{\Cm_{\mathrm{E}}}

\def\Qm{\boldsymbol{Q}}
\def\Qbar{\bar{\Qm}}
\def\Rm{\boldsymbol{R}}
\def\Sm{\boldsymbol{S}}
\def\Um{\boldsymbol{U}}
\def\Xm{\boldsymbol{X}}

\def\Wm{\boldsymbol{W}}
\def\Vm{\boldsymbol{V}}
\def\Zm{\boldsymbol{Z}}
\def\Ap{\Am_\plant}
\def\Bp{\Bm_\plant}
\def\Cp{\Cm_\plant}
\def\Ac{\Am_\controller}
\def\Bc{\Bm_\controller}
\def\Cc{\Cm_\controller}
\def\Dc{\Dm_\controller}
\def\Gammam{\matrixsym{\Gamma}}
\def\Phim{\matrixsym{\Phi}}
\def\Gammac{\Gammam_\controller}
\def\Gammap{\Gammam_\plant}
\def\Phic{\Phim_\controller}
\def\Phip{\Phim_\plant}
\def\Lm{\boldsymbol{L}}
\def\Om{\boldsymbol{O}}

\def\dk{{\kappa}}
\def\LC{L_\mathrm{C}}
\def\LD{L_\mathrm{D}}

\def\Nk{\changed{\Nm(\dk)}}
\def\Qk{\changed{\Qm(\dk)}}

\def\NkCE{\begin{bmatrix}\Nk \\ \CE\end{bmatrix}}
\def\NkCEt{\begin{bmatrix}\Nk\tran & \CE\tran\,\end{bmatrix}}

\def\e{\mathrm{e}}

\def\Fw{\Fm_\disturbance}
\def\Fv{\Fm_\noise}
\def\Rw{\Rm_\disturbance}
\def\Rv{\Rm_\noise}
\def\Qw{\Qm_\disturbance}
\def\Qv{\Qm_\noise}
\def\cvw{c_{\noise\disturbance}}
\def\cvv{c_{\noise}}
\def\Cw{\Cm_\disturbance}
\def\Cv{\Cm_\noise}

\def\eoo{\begin{bmatrix}\ev \\ \O \\ \O\end{bmatrix}}
\def\ppluse{\left(\!\ptilde+\eoo\right)}
\def\ppluset{{\ppluse\!}\tran}

\newcommand{\changed}[1]{{\color{blue} #1}}
\renewcommand{\changed}[1]{#1}
\newcommand{\changedtwo}[1]{#1}

\pgfplotsset{
    every axis legend/.append style={font=\scriptsize},
}

\title{Self-Triggered Output-Feedback Control of LTI Systems Subject to Disturbances and Noise\protect\thanks{© 2020. This manuscript version is made available under the CC-BY-NC-ND 4.0 license \protect\url{http://creativecommons.org/licenses/by-nc-nd/4.0/}. This work is supported by the European Research Council through the SENTIENT project (ERC-2017-STG \#755953).}}

\author[$\star$]{Gabriel de A. Gleizer} 
\author[$\star$]{Manuel Mazo Jr.} 

\affil[$\star$]{\footnotesize Delft Center for Systems and Control, TU Delft
	, The Netherlands (e-mail: \{g.gleizer, m.mazo\}@tudelft.nl)}

\begin{document}

\maketitle

\begin{abstract}                          
Self-triggered control (STC) and periodic event-triggered control (PETC) are aperiodic sampling techniques aiming at reducing control data communication when compared to periodic sampling. In both techniques, the effects of measurement noise in continuous-time systems with output feedback are unaddressed. 
In this work we prove that additive noise does not hinder stability of output-feedback PETC of linear time-invariant (LTI) systems. Then we build an STC strategy that estimates PETC's worst-case triggering times. To accomplish this, we use set-based methods, more specifically ellipsoidal sets, which describe uncertainties on state, disturbances and noise. Ellipsoidal reachability is then used to predict worst-case triggering condition violations, ultimately determining the next communication time. The ellipsoidal state estimate is recursively updated using guaranteed state estimation (GSE) methods. The proposed STC is designed to be computationally tractable at the expense of some added conservatism.
It is expected to be a practical STC implementation for a broad range of applications.
\end{abstract}

\section{Introduction}

Event\hyp{}Triggered Control (ETC) and Self\hyp{}Triggered Control (STC) are possibly the two dominant aperiodic sampling techniques of the past couple of decades. \changed{ETC, proposed independently and with different strategies by \cite{tabuada2007event} and \cite{aastrom2002comparison}, implements a state-dependent sampling mechanism, where the current measurements are monitored continuously (or periodically, as in Periodic ETC, PETC \cite{heemels2013periodic}) only on the sensor side, and the decision to close the loop is triggered upon the occurrence of a significant event. Its close relative STC \cite{velasco2003self} has the controller determining when to sample next, often by predicting when an ETC event would occur \cite{anta2008self,mazo2008event,mazo2010iss}.} Both methods promise to significantly reduce network usage on Networked Control Systems (NCSs) by having input and output data communicated only when needed. ETC provides the largest savings and has a straightforward implementation --- a simple triggering mechanism ---, but its actual usage in NCSs is challenging as it needs dedicated hardware \cite{anta2008self} and its communication times are difficult to predict \cite{kolarijani2016formal}. Such prediction is particularly important to avoid communication collisions when multiple control loops share the network.

In STC, the controller determines the next sampling time based on available information, thus its communication is one-step predictable by design. Its sampling time computation is generally based on conservative estimates of when an ETC would trigger, and most of the STC literature considers state-feedback with noiseless measurement. For example, in \cite{mazo2010iss}, disturbances may be present but are not considered in the event prediction. While this method guarantees stability and a finite $\Linf$-gain, its disturbance rejection is poorer than ETC's, since event-triggering naturally takes disturbances into account. To improve disturbance attenuation, \cite{gleizer2018selftriggered} recently proposed an STC that considers disturbances within the prediction; this way STC has the same performance as ETC, although yielding more frequent communication.

Unfortunately, most practical control systems are not state-feedback regulators, but take the output feedback form. Moreover, measurement noise is always present, which can significantly affect the event predictions that are inherent to STC. When not all states are measured, few approaches are available in the literature. In \cite{almeida2014self}, an observer was developed for self-triggered state-feedback control of LTI systems. For general dynamic output-feedback controllers, still noiseless, \cite{gleizer2018selftriggered} developed a self-triggered mechanism, where an open-loop ellipsoidal observer was employed. One of its drawbacks is that, as for any open-loop observer, there is no control on its convergence. Also in \cite{gleizer2018selftriggered}, matrix norms were used for disturbance-related reachability, leading to excessive conservativeness. In this work, tighter ellipsoidal reachability \cite{kurzhanski1997ellipsoidal} is instead used to compute disturbance-related reachable sets.

\changedtwo{Set-based methods have also been employed for ETC and STC on recent works, such as observer-based state feedback ETC in \cite{moreira2019observer}, and ETC and STC for discrete-time systems subject to disturbances and noise in \cite{brunner2019event}. Conceptually, the latter is the most similar to our work, aside from \cite{gleizer2018selftriggered}, because of the usage of set-based methods for the disturbance reachability. The major differences are the following: (i) their stability results are for discrete-time systems, which do not immediately provide guarantees for continuous-time systems; (ii) they invoke the novel notion of $\theta$-uniform global asymptotic stability ($\theta$-UGAS), a system theoretic property weaker than input-to-state stability, which is what we use in this paper; (iii) their output-feedback controllers are restricted to observer-based state feedback; and (iv) they introduce new set-based events, while in this work we employ well-established event-triggering mechanisms. In addition, our work is particularly focused on implementation and computational efficiency, aspects that are very briefly touched upon in \cite{brunner2019event}. In summary, to the best of our knowledge, no available STC strategy takes measurement noise into account for continuous-time systems, nor is it prepared for general forms of output-feedback controllers.}

This work has two main contributions: first, \changed{we prove that, if a PETC or STC closed-loop LTI system is globally exponentially stable, then it is input-to-state stable with respect to disturbances, measurement noise, and additive perturbations in the triggering condition;} second, we devise a method to build self-triggered implementations of controllers subject to unknown but bounded disturbances and measurement noise. The stability results make use of the notion of homogeneous hybrid systems from \cite{nesic2013finite}. The STC design is an improvement and extension of \cite{gleizer2018selftriggered} for the noisy case, which consists of computing a lower bound to the triggering times of the PETC strategy from \cite{heemels2013periodic}. Here we use set-theoretic methods for control, namely set-valued reachability (SVR) and guaranteed state estimation (GSE). The state estimator keeps track of a set that contains all possible states in which the plant and controller could be. Reachable sets from the observer state set are then computed for a given sequence of elapsed time instants. At each of these instants, an algorithm checks if there is a point in the reachable set that violates a designed triggering condition. Such a check is conservative but computationally efficient. We hereafter refer to this method as Preventive Self-Triggered Control (PSTC), since it is designed to prevent triggers later than the reference PETC. The separation properties of linear systems allow for most of the computations to be carried out offline. Like in \cite{gleizer2018selftriggered}, we choose ellipsoids for the description of sets, even though other descriptions have been shown to be more effective for general-purpose SVR and GSE (e.g., constrained zonotopes in \cite{scott2016constrained}). One reason is that the considered triggering functions are quadratic, which simplifies computations when ellipsoids are used. In any case, efficient ellipsoidal SVR and GSE methods are available for linear systems: for SVR we use \cite{kurzhanski1997ellipsoidal} and \cite{kurzhanski2006ellipsoidal}; for GSE, we adapt the results from \cite{schweppe1968recursive}, \cite{ros2002ellipsoidal} and \cite{scott2016constrained}. The final algorithm attains similar control performance as PETC, while keeping the advantages of STC and reasonably small computational costs; thus, it is likely to fit most linear control applications.

\subsection{Notation}

Throughout the paper, bold letters are used for vectors and matrices, or vector-valued and matrix-valued functions; and calligraphic letters are used for sets or set-valued functions. Signals are denoted with greek letters, while points are denoted with roman letters.
 
We denote by $\No$ the set of natural numbers including 0, $\N \coloneqq \No\setminus\{0\}$, and $\R_+ \coloneqq \{x \in \R: x \geq 0\}$. The floor function on $x\in\R$ is denoted by $\floor{x}$. For a vector $\xv \in \R^n$ we denote by $\norm[\xv]$ its 2-norm
. The canonical vector, denoted by $\cv_i$, has its $i$-th entry equal to 1 and the rest equal to zero. For a matrix $\Am \in \R^{n \times m}$ we denote by $\Am\tran$ its transpose, by $\rank(\Am)$ its rank, \changed{by $\lambda(\Am)$ its eigenvalues, by $\lambda_{\max}(\Am)$ ($\lambda_{\min}(\Am)$) its maximum(minimum)-in-real-part eigenvalue}, by $|\Am|$ its 2-induced norm, by $\Tr(\Am)$ its trace, and by $\Am^\dagger$ its pseudoinverse. %
We denote $\Am|_{\Is,\Js}$ the sub-matrix of $\Am$ indexed by the row index set $\Is \subseteq \{1,...,n\}$ and the column index set $\Js \subseteq \{1,...,m\}$. If $\Is = \{1,\ldots,n\}$ or $\Js = \{1,\ldots,m\}$ we use $\Am|_{\bullet,\Js}$ or $\Am|_{\Is,\bullet}$, respectively. For a symmetric square matrix $\Sm \in \R^{n \times n}$, the statements $\Sm \succ \O$ and $\Sm \succeq \O$ denote that $\Sm$ is positive definite or positive semidefinite, respectively. We denote by $\S^n \coloneqq \{\Sm \in \R^{n\times n}| \Sm=\Sm\tran\,\}, \S^n_+\coloneqq \{\Sm\in\S^n|\Sm\succeq\O\},$ and $\S^n_{++}\coloneqq\{\Sm\in\S^n|\Sm\succ\O\}$ the sets of symmetric, symmetric positive semidefinite, and symmetric positive definite, respectively. %
The set $\Bs(r)$ is a ball of radius $r\geq0$. For two sets $\Xs_1$ and $\Xs_2$ we denote their Minkowski sum as $\Xs_1 + \Xs_2$.  We often denote a singleton $\{\xv\}$ as $\xv$ when it is in an operation between sets.

\section{Preliminaries}

\subsection{Hybrid Dynamical Systems}

For stability results, we will model the STC closed-loop system as a hybrid system, which allows states to flow on continuous time and/or to jump instantly. In this modeling framework, solutions are defined on the \emph{hybrid time domain}, which is a subset of $\R_+ \times \N$ that can be written as $\cup_{i\in\{0,...,J\}}([t_i, t_{i+1}] \times \{i\}),$ where $J \in \N$ and $0 = t_0 \leq t_1 \leq ... \leq t_{J+1}$, with $J$ and/or $t_{J+1}$ possibly $\infty$. A \emph{hybrid signal} $\chiv$ is a function defined on a hybrid domain. %
A \emph{hybrid system} is described as follows:
\begin{equation}\label{eq:hybrid}
\begin{cases}
\dot\chiv = \fv(\chiv,\deltav), & (\chiv(t,j),\deltav(t,j)) \in \Cs \\
\chiv^+ = \gv_i(\chiv,\deltav), & (\chiv(t,j),\deltav(t,j)) \in \Ds_i\\
\psiv = \hv(\chiv,\deltav), &
\end{cases}
\end{equation}
with $i \in \{1,...,I\},$ where $\chiv(t,j) \in \R^n$ is the state vector, $\deltav(t,j) \in \R^\nd$ is an exogenous input, $\psiv(t,j) \in \R^\ny$ is the output vector, $\fv, \gv_i $ and $\hv$ are continuous functions with inputs and outputs of appropriate dimensions, and $\Cs \subseteq \R^{n+\nd}$ and $\Ds_i\subseteq \R^{n+\nd}$ are closed sets.
Following \cite{cai2009characterizations} and \cite{nesic2013finite}, we say that a pair $(\chiv,\deltav)$ is a solution to (\ref{eq:hybrid}) if $\dom\chiv = \dom\deltav$ and
\begin{itemize}
	\item for all $j \in \N$ and almost all $t$ such that $(t,j) \in \dom\chiv,$ the pair satisfies $(\chiv(t,j),\deltav(t,j))\in\Cs$ and $\dot\chiv(t,j) = \fv(\chiv(t,j),\deltav(t,j))$;
	\item for all $i \in \{1,...,I\}$ and all $(t,j) \in \dom\chiv$ such that $(t,j+1)\in\dom\chiv$, the pair satisfies $(\chiv(t,j),\deltav(t,j))\in\Ds_i$ and $\chiv(t,j+1) = \gv_i(\chiv(t,j),\deltav(t,j))$.
\end{itemize}

\begin{defn}[$\Ls_p$ norm, \cite{nesic2013finite}] For a hybrid signal $\psiv$, with domain $\dom\psiv$, and a scalar $T \in \R_+$, the $T$-truncated $\Ls_p$-norm of $\psiv$ is given by\footnote{As a convention, $\sum_{i=1}^0f(i) = 0$.}
	\begin{equation}\label{eq:lpnormtrun}
	\|\psiv_{[T]}\|_p \coloneqq\! \left( \sum_{i=1}^{j(T)}\norm[\psiv(t_i,i-1)]^p +\! \sum_{i=0}^{j(T)}\int_{t_i}^{\sigma_i}\!\!\norm[\psiv(s,i)]^p\d s\!\right)^{\!\!\frac{1}{p}}\!\!\!,
	\end{equation}
	where $j(T) \coloneqq \max\{k: (t,k) \in \dom\psiv, t+k\leq T\},$ and $\sigma_i \coloneqq \min(t_{i+1},T-i)$. From (\ref{eq:lpnormtrun}), the $\Ls_p$-norm of $\psiv$ is defined as
	\begin{equation}\label{eq:lpnorm}
	\|\psiv\|_p \coloneqq \lim_{T\to T^*}\|\psiv_{[T]}\|_p,
	\end{equation}
	\changed{where $T^* = \sup\{t+j: (t,j) \in \dom \psiv\}$ (possibly infinity).} The $\Linf$ norm is taken by replacing the sums (integrals) in (\ref{eq:lpnormtrun}) by the (essential) suprema.
\end{defn}	

\begin{defn} {\bf(Global Exponential ISS, \cite{nesic2013finite})}~ System (\ref{eq:hybrid}) is exponentially \changed{finite-gain} input-to-state stable from $\deltav$ if there exist positive scalars $k, a, $ and $\gamma$ such that, for any initial condition $\xv$ and any $\deltav \in \Linf$, all solutions to (\ref{eq:hybrid}) satisfy
	\begin{equation}\label{eq:iss}
	|\chiv(t,j)| \leq \max\left\{k\e^{-a(t+j)}|\xv|, \gamma\|\deltav\|_\infty \right\}
	\end{equation}
	for all $(t,j) \in \dom\chiv$. Moreover, the origin is globally exponentially stable (GES) if (\ref{eq:iss}) holds with $\deltav \equiv \O$.
\end{defn} 
\begin{defn}[$\Ls_p$ stability, \cite{nesic2013finite}] Given $p \in [1,+\infty)$, system (\ref{eq:hybrid}) is $\Ls_p$ stable from $\deltav$ to $\psiv$ with gain (upper bounded by) $k_p \geq 0$ if there exists a scalar $\beta \geq 0$ such that any solution to (\ref{eq:hybrid}) satisfies
	\begin{equation}\label{eq:lpgain}
	\|\psiv\|_p \leq \beta|\xv| + k_p\|\deltav\|_p
	\end{equation}
	for any initial condition $\xv\in\R^n$ and any $\deltav \in \Ls_p$.
\end{defn}

The last definition we need is that of homogeneous hybrid systems of degree zero:
\begin{defn}{\bf(Homogeneous hybrid system, \cite{nesic2013finite})}~ \label{def:homo} The system (\ref{eq:hybrid}) is homogeneous of degree zero if\changed{, for any scalar $\lambda > 0,$ we have}
	\begin{align}
	\myalignalign{} &
	\myalignalign*{}
	\begin{aligned}\label{eq:fghomo}
	\!\fv(\lambda\chiv,\O) &= \lambda\fv(\chiv,\O),   \forall \chiv(t,j)\in\Cs_0,\!\\
	\!\gv_i(\lambda\chiv,\O) &= \lambda\gv_i(\chiv,\O),   \forall \chiv(t,j)\in\Ds_{i0}, i \in \{1,...,I\},\!
	\end{aligned} \\
	\myalignalign{} &
	\myalignalign*{}
	\begin{aligned}\label{eq:setshomo}
	\chiv \in \Cs_0 &\implies \lambda\chiv \in \Cs_0,\\
	\chiv \in \Ds_{i0} &\implies \lambda\chiv \in \Ds_{i0}, \forall i \in \{1,...,I\},
	\end{aligned}
	\end{align}
	where closed sets $\Cs_0, \Ds_{i0}$ are projections of $\Cs$ and $\Ds_i$ when $\deltav \equiv \O$. 
\end{defn}

\changed{We are particularly interested in homogeneous systems that satisfy the following assumption:
	
\begin{assum} {\bf(Flow and jump sets,  \cite{nesic2013finite})}\label{assum:setpert}~
	For system (\ref{eq:hybrid}), there exist scalars $\LC$ and $\LD$ such that, for all $(\xv,\dv) \in \R^{n+\nd},$
	\begin{subequations}
		\begin{align}
		(\xv,\dv) \in \Cs &\implies \xv \in \Cs_0 + \LC\Bs(|\dv|)\label{eq:flowsetbound}\\
		(\xv,\dv) \in \Ds_i &\implies \xv \in \Ds_{i0} + \LD\Bs(|\dv|).\label{eq:jumpsetbound}
		\end{align}
	\end{subequations}
\end{assum}
}
Homogeneous systems \changed{satisfying Assumption \ref{assum:setpert}} have a powerful stability property:\footnote{This result was proven for a single pair of jump map and set, i.e., $I=1$. However, the proofs could incorporate multiple jump maps and sets, with the results remaining valid.}:
\begin{thm}[\cite{nesic2013finite}]
	\label{thm:nesic}%
	Let system (\ref{eq:hybrid}) be homogeneous in the sense of Definition \ref{def:homo} \changed{and Assumption \ref{assum:setpert} hold}; then, the following statements are equivalent:
	\begin{itemize}
		\item the origin of system (\ref{eq:hybrid}) is GES if $\deltav \equiv \O$;
		\item system (\ref{eq:hybrid}) is \changed{globally exponentially} ISS;
		\item system (\ref{eq:hybrid}) is $\Ls_p$ stable from $\deltav$ to $\psiv$.
	\end{itemize}
\end{thm}

\subsection{Recursive Guaranteed State Estimator}\label{ssec:rgse}
Consider an LTI system of the form:
\begin{equation}\label{eq:linplant}
\begin{aligned}
\dot\xip(t) &= \Ap\xip(t) + \Bp\hat{\upsilonv}(t) + \Em\omegav(t), \\
\psiv(t) &= \Cp\xip(t) + \nuv(t), \\
\xip(0) &= {\xp},
\end{aligned}
\end{equation}
\changed{where the sub-index $\plant$ is used to denote plant variables,} with $\xip(t) \in \R^\np$ as its state, $\hat{\upsilonv}(t) \in \Us \subset \R^\nup$ as its received control input, $\omegav(t) \in \Ws \subset \R^\nw$ as the unknown disturbances, $\psiv(t) \in \R^\ny$ as the measured output, $\nuv(t) \in \Vs \subset \R^\ny$ as the unknown measurement noise, and $\xp \in \Xs_0 \subset \R^\np$ as its initial state. The following assumptions hold:

\changed{\begin{assum}\label{assum:obs} Sets $\Us, \Ws,$ and $\Vs$ are compact, and the pair $(\Ap,\Cp)$ is observable.\footnote{\changed{$(\Ap,\Cp)$ could be relaxed to be detectable. The unobservable but stable subspace does not affect the controller, thus one should only consider the observable subspace when implementing the results in this paper.}} \end{assum}}

Let $\Fsu$ (resp. $\Fsw$) be the set of essentially bounded piecewise continuous functions from $\R_+$ to $\Us$ (resp. $\Ws$). We denote a solution of system (\ref{eq:linplant}) for initial state $\xp$, input $\hat\upsilonv \in \Fsu$, and disturbance $\omegav \in \Fsw$ by
\begin{equation*}
\xixuw(t) = \e^{\Ap t}\xp +\! \int_0^t{\!\!\e^{\Ap(t-\tau)}(\Bp\hat{\upsilonv}(\tau) + \Em\omegav(\tau))}\d \tau.
\end{equation*}

We are interested in computing the set of possible solutions to system (\ref{eq:linplant}) for sets of initial states, control input trajectories and disturbance trajectories. For that, the following definitions are necessary.
\begin{defn}[Reachability operator] Given an initial time $t_1$, a final time $t_2$, an initial state set $\Xs$ and the sets $\Us$ and $\Ws$, the reachability operator $\reach(\cdot)$ is defined as $\reach(t_1,t_2,\Xs,\Us,\Ws) \!\coloneqq\! \{\xixuw(t_2): \xixuw(t_1) \allowbreak \in \Xs, \upsilonv \in \Fsu, \omegav \in \Fsw\}$. Moreover, the output of this operator is denoted as the reachable set.
\end{defn}
A recursive GSE is a set-valued version of a general recursive state estimation and, as such, it follows the same principles. A GSE requires that bounds to input, disturbance and noise signals are known in the form of sets:
\begin{assum} There exist known compact sets $\tilde{\Us},\tilde{\Ws}$ and $\tilde{\Vs}$ such that  $\Us \subseteq \tilde{\Us}, \Ws \subseteq \tilde{\Ws}$ and $\Vs \subseteq \tilde{\Vs}$. \end{assum}
\begin{defn} {\bf(Recursive GSE, {\cite[Chap.~11]{blanchini2008set}})} Let $\tilde{\Xs}(t_1|t_1) \ni \changed{\xip}(t_1)$ be an available set estimate of the current state at time $t_1$. Let $\yv \coloneqq \psiv(t_2)$ be an output measurement obtained at $t_2$. A recursive GSE has the form
	\begin{subequations}
		\begin{align}
		\tilde{\Xs}(t_2|t_1) &= \reach(t_1,t_2,\tilde{\Xs}(t_1|t_1),\tilde{\Us},\tilde{\Ws}), \label{eq:generalprediction}\\
		\Xsy(t_2) &= \{\xp\in\R^\np| \exists \vv\in\tilde{\Vs}: \Cp\xp+\vv = \yv\}, \label{eq:outputcoherent}\\
		\tilde{\Xs}(t_2|t_2) &= \tilde{\Xs}(t_2|t_1) \cap \Xsy(t_2). \label{eq:generalcorrection}
		\end{align}
	\end{subequations}
\end{defn}
Eq.~(\ref{eq:generalprediction}) is the prediction step, simply a reachability operation. Eq.~(\ref{eq:generalcorrection}) is the update step, where the predicted set is intersected with $\Xsy(t_2)$, the set of all possible states that are coherent with the measurement. By construction, $\tilde{\Xs}(t_2|t_2) \owns \changed{\xip}(t_2)$. The sets above can have arbitrary complexity. Hence, it is common to replace the equalities above with superset operations, then restricting the set families to computationally tractable ones.

Throughout this paper, the aforementioned sets will be (outer-approximated by) ellipsoids. This idea dates back to 1968 \cite{schweppe1968recursive}, when possibly the first GSE was proposed. Ellipsoids are described by few parameters -- one vector and one symmetric matrix -- and are bounded. Since they may be described as quadratic inequalities, they also harmonize well with the quadratic triggering functions generally employed for ETC of LTI systems. Some definitions follow:
\begin{defn} {\bf(Ellipsoid, {\cite[Chap.~2]{kurzhanski1997ellipsoidal}})}\label{def:ell}~ Let $\mv \in \R^n$ and $\Mm \in \S^n_+$. An ellipsoid is defined in terms of its support function:
	\begin{equation*}\label{eq:ellipsoidsupport}
	\Es(\mv,\Mm) \!\coloneqq\! \{\xv \in \R^n: \lv\tran\xv \leq \lv\tran\mv + (\lv\tran\Mm\lv)^{\nicefrac{1}{2}}, \forall \lv \in \R^n\}.
	\end{equation*}
\end{defn}
\begin{rem} In case the ellipsoid is not degenerate ($\Mm \succ \O$), it can be described in the well-known inequality form $\Es(\mv,\Mm) = \{\xv \in \R^n: (\xv-\mv)\tran\Mm^{-1}(\xv-\mv) \leq 1\}$. The degenerate case is flat on some of its semi-axes. \end{rem}
A closely related set is the elliptical cylinder. The following definition comes from \cite{ros2002ellipsoidal}, with a small change in notation:

\begin{defn}[Elliptical Cylinder]\label{def:cyl} Let $\Mm\in \S^\changed{m}_{++},$ $\Cm \in \R^{m \times n}, m \leq n,$ and $\rank(\Cm) = m$. An Elliptical Cylinder is defined as
	$$\Ks(\yv,\Mm,\Cm) \coloneqq \{\xv \in \R^n\!: (\Cm\xv-\yv)\tran\Mm^{-1}(\Cm\xv-\yv) \leq 1\}.$$
\end{defn}
\begin{rem} If $m<n$, the elliptical cylinder is unbounded. If $m=n$, it trivially resolves to the ellipsoid $\Es(\Cm^{-1}\yv, \Cm^{-1}\Mm\Cm\traninv\,)$. \end{rem}

We use some operations on ellipsoids, namely affine transformations, intersections and Minkowski sums. An affine transformation on an ellipsoid is also an ellipsoid: $\Am\Es(\mv,\Mm) + \bv = \Es(\Am\mv+\bv,\Am\Mm\!\Am\tran\,)$. Even though ellipsoids are not closed under Minkowski sums and intersections, there are methods to tightly outer\hyp{}approximate them with ellipsoids. Here we use trace-optimal outer\hyp{}approximations. For the Minkowski sum, one has \cite[Chap.~2]{kurzhanski1997ellipsoidal}:
\begin{equation}\label{eq:ellipsoidmink}
\begin{aligned}
\Es(\mv^*,\Mm^*) &\supseteq \Es(\mv_1,\Mm_1) + \Es(\mv_2,\Mm_2)\\
\mv^* &\coloneqq \mv_1 + \mv_2 \\
\Mm^* &\coloneqq (1 + p^{-1})\Mm_1 + (1 + p)\Mm_2\\
p &\coloneqq \sqrt{\Tr(\Mm_1)\Tr(\Mm_2)^{-1}}.
\end{aligned}
\end{equation}
If not empty, the intersection may be outer-approximated by a fusion (see below). We particularly need to compute the intersection between an ellipsoid and an elliptical cylinder. 
\begin{defn}[Fusion] (Adapted from \cite{ros2002ellipsoidal}) A fusion between the ellipsoid $\Es(\mv_1,\Mm_1)$ and the elliptical cylinder $\Ks(\yv,\Mm_2,\Cm)$ is the ellipsoid $\Es_\lambda(\mv,\Mm)$ defined over a parameter $\lambda \in [0,1)$, such that:
	\begin{equation}\label{eq:ellipsoidfusion}
	\begin{aligned}
	&\Es_\lambda(\mv,\Mm) \supseteq \Es(\mv_1,\Mm_1) \cap \Ks(\yv,\Mm_2,\Cm)\\
	&\Mm = z\Zm^{-1} \\
	&\Zm = \lambda\Mm_1^{-1} + (1-\lambda)\Cm\tran\Mm_2^{-1}\Cm \\
	&\ev = \yv-\Cm\mv_1 \\
	&z = 1 - \lambda(1-\lambda)\ev\tran(\lambda\Mm_2 + (1-\lambda)\Cm\Mm_1\Cm\tran)^{-1}\ev \\
	&\mv = \Zm^{-1}(\lambda\Mm_1^{-1}\mv_1 + (1-\lambda)\Cm\tran\Mm_2^{-1}\yv).
	\end{aligned}
	\end{equation}
\end{defn}
The parameter $\lambda$ controls how close the output ellipsoid is to either of its inputs. For $\lambda = 1$, $\Es_0(\mv,\Mm) = \Es(\mv_1,\Mm_1)$; when $\lambda$ gets close to 0, the output tends to be close to $\Ks(\yv,\Mm_2,\Cm)$.

\begin{rem} The trace of the matrix $\Mm$ is convex over $\lambda$, since the trace of the inverse is a convex function \cite{boyd2004convex} and $z \in [0,1]$ provided the intersection is not empty \cite{ros2002ellipsoidal}. This allows the use of bisection or golden search methods to compute $\lambda$ that minimizes the fusion trace.\end{rem}

\begin{figure}
	\begin{center}
		\input{reach_ellipsoids.tex}
		\caption{\label{fig:reach} {Illustration of a reachable set of the disturbance response $\Xw(t)$ and an ellipsoidal outer-approximation $\Xwtilde(t)$. }}
	\end{center}
\end{figure}
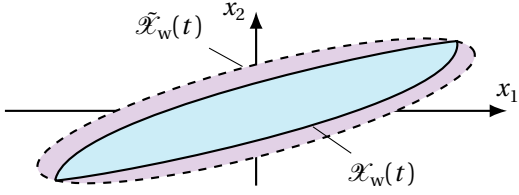

\changed{\subsubsection{Ellipsoidal reachability}\label{ssec:ellipreach}}
For linear systems with ellipsoidal descriptions of $\Xs, \Us, \Ws,$ and $\Vs$, ellipsoidal reachability can be used. The concept and techniques are thoroughly explained in \cite[Chap.~3]{kurzhanski1997ellipsoidal}. Its authors developed the Ellipsoidal Toolbox \cite{kurzhanski2006ellipsoidal}, which contains operations to compute reachable sets. In this paper we use the reachable set for the disturbance response $\Xs_\disturbance(t) \coloneqq \reach(0,t,\O,\O,\Ws)$. The Ellipsoidal Toolbox has the tools to compute outer-approximations of $\Xs_\disturbance(t)$, denoted by $\bar{\Xs}_\disturbance(t,\lv)$, that are tight along the ray supported by a given vector $\lv \in \R^{\np}$, i.e., $\forall \alpha \in \R, \alpha\lv \in \bar{\Xs}_\disturbance(t,\lv) \iff \alpha\lv \in \Xs_\disturbance(t)$. Overall tighter over-approximations can be obtained by computing $\bar{\Xs}_\disturbance(t,\lv_i)$ for different input vectors $\lv_i$ and taking an ellipsoidal outer-approximation of the intersection, offering a trade-off between accuracy and precision. Let $\Ls$ be a pre-specified set of the said vectors. The outer-approximation $\tilde{\Xs}_\disturbance(t)$ satisfies $\tilde{\Xs}_\disturbance(t) \supseteq \cap_{\lv\in\Ls}\bar{\Xs}_\disturbance(t,\lv)$. Figure \ref{fig:reach} depicts the sets $\Xw(t)$ and $\Xwtilde(t)$ for a given instant. The Ellipsoidal Toolbox is used to compute the intersection outer-approximation.

\section{Problem definition and stability results}
Consider a controller for system (\ref{eq:linplant}) of the form
\begin{equation}\label{eq:lincont}
\begin{aligned}
\xic(k+1) &= \Ac\xic(k) + \Bc\hat{\psiv}(k),\\
\upsilonv(k) &= \Cc\xic(k) + \Dc\hat{\psiv}(k),
\end{aligned}
\end{equation}
where $\xic(k) \in \R^{\nc}$ is the controller state, $\upsilonv(k) \in \R^\nup$ is the computed control command and $\hat{\psiv}(k) \in \R^\ny$ is the available plant output measurement. The controller runs with period $h$, so that $t = hk$. The feedback loop is of sample-and-hold form. For two consecutive sampling times $k_b$ and $k_{b+1}$, $\hat{\upsilonv}(t) = \upsilonv(k_b), \forall t \in [hk_b,hk_{b+1})$ and $\hat{\psiv}(k) = \psiv(hk_b), \forall k \in \{k_b,k_b+1,...,k_{b+1}-1\}$. The closed-loop system is depicted in Fig.~\ref{fig:blockdiagram}. We pose the PSTC problem as follows:
\begin{figure}
	\begin{center}
		\input{block_diagram_stc.tex}
		\caption{\label{fig:blockdiagram} Block diagram of a plant controlled with STC. ZOH stands for zero-order hold.}
	\end{center}
\end{figure}
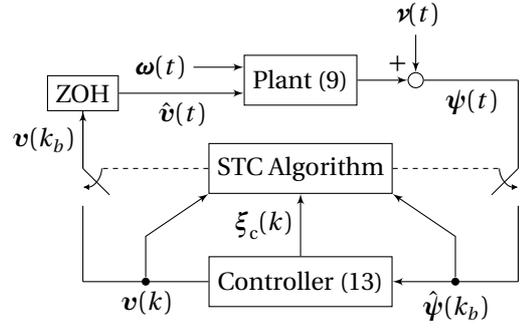
\begin{prob}\label{prob:theprob}
	Let the plant (\ref{eq:linplant}) and controller (\ref{eq:lincont}) models be known and suppose that (conservative estimates of) the sets $\Xs_0, \Ws, \Vs$ are known. Design an algorithm that computes $\dk_b \coloneqq k_{b+1}-k_b$ at time $k_b$ based on (historical values of) $\hat{\upsilonv}, \hat{\psiv}$ and other available information, e.g., $\xic(k_b)$. The closed-loop system must be \changed{globally exponentially ISS w.r.t.}~bounded disturbances and noise.
\end{prob}
\changed{\begin{rem}\label{rem:x0}
	 A compact set $\Xs_0$ is required for the STC strategy we develop in Sec.~\ref{sec:main}. A large enough set may be easily estimated in most applications. For $\Xs_0 = \R^{\np}$, we provide an initialization algorithm in Appendix \ref{ap:init}.
\end{rem}}

\subsection{Triggering mechanism and stability results}
\changed{In the spirit of \cite{gleizer2018selftriggered}, we design an algorithm that computes worst-case triggering times of PETC.} For compactness of expressions, denote the auxiliary vectors 
$$\zetav(t) \coloneqq \begin{bmatrix}\psiv(t) \\ \upsilonv(\floor{t/h})\end{bmatrix} \ \text{ and } \ \hat{\zetav}(t) \coloneqq \begin{bmatrix}\hat{\psiv}(\floor{t/h}) \\ \hat{\upsilonv}(t)\end{bmatrix}$$
as the updated output/input and the held output/input, respectively.
We start with a centralized output-based PETC triggering mechanism from \cite{heemels2013periodic}, which for STC means that all inputs and outputs are updated at the same time:
\begin{subequations}\label{eq:trigger}
	\begin{align}
	t_{b+1} &= \inf_{t\in\Ts_b}\eta(\zetav(t),\hat{\zetav}(t)) > \epsilon^2, \label{eq:trigcond}\\
	\eta(\zetav(t),\hat{\zetav}(t)) &\coloneqq \shortnorm[\zetav(t)-\hat{\zetav}(t)]^2 - \sigma^2\norm[\zetav(t)]^2, \label{eq:lintrig}
	\end{align}
\end{subequations}
where $\Ts_b = \{t_b+hk, t_b+2hk,...,t_b+h\bar{\kappa}\}$, $0 \leq \sigma < 1$ is the designed triggering parameter, $\bar{\kappa}$ is a specified maximum inter-event discrete time\changed{\footnote{\changed{This parameter often arises naturally in ETC (see \cite{gleizer2018selftriggered}) or can be specified by the user in order to establish a heart beat of the system. It is necessary for STC, in order to impose a finite number of steps to be calculated. It does not hinder stability because it only causes early triggers w.r.t.~PETC.}}}, and $\epsilon \geq 0$ is a margin parameter.\footnote{\changed{When $\epsilon > 0$, Eq.~\eqref{eq:trigger} is called mixed-triggering \cite{borgers2014event}, which is often used in practice to improve sampling performance around the origin. When $\sigma = 0$, it is known as Lebesgue sampling \cite{aastrom2002comparison}.}}%

\changed{Unfortunately, there are no results in the literature for whether the closed-loop PETC system is ISS w.r.t.~measurement noise or a positive value of $\epsilon$. Thus, first we prove that this is the case; i.e.~when the PETC (or any mechanism that triggers earlier) is GES, then it is ISS and $\Ls_p$ stable w.r.t.~additive disturbances, measurement noise, and the $\epsilon$ parameter. These results are relevant not only for the current STC work, but also for PETC.} %

We first model the plant (\ref{eq:linplant}) controlled with (\ref{eq:lincont}) under the PETC triggering rule (\ref{eq:trigger}) \changed{with $\bar{\kappa} = \infty$} as a hybrid system \eqref{eq:hybrid} equipped with a timer, with $\chiv\tran \coloneqq [\xiv_p\tran \ \ \xiv_c\tran \ \ \hat{\psiv}\tran \ \ \hat{\upsilonv}\tran \,]$ and $\deltav\tran \coloneqq [\omegav\tran \ \ \nuv\tran \ \ \epsilon \,]$; the model is
\begin{subequations}\label{eq:impulsive}
\begin{align}
\begin{bmatrix}\dot{\chiv} \\ \dot{\tau}\end{bmatrix} &= \begin{bmatrix}\bar{\Am}\chiv + \bar{\Bm}\omegav \\ 1 \end{bmatrix}, \quad \tau \in [0, h], \label{eq:impulsiveflow}\\
\begin{bmatrix}\chiv^+ \\ \tau^+\end{bmatrix}\! &= \!\!
	\begin{cases}
	{\thickmuskip=2mu \medmuskip=2mu
	    \!\begin{bmatrix}\changed{\Jm_1\chiv\!+\!\Lm\nuv} \\ 0 \end{bmatrix}\!\!, \begin{array}{l} \tau = h,\\ (\bar{\Fm}\chiv\! +\! \bar{\Gm}\nuv)\tran\bar{\Qm}(\bar{\Fm}\chiv\! +\! \bar{\Gm}\nuv)\geq \epsilon^2\end{array}  \begin{array}{c}\text{\!(\ref{eq:impulsive}b)} \\ \text{(\ref{eq:impulsive}c)}\end{array}} \\
	    \!\begin{bmatrix}\Jm_2\chiv \\ 0 \end{bmatrix}\!, \begin{array}{l} \tau = h,\\ (\bar{\Fm}\chiv + \bar{\Gm}\nuv)\tran\bar{\Qm}(\bar{\Fm}\chiv + \bar{\Gm}\nuv)\leq \epsilon^2\end{array}  \,\,\,\,\,\,\, \begin{array}{r}\text{(\ref{eq:impulsive}d)} \\ \text{(\ref{eq:impulsive}e)}\end{array}
	\end{cases}\notag\\
\psiv &= \bar{\Cm}\chiv + \nuv, \label{eq:impulsiveoutput}\tag{\ref{eq:impulsive}f}
\end{align}
\end{subequations}
where
\begin{align}
\!\!\!\bar{\Am}& = \begin{bmatrix}\Ap & \O & \O & \Bp \\ \O & \O & \O & \O \\ \O & \O & \O & \O \\ \O & \O & \O & \O \end{bmatrix}\!\!, \, \bar{\Bm} = \begin{bmatrix} \Em \\ \O \\ \O \\ \O \end{bmatrix}\!\!, \, \bar{\Cm} = \begin{bmatrix}\Cp & \O & \O & \O\end{bmatrix}\!,\!\!\! \notag\\
\!\!\!\Jm_1\!& = \begin{bmatrix}\I & \O & \O & \O \\ \Bc\Cp & \Ac & \O & \O \\ \Cp & \O & \O & \O \\ \changed{\Dc\Cp} & \Cc & \changed{\O} & \O \end{bmatrix}\!\!, \,
\Jm_2 = \!\begin{bmatrix}\I & \O & \O & \O \\ \O & \Ac & \Bc & \O \\ \O & \O & \I & \O \\ \O & \O & \O & \I \end{bmatrix}\!\!, \notag\\
\Lm &= \!\begin{bmatrix}\O \\ \Bc \\ \I \\ \Dc\end{bmatrix}
 \,
\Qbar = \begin{bmatrix} (1 - \sigma^2)\I & -\I \\ -\I & \I \end{bmatrix}\!\!, \label{eq:augmented}\\ 
\bar{\Fm} &= \!\begin{bmatrix} \Cp & \O & \O & \O \\ \changed{\!\Dc\Cp\!} & \Cc & \changed{\O} & \O \\ \O & \O & \I & \O \\ \O & \O & \O & \I\end{bmatrix}\!\!, \, 
\bar{\Gm} = \!\begin{bmatrix}\I \\ \changed{\Dc} \\ \O \\ \O\end{bmatrix}\!\!, \notag
\end{align}
where $\bar{\Qm}$ is partitioned according to $(\zetav, \hat{\zetav})$. The jump map matrices represent the update of input and output ($\Jm_1$) or no update except for the controller state ($\Jm_2$). The quadratic inequalities represent the triggering condition (\ref{eq:trigcond}), where condition (\ref{eq:impulsive}c) is present for the PETC, but absent for an STC that triggers no later than PETC. For absent noise ($\nuv \equiv \O$) and $\epsilon=0$, LMI conditions for verifying stability are available in \cite{heemels2013periodic} for PETC and in \cite{gleizer2018selftriggered} for STC. The main result of this Section is that the system is homogeneous in the sense of Definition \ref{def:homo}, which implies that \changed{it is input-to-state and $\Ls_p$ stable w.r.t.~noise and the $\epsilon$ parameter.}
\changed{\begin{rem}\label{rem:nondet}
	The choice of non-strict inequalities in Eq.~(\ref{eq:impulsive}c) and Eq.~(\ref{eq:impulsive}e) renders the system non-deterministic. This choice was made for mathematical convenience: the proofs using Eq.~\eqref{eq:impulsive} are valid across the non-determinism, and thus cover both choices of making strict either inequality.
\end{rem}}
\begin{lem}\label{lem:petcishomo}
System (\ref{eq:impulsive}) is homogeneous in the sense of Definition \ref{def:homo} \changed{and satisfies Assumption \ref{assum:setpert}.}
\end{lem}
\changed{Homogeneity is trivial; for Assumption \ref{assum:setpert}, the proof is found in Appendix \ref{ap:proof_lemma_homo}.} The following result follows from Theorem \ref{thm:nesic} and Lemma \ref{lem:petcishomo}.
\begin{thm}\label{thm:noiseisok}
	If the system (\ref{eq:linplant}) with controller (\ref{eq:lincont}), using triggering mechanism (\ref{eq:trigger}) (or triggering earlier) is GES when $\omegav \equiv \O, \nuv \equiv \O$ and $\epsilon = 0$, then it is ISS and $\Ls_p$-stable if $\omegav \neq \O, \nuv \neq \O$ and $\epsilon \neq 0$.
\end{thm}
\begin{rem} Lemma \ref{lem:petcishomo} and Theorem \ref{thm:noiseisok} are valid for any quadratic triggering function of the form
	$ \eta(\zetav(t),\hat{\zetav}(t)) = \begin{bmatrix}\zetav(t)\tran & \hat{\zetav}(t)\tran\end{bmatrix}\bar{\Qm}\begin{bsmallmatrix}\zetav(t) \\ \hat{\zetav}(t)\end{bsmallmatrix}, $
as long as $\bar{\Qm}$ renders the closed-loop GES. We focus on the triggering function \eqref{eq:lintrig} because for this case there are design procedures available (e.g., \cite{heemels2013periodic}).

\end{rem}

\section{Self-triggered control implementation}\label{sec:main}

In this section, we devise a method to compute a lower bound of the PETC triggering time $t_{b+1}$ from the available information at $t_b$. This lower bound becomes the STC triggering time. \changed{Throughout this section, we denote $\zv \coloneqq \zetav(t_b)$ and $\uv \coloneqq \upsilonv(t_b)$.} A way of computing such worst-case (earliest) time is by checking, for increasing values of $\kappa \in \N, \kappa \leq \bar{\kappa}$, whether $\eta(\zetav(t_b+h\kappa),\zv)$ can be greater than $\epsilon^2$ given the available information. This leads to the following subproblem:
\begin{sprob}\label{prob:canitcross} Let (supersets of) $\Xs(t_b)$ and $\Ws$ be known. For a given $\kappa \in \{1,...,\bar{\kappa}\}$, determine, in a conservative but computationally efficient way, if \changed{there exist 
	$\xp' \!\in \reach(t_b,t_b+h\kappa, \Xs(t_b), \uv, \Ws)$ and $\vv \in \Es(\O, \Vm)$ such that $ \eta\left(\begin{bmatrix}\Cp\xp'+\vv & \ \upsilonv(t_b+h\kappa)\end{bmatrix}\tran\!\!,\zv\right) > \epsilon^2.$}
\end{sprob}
\changed{In the subproblem above, conservative means that, if the exact answer cannot be established, the answer is assumed to be true. Note that it requires the state set $\Xs(t_b)$, which ideally would be a single point. The larger this set is, the more conservative our solution is. This brings us the following subproblem:}
\begin{sprob}\label{prob:getx} Given a superset of $\Xs_0$, historical values of $\hat{\zetav}$, and $\xic(k)$, determine a small outer-approximation of $\Xs(t_b)$.
\end{sprob}
\changed{In order to use ellipsoidal methods, we assume initial set estimates to be ellipsoids}:
\begin{assum}\label{assum:sets} Matrices $\Xm_0 \in \S^{\np}_{++}, \bar{\Wm}\in\S^{\nw}_{++},$ and $\Vm\in\S^{\ny}_{++}$ are known, such that $\tilde{\Xs}_0 = \Es(\O,\Xm_0) \supseteq \Xs_0, \tilde{\Ws} = \Es(\O,\bar{\Wm}) \supseteq \Ws,$ and $\tilde{\Vs} = \Es(\O,\Vm) \supseteq \Vs$. \end{assum}
Let us solve Subproblems \ref{prob:canitcross} and \ref{prob:getx} recursively. Suppose that, at time $k_b$, an ellipsoid $\Xstilde(k_b|k_{b-1}) \coloneqq \Es(\xiptilde(k_{b-1}),\Xm_{b|b-1}) \owns \xiv_p(hk_b)$ is known. First the state estimate $\Xstilde$ is updated with the newly acquired information $\yv$. That is achieved through the intersection operation in (\ref{eq:generalcorrection}), which returns $\Xstilde(k_b|k_b)$: in this case, $\Xsy(t_b) = \Ks(\yv,\Vm,\Cp)$ and therefore the trace-optimal Fusion in Eq.~(\ref{eq:ellipsoidfusion}) is used.\footnote{Only a scalar parameter needs to be optimized and, since the function is convex, a golden search can be used up to a given precision. Nonetheless, this may be computationally too expensive depending on the application. In that case, a fixed $\lambda$ can be picked, improving computation speed at the expense of larger ellipsoids and more frequent triggering.} From this point, denote the center of the state estimate as $\xptilde\in\R^{\np}$ and its shape matrix as $\changed{\Xm}\in\S_{++}^\np$; thus, $\Xstilde(k_b|k_b) = \Es(\xptilde,\changed{\Xm}).$

We can now compute the reachable sets for the controller and plant states. %
First define the transition matrices:
\begin{subequations}\label{eq:transmat}
	\begin{align}
	\Phip(\dk) &\coloneqq \e^{\Ap h\dk}, &\Gammap(\dk) &\coloneqq \!\int_0^{h\dk}{\!\!\e^{\Ap s}\Bp}\d s, \label{eq:transplant}\\
	\Phic(\dk) &\coloneqq \Ac^{\dk}, &\Gammac(\dk) &\coloneqq \sum_0^{\dk-1}{\Ac^{\dk}\changed{\Bc}}, \label{eq:transcontrol}
	\end{align}
\end{subequations}
Due to linearity, we can separate the reachable set $\Xs(t_b+h\dk|t_b)$ between the contribution of state and control input, and that of the unknown disturbances:
\begin{subequations}\label{eq:linreach}
	\begin{align}
	\!\!\!&\Xstilde(t_b\!+\!h\dk|t_b) \!=\! \Phip(\dk)\Xstilde(k_b|k_b) \!+\! \Gammap(\dk)\uv \!+\! \Xwtilde(\dk),\!\!\!\label{eq:linreachmain}\\
	\!&\Xwtilde(\dk)\supseteq \Xw(\dk) = \!\!\bigcup_{\omegav \in \Fsw}{\int_0^{h\dk}{\!\!\e^{\Ap (h\dk-s)}\Em\omegav(s)}\d s}. \label{eq:distreach}
	\end{align}
\end{subequations}
\begin{rem}
	The computation of supersets $\Xwtilde(\dk) \supseteq \Xw(\dk)$ can be done off-line for all $\kappa\in\{1,...,\bar{\kappa}\}$ using the method described in \changed{Section \ref{ssec:ellipreach}}.
\end{rem}

We are ready to solve Subproblem \ref{prob:canitcross}. Denote $\changed{\Wm(\dk)}$ as the shape matrix of $\Xwtilde(\dk)$, i.e., $\Xwtilde(\dk) \coloneqq \Es(\O,\changed{\Wm(\dk)})$; also, let $\pv\tran \coloneqq [\xp\tran \ \ \xc\tran \ \ \yv\tran\,]$ and
\begin{gather*}
\CE \coloneqq \begin{bmatrix} \O & \O & \I \\ \O & \Cc & \Dc \end{bmatrix}\!,\\
\Nk \coloneqq \begin{bmatrix}\Cp\Phip(\dk) & \Cp\Gammap(\dk)\Cc & \Cp\Gammap(\dk)\Dc \\
\O & \Cc\Phic(\dk) & \Cc\Gammac(\dk) + \Dc\end{bmatrix}.
\end{gather*}
Note that, if there exists $\zv'$ yielding $\eta(\zv',\zv) > \epsilon\changed{^2}$, then $\max_{\zv'}\eta(\zv',\zv) > \epsilon\changed{^2}$. This means that we can pose Subproblem \ref{prob:canitcross} as an optimization problem: %
\begin{sprob}\label{prob:optproblem} \changed{From existing information on the controller, determine the worst-case triggering function value at a given time instant.} That is, given $\xptilde, \Xm, \xc$ and $\yv$, determine, for a given $\kappa$,
	\begin{subequations}\label{eq:optquad}
		\begin{align}
		\max_{\zv',\zv,\xp,\dv,\vv'}{} \quad &\eta(\zv',\zv) = [{\zv'}\tran \ \ \zv\tran\,]\Qbar\begin{bmatrix}\zv' \\ \zv\end{bmatrix} \label{eq:optquadmin}\\
		\mathrm{subject~to} \quad & \zv' = \Nm_\dk\pv + \begin{bmatrix}\vv' \\ \O\end{bmatrix} + \begin{bmatrix}\Cp\dv \\ \O\end{bmatrix}\!, \label{eq:optquadzeta}\\
		& \zv = \CE \,\pv,\label{eq:optquadz}\\
		& (\xp - \xptilde)\tran\changed{\Xm}^{-1}(\xp-\xptilde)\leq 1,\label{eq:optquadx}\\
		& \dv\tran\,\changed{\Wm(\dk)}^{-1}\dv \leq 1, \label{eq:optquadtheta}\\
		& {\vv'}\tran\Vm^{-1}\vv' \leq 1, \label{eq:optquadv}
		\end{align}
	\end{subequations}
\end{sprob}
The decision variables are $\zv'$ representing the possible values of $\zetav(t_b+h\kappa)$; $\zv$; $\xp$ which is the unknown value of $\xip(t_b)$; $\dv$ as the contribution from the unknown disturbances to states at $t_b+h\kappa$; and $\vv'$ as the unknown future noise $\nuv(t_b+h\kappa)$. The objective function (\ref{eq:optquadmin}) is the triggering function and the constraints are: (\ref{eq:optquadzeta}) for the dynamics of $\zetav$; (\ref{eq:optquadz}) as its initial condition; and (\ref{eq:optquadx}), (\ref{eq:optquadtheta}) and (\ref{eq:optquadv}) as the ellipsoidal constraints for the state estimate, $\dv$ and $\vv'$, respectively.
This problem is solved for increasing values of $\dk \in \{1,...,\bar{\kappa}\}$, until one yields a value greater than $\epsilon$. 

\begin{rem}\label{rem:QCQP} Subroblem \ref{prob:optproblem} is a non-convex Quadratically Constrained Quadratic Programming (QCQP) problem. Its constraints are convex but the objective function is non-convex since $\Qbar$ is not definite. \changed{Nevertheless, it is always feasible: one solution is obtained by taking $\dv=\O, \vv=\O, \xp = \xptilde$, and using these values to determine $\zv'$ and $\zv$ in Eqs.~\eqref{eq:optquadzeta} and \eqref{eq:optquadz}.} \end{rem}

The remark above discourages solving the actual optimization problem. Instead, we propose computing a conservative upper bound for it like in \cite{gleizer2018selftriggered}.
Let $\ptilde\tran \coloneqq [\xptilde\tran \ \ \xc\tran \ \ \yv\tran\,]$ be the vector of available information, $\Ns \coloneqq \{1,2,...,\np\}$, and
\begin{gather}
{\thickmuskip=3mu
\Qk \coloneqq {\NkCE\!}\tran\Qbar\NkCE, \ \Cw \coloneqq \begin{bmatrix}\Cp \\ \O\end{bmatrix}, \ \Cv \coloneqq \begin{bmatrix}\I \\ \O\end{bmatrix},} \notag\\
{\thickmuskip=1mu \medmuskip=1mu
\!\Fw(\dk) \coloneqq \NkCEt\! \Qbar \Cw, \ \Fv(\dk) \coloneqq \NkCEt\! \Qbar \Cv,} \notag\\
\Rw(\dk) \coloneqq \Fw(\dk) \changed{\Wm(\dk)} \Fw(\dk)\tran,  \ \Rv(\dk) \coloneqq \Fv(\dk) \Vm \Fv(\dk)\tran, \notag\\
\Qw \coloneqq \Cw\tran\Qbar\Cw, \ \Qv \coloneqq \Cv\tran\Qbar\Cv, \ \cvv \coloneqq \lambda_{\max}(\Vm \Qv),\notag\\
\cvw(\dk) \coloneqq \sqrt{\lambda_{\max}(\Cv\tran\Qbar\Cw\changed{\Wm(\dk)}\Cw\tran\Qbar\Cv\Vm)}. \label{eq:offlinematrices}
\end{gather}
Note that all of the matrices and scalars above can be computed off-line for $\kappa\in\{1,...,\bar{\kappa}\}$. Define the estimate of the triggering function
\begin{multline*}
\bar{\eta}(\kappa,\ptilde,\Xm) \coloneqq \ptilde\tran\Qk\ptilde + 2\sqrt{\ptilde\tran\changed{\Qk|_{\bullet,\Ns}} \Xm \changed{\Qk|_{\bullet,\Ns}\tran}\ptilde}\\ + \lambda_{\max}(\Xm \Qk|_{\Ns,\Ns}) 
+ 2\sqrt{\ptilde\tran\Rv(\dk)\ptilde}\\ + 2\sqrt{\lambda_{\max}(\Rv(\dk)|_{\Ns\!,\Ns}\Xm)} + 2\sqrt{\ptilde\tran\Rw(\dk)\ptilde}\\ + 2\sqrt{\lambda_{\max}(\Rw(\dk)|_{\Ns\!,\Ns}\Xm)}\\ + 2\cvw(\dk) + \cvv + \lambda_{\max}(\changed{\Wm(\dk)}\Qw).
\end{multline*} 
\changed{All eigenvalues in Eq.~\eqref{eq:offlinematrices} and in $\bar{\eta}$ are real, because their arguments are either symmetric matrices or products of symmetric matrices. We have the following result, whose proof is found in Appendix \ref{ap:proof_theorem_bound}}.
\begin{thm}\label{thm:bigbound}
	$\bar{\eta}(\kappa,\ptilde,\Xm)$ provides an upper bound for the solution of Subproblem \ref{prob:optproblem}. That is, $$\bar{\eta}(\kappa,\ptilde,\Xm) \geq \eta(\zv', \zv)$$ for all $\zv',\zv,\xp,\dv,\vv'$ satisfying constraints \eqref{eq:optquadzeta}--\eqref{eq:optquadv}.
\end{thm}
The controller selects $\kappa^* = \inf_{\kappa}\bar{\eta}(\kappa,\ptilde,\changed{\Xm})>\epsilon^2$, if $\bar{\eta}>\epsilon^2$  for some $\kappa  \leq \bar{\kappa}$, or $\kappa^* = \bar{\kappa}$ otherwise. Finally, step (\ref{eq:generalprediction}) of the observer is executed using Eq.~(\ref{eq:linreachmain}). Its operations are the affine transformation $\Gammap(\dk^*)\Xstilde(t_b|t_b) + \Phip(\dk^*)\uv$ followed by a Minkowski sum with $\Xwtilde(\dk^*)$, which is outer-approximated through Eq.~(\ref{eq:ellipsoidmink}).

\begin{figure}[tb]
	\begin{center}
		\input{gse.tex}
		\caption{\label{fig:gse} {Steps of the ellipsoidal GSE in Alg.~\ref{alg}: step 11 (top right), step 12 (bottom left) and step 2 (bottom right).}}
	\end{center}
\end{figure}
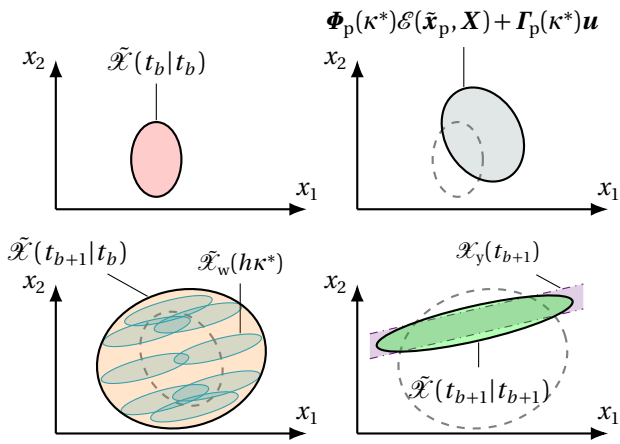

Algorithm \ref{alg} summarizes the steps performed at every instant $k_b$ for both updating the state estimate and computing $\kappa^*$. The operations ``fusion'' and ``minksum'' represent the ellipsoidal outer-approximations from Eqs.~(\ref{eq:ellipsoidfusion}) and (\ref{eq:ellipsoidmink}), respectively. The ellipsoidal GSE (steps 2, 11 and 12) is depicted in Fig.~\ref{fig:gse}.
\begin{algorithm}[tb]\caption{\label{alg}PSTC Algorithm}
\hspace*{\algorithmicindent} \textbf{Input:} $\xc, \yv$ \\
\hspace*{\algorithmicindent} \textbf{Output:} $\uv, \kappa^*$
\begin{algorithmic}[1]
	\State $\uv \gets \Cc\xc + \Dc\yv$
	\State $\Es(\xptilde,\Xm) \!\gets\! \text{fusion}\left(\Es(\xptilde,\Xm),\Ks(\yv,\Vm,\Cp)\right)$ (Eq.~\ref{eq:ellipsoidfusion})
	\State $\ptilde \gets [\xptilde\tran \ \ \xc\tran \ \ \yv\tran\,]\tran$
	\State $\kappa^* \gets 1$
	\While {$\kappa^* < \bar{\kappa}$}
		\If {$\bar{\eta}(\kappa^*,\ptilde,\Xm) > \epsilon^2$}
			\State break
		\EndIf
		\State $\kappa^* \gets \kappa^* + 1$
	\EndWhile
	\State $\Es(\xptilde,\Xm) \gets \Phip(\kappa^*)\Es(\xptilde,\Xm) + \Gammap(\kappa^*)\uv$
	\State $\Es(\xptilde,\Xm) \!\gets\! \text{minksum}(\Es(\xptilde,\Xm),\Es(\O,\Wm_{\kappa^*}\!))\!$ (Eq.~\ref{eq:ellipsoidmink})
\end{algorithmic}
\end{algorithm}

\def\xpbar{\bar{\xv}_\plant}
\begin{rem} For the noiseless case ($\Vm=\O$), we need to modify step 2 of Alg.~\ref{alg}, because in this case the elliptical cylinder $\Ks(\yv,\Vm,\Cp)$ degenerates to a hyperplane. The intersection between an ellipsoid and a hyperplane has an exact ellipsoidal solution (see \cite[Appendix IV]{schweppe1968recursive}). \end{rem}

\changed{
\begin{rem} The complexity of Algorithm \ref{alg} is $\mathcal{O}(\bar{\dk}\allowbreak\max(\np, \nw, \ny)^3)$. It is dominated by the iterative procedure to compute $\bar{\eta}$ (line 6), which involves matrix multiplications and eigenvalue computations on matrices whose sizes depend on $\np, \nw$ and $\nw$.\footnote{\changed{Computing eigenvalues has been proven to have the same big-O complexity as matrix multiplication in \cite{demmel2007fast}. The actual complexity of the matrix multiplication is unknown, the best known being $\mathcal{O}(n^{2.37})$. We chose to use the exponent of 3 because most practical algorithms for small matrices have this complexity.}}
\end{rem}
}

\section{Numerical example\protect\footnote{Code to reproduce this paper's numerical results is available in \protect\url{https://github.com/ggleizer/pstc}.}}

\changed{
Consider the perturbed, unstable linearized batch plant with a PI controller taken from \cite{walsh2001scheduling}\footnote{\changed{The controller was discretized using forward-Euler.}}:
\begin{gather*}
 \Ap = \begin{bmatrix}1.38 & -0.208 &  6.715 & -5.676 \\
						-0.581 & -4.29 &  0     &  0.675 \\
						1.067 &  4.273 & -6.654 &  5.893 \\
						0.048 &  4.273 & 1.343  &-2.104\end{bmatrix}\!, \\
	\Bp = \begin{bmatrix}0 & 0 \\
						 5.679 & 0 \\
						  1.136 & -3.146 \\
				1.136 & 0\end{bmatrix}\!, \
\Cc = \begin{bmatrix}1 & 0 & 1 & -1 \\
                        0 & 1 & 0 & 0\end{bmatrix}\!, \ 
                        \Em = \begin{bmatrix}1 \\ 0 \\ 0 \\ 0\end{bmatrix}, \\ 
\Ac = \begin{bmatrix}1 & 0\\0 & 1\end{bmatrix}\!\!, \ \Bc = \begin{bmatrix}0 & h\\h & 0\end{bmatrix}\!\!, \ \Cc = \begin{bmatrix}-2 & 0\\0 & 8\end{bmatrix}\!\!, \ \Dc = \begin{bmatrix}0 & -2\\5 & 0\end{bmatrix}\!\!,
\end{gather*}
with} $h=0.01$, $\xip(0)=10[1 \ -\!1 \ -\!1 \ 1]\tran$ and $\xic(0) = \O$. The triggering parameter was set to $\sigma = 0.1$. We set $\bar{\kappa} = 25$ and computed $\changed{\Wm(\dk)}$ using the procedure described in \changed{Sec.~\ref{ssec:ellipreach}, with $\Xs_\disturbance(0)$} $= \Es(\O, 10^{-4}\I)$ and $\Ls = \{\cv_i|i \in \{1,2,...,\np\}\}$. The simulated disturbance was the same as the one in \cite{gleizer2018selftriggered}: $\omega(t) = 0.1, \text{ if } t \leq 5; 0$ otherwise. Simulations were run using Matlab R2018a on a MacBook Pro with a 3.1 GHz Intel Core i5 and 8 GB, 2133 MHz LPDDR memory. Noise was simulated through pseudo-random numbers between -0.01 and 0.01, which were pre-generated for all simulation steps with seed 1907. The optimal fusions from Eq.~(\ref{eq:ellipsoidfusion}) were computed with the function \texttt{fminbnd} with default options. We set $\Wm = 0.1^2$ and $\Vm = 2\cdot0.011^2\I$, with the observer starting with $\Xstilde_0 = \R^\np$.

\begin{figure}[tb]
	\begin{center}
		\input{automatica_vs_necsys.tex}
		\caption{\label{fig:quiet} {Simulation results without noise for PSTC, STC from \cite{gleizer2018selftriggered} (GM18-STC), and PETC: state norm $|\xiv(t)|$ (top) and inter-event times $\kappa^*$ (bottom).}}
	\end{center}
\end{figure}
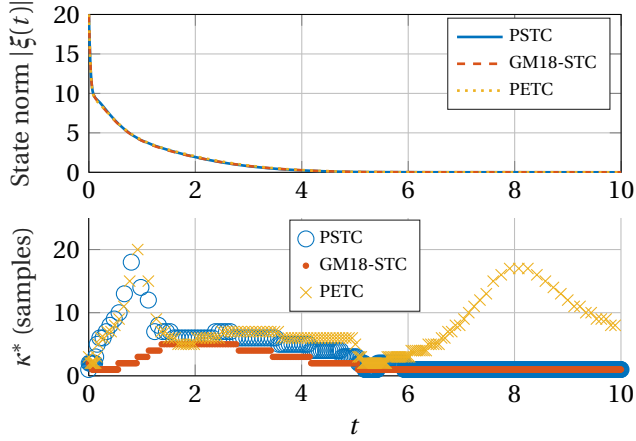

We first simulated the closed-loop STC without noise with $\epsilon = 0$, comparing its control and sampling performances with the method from \cite{gleizer2018selftriggered} and PETC (Fig.~\ref{fig:quiet}). The state norms of all cases converge to zero at virtually the same rate, while, especially at the first two time units, PSTC yields higher sampling times than the STC from \cite{gleizer2018selftriggered}. This improvement is due to the intersection step from Eq.~(\ref{eq:generalcorrection}), which provides faster observer convergence, and to the increased tightness of the disturbance ellipsoids $\Ws_\dk$, when compared to the norm-based bounds of \cite{gleizer2018selftriggered}. Nevertheless, for both STC cases, the triggering times tend to 1 as the state approaches the origin because $\bar{\eta}(\kappa,\O,\Xm) > 0$ for any $\kappa,\Xm$.

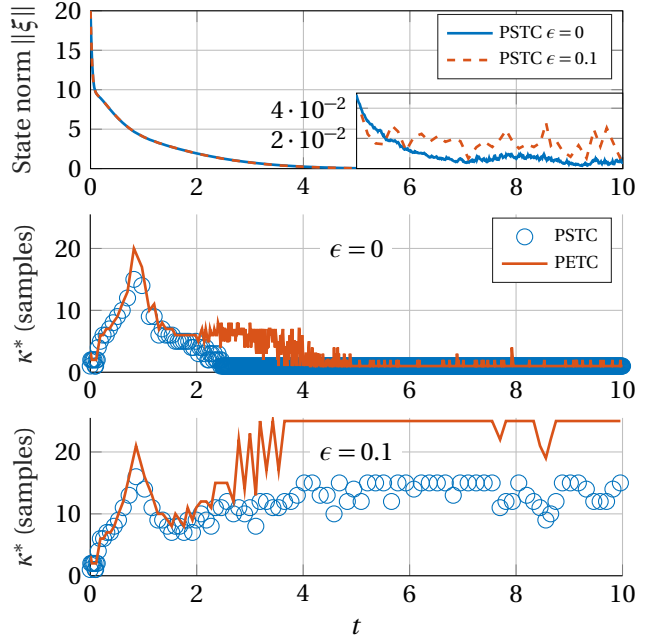
\begin{figure}[tb]
	\begin{center}
		\input{stc_vs_etc_automatica.tex}
		\caption{\changed{\label{fig:noisy} {Simulation results with noise. State norm $|\xiv(t)|$ with PSTC with $\epsilon \in \{0, 0.1\}$ (top); inter-event times $\kappa^*$ from PSTC and PETC with $\epsilon = 0$ (middle) and $\epsilon = 0.1$ (bottom).}}}
	\end{center}
\end{figure}

For the scenario with measurement noise, Fig.~\ref{fig:noisy} (top) displays the triggering times from PSTC. These are compared to the times triggered by the PETC logic (\ref{eq:lintrig}) at each PSTC step. As expected, the PSTC times constitute lower bounds for the PETC ones. It is also clear how the sampling performances of both PSTC and PETC are affected by the noise: as the inputs get close enough to zero, noise alone can provoke a trigger. Due to that, we also simulated a case with $\epsilon = 0.1$, depicted in the bottom plot of Fig.~\ref{fig:noisy}. The resulting triggering times got significantly higher at a small cost in steady state error.%

The on-line CPU time statistics of Alg.~\ref{alg} are displayed in Table \ref{table}. These numbers were obtained for the case with noise with $\epsilon=0$, after ten consecutive runs of the main script to mitigate the overhead from, e.g., just-in-time compilation and process management of the operating system. The initialization step time (Appendix \ref{ap:init}) was 0.03 ms. The figures show that the computations are fast, despite involving an optimization step for the fusion. The most expensive step was the calculation of $\bar{\eta}$, mainly due to the computation of eigenvalues and matrix multiplications.  
The off-line computations totaled 623.46 ms, out of which 609.26 ms were spent on the reachability ($\changed{\Wm(\dk)}$) and 14.19 ms on the remaining matrices and scalars (Eq.~\ref{eq:offlinematrices} and the ones in Appendix \ref{ap:init}).

\begin{table}[htb]
\begin{center}\caption{\label{table}CPU times of Alg.~\ref{alg} for the numerical example.}
	\begin{footnotesize}
\begin{tabular}{|c|c|c|c|}
	\hline 
	& \multicolumn{3}{c|}{\bfseries Time (ms)} \\ 
	\hline 
	\bfseries Phase (line(s) in Alg.~\ref{alg}) & \bfseries Min. & \bfseries Mean & \bfseries Max. \\ 
	\hline 
	Fusion (line 2) & 0.39 & 0.49 & 1.71 \\ 
	Calculation of $\bar{\eta}$ (line 6) & 0.50 & 0.60 & 1.90 \\ 
	Prediction (lines 11 and 12) & 0.02 & 0.02 & 0.08 \\ 
	\bfseries Full PSTC cycle & \bfseries 1.01 & \bfseries 1.27 & \bfseries 8.49 \\ 
	\hline 
\end{tabular} 
\end{footnotesize}
\end{center}
\end{table}

\begin{rem}
	Qualitatively comparing with \cite{brunner2019event}, the issue of eventually triggering always when $\epsilon = 0$ also happens with $\theta=1$ and $\gamma=1$ in their STC. In this setting, one would achieve UGAS of the minimal robustly positive invariant subset associated with periodic control with disturbances. Increasing $\theta$ and $\gamma$ enlarges the terminal set, in a similar way $\epsilon>0$ does.
\end{rem}

\section{Conclusions}

We presented a self-triggered strategy for output-feedback control of linear systems subject to bounded disturbances and noise, named PSTC.
It is, to our knowledge, the first self-triggered implementation of such a general control structure, improving the results and broadening the applicability of \cite{gleizer2018selftriggered}. We first proved that the introduction of noise or mixed triggering does not hinder stability of neither PETC nor PSTC, then developed an algorithm that uses set-based methods for a viable self-triggered implementation. PSTC achieves virtually the same control performance as PETC, with slightly smaller inter-sample times. It is expected to be fast enough for most applications, as each step CPU time averaged 1 ms for the simulated four-state plant; and it scales well with the state-space dimension, since the few online optimization and line search operations are done on scalars, while higher-dimension computations are handled with simple linear algebra.

PSTC was developed for linear plants with linear controllers, which presents a limitation to its applicability. Some classes of nonlinearities could be handled by considering them as disturbances; since we assume that they are bounded, one would have to determine a compact set on which the states lie in order to compute the proper bounds. For locally linearizable systems, other types of unknown-but-bounded uncertainty descriptions are more suitable, such as parametric model uncertainty. In this case, the ellipsoidal estimator in \cite{el2001robust} could be used as a starting point. %
There are also opportunities for improving the PSTC performance for linear systems. Aiming at a small computation complexity, we chose ellipsoids as set descriptors and devised simple upper bounds to the solution of online non-convex QCQP problems; however, these choices probably bring additional conservatism and hence increased communication frequency. From our simulations, this seems to be particularly relevant when the state approaches the origin and when disturbances are significantly smaller than their estimated bounds. A few alternatives might reduce conservativeness: for example, (constrained) zonotopes \cite{scott2016constrained} could replace ellipsoids; note, however, that this would require reformulating the optimization problem. Another possibility would be deriving tighter bounds for the non-convex QCQP. %
Finally, the methods proposed in this paper are not restricted to STC. For example, we are extending this work to ETC communication scheduling, by employing the PSTC algorithm as a generator of triggering times' lower bounds.

\bibliographystyle{plain}        
\bibliography{mybib}           

\appendix
\section{Proof of Lemma \ref{lem:petcishomo}}\label{ap:proof_lemma_homo}

	\changed{Before approaching the proof, one remark must be made: system \eqref{eq:impulsive} is equipped with a timer, with jumps only occurring after a certain time; this specializes it to what is defined in \cite[Section 5]{nesic2013finite} as a system with average dwell time, with $N = 1, \delta = 1/h,$ and $\zeta$ arbitrarily small. This actually relaxes the Lyapunov stability conditions presented therein \cite[Proposition 2]{nesic2013finite}. Theorem 2 of \cite{nesic2013finite} states that homogeneous systems with average dwell time satisfy Theorem \ref{thm:nesic} with $\psiv = \chiv.$ Remark 16 of \cite{nesic2013finite} argues that the same Propositions that build the proof of \cite[Theorem 2]{nesic2013finite} can be derived with $\psiv \neq \chiv$. Thus, the timer does not play a significant role in our proofs; as an additional benefit, the results without timer can be applied to continuous ETC.}
	
	For analysis purposes, even though $\epsilon$ is a design parameter, we can treat it as a disturbance on the jump set. With that, let $n\coloneqq\np+\nc+\nup+\ny$ \changed{and the collected vector of exogenous signals $\dv\tran \coloneqq \begin{bmatrix}\wv\tran & \vv\tran & \epsilon\end{bmatrix}$, giving $\nd\coloneqq\nw+\ny+1.$} The flow sets are $\Cs = \R^n$ and $\Cs_0 = \R^{n+\nd}$, and the jump sets are 
	\changed{
	\begin{gather}
	\begin{multlined}[t][0.8\linewidth]\label{eq:d1}
	\Ds_1 = \{\xv \in \R^n, [\wv\tran \ \vv\tran \ \epsilon]\tran \in \R^\nd: \\ (\bar{\Fm}\xv + \bar{\Gm}\vv)\tran\bar{\Qm}(\bar{\Fm}\xv + \bar{\Gm}\vv)\geq \epsilon^2\},
	\end{multlined}\\ 
	\begin{multlined}[t][0.8\linewidth]\label{eq:d2}
	\Ds_2 = \{\xv \in \R^n, [\wv\tran \ \vv\tran \ \epsilon]\tran \in \R^\nd: \\ (\bar{\Fm}\xv + \bar{\Gm}\vv)\tran\bar{\Qm}(\bar{\Fm}\xv + \bar{\Gm}\vv)\leq \epsilon^2\},
	\end{multlined}
	\end{gather}
	and their projections with $\dv=\O$ are
	\begin{gather}
	\Ds_{10} = \{\xv \in \R^n: \xv\tran\bar{\Fm}\tran\bar{\Qm}\bar{\Fm}\xv \geq 0\},\nonumber\\ \Ds_{20} = \{\xv \in \R^n: \xv\tran\bar{\Fm}\tran\bar{\Qm}\bar{\Fm}\xv \leq 0\}.\nonumber
	\end{gather}
	}
	Since sets $\Ds_{i0}$ are conic and the flow and jump maps in \eqref{eq:impulsive} are linear, properties (\ref{eq:fghomo}) and (\ref{eq:setshomo}) hold; also, condition (\ref{eq:flowsetbound}) is trivially satisfied because $\Cs$ and $\Cs_0$ are the entire Euclidean space. 
	
	What remains to be verified is condition (\ref{eq:jumpsetbound}). Note that the only components of $\dv$ that enter the jump sets are $\vv$ and $\epsilon$. 
	Rewriting the set sum on the LHS of (\ref{eq:jumpsetbound}) gives
	\begin{multline*}
	\Ds_{i0} + \LD\Bs(\changed{|\dv|}) = \{\xv'+\xv'': \xv'\in\Ds_{i0}, \xv''\in\LD\Bs(|\dv|)\} \\
	= \{\xv: (\xv-\xv'')\tran\bar{\Fm}\tran\bar{\Qm}\bar{\Fm}(\xv-\xv'') \sim_i 0, \\{\xv''}\tran\xv'' \leq \LD^2(\vv\tran\vv+\epsilon^2)\},
	\end{multline*}
	where $\sim_1$ is $\geq$ and $\sim_2$ is $\leq$. Thus, \eqref{eq:jumpsetbound} can be restated as
	\begin{multline}\label{eq:newjumpcond}
	\!\!\!\!\forall \xv\in\R^n, \vv\in\R^\nv\!, \epsilon\in\R\!: (\bar{\Fm}\xv +\bar{\Gm}\vv)\tran\bar{\Qm}(\bar{\Fm}\xv + \bar{\Gm}\vv) \sim_i \epsilon^2\!,\\
	\exists \xv'' \in \R^n: (\xv-\xv'')\tran\bar{\Fm}\tran\bar{\Qm}\bar{\Fm}(\xv-\xv'') \sim_i 0, \\ {\xv''}\tran\xv''\leq \LD^2(\vv\tran\vv+\epsilon^2).
	\end{multline}
	Since the pair $(\Ap,\Cp)$ is observable, we can assume system (\ref{eq:linplant}) is in its canonical observable form; thus, taking $\Cp = \begin{bmatrix}\I & \O\end{bmatrix}$, we can partition $\bar{\Fm}$ as
	$$ \bar{\Fm} = \begin{bmatrix}\begin{array}{c|cccc}
	\I & \O & \O & \O & \O \\
	\changed{\Dc} & \O & \Cc & \changed{\O} & 0 \\
	\O & \O & \O & \I & \O \\
	\O & \O & \O & \O & \I
	\end{array}\end{bmatrix} = \begin{bmatrix}\begin{array}{c|c}\bar{\Gm} & \bar{\Hm}\end{array}\end{bmatrix},
	$$
	where $\xv\tran$ is partitioned accordingly as $\begin{bmatrix}\yv\tran & \bar{\xv}\tran\end{bmatrix}$, with $\bar{\xv}$ containing all the remaining state components, \changed{obtaining $\bar{\Fm}\xv = \bar{\Gm}\yv + \bar{\Hm}\bar{\xv}$.} We now divide the proof in two parts: $i=1$ and $i=2$.%
	
	\changed{To show \eqref{eq:newjumpcond} for $i=1$, let us construct one $\xv''$ that satisfies it for every $\xv, \vv, \epsilon:$	take} ${\xv''}\tran = \begin{bmatrix}-\vv\tran & \O\end{bmatrix}$. Then obviously ${\xv''}\tran\xv'' = \vv\tran\vv \leq \changed{\LD^2}(\vv\tran\vv+\epsilon^2)$ with $L_D=1$ and
	\begin{multline*}
	(\xv-\xv'')\tran\bar{\Fm}\tran\bar{\Qm}\bar{\Fm}(\xv-\xv'') = (\bar{\Fm}(\xv-\xv''))\tran\bar{\Qm}\bar{\Fm}(\xv-\xv'')\\
	= \left(\bar{\Fm}\begin{bmatrix}\yv+\vv \\ \bar{\xv}\end{bmatrix}\right)\tran \bar{\Qm} \left(\bar{\Fm}\begin{bmatrix}\yv+\vv \\ \bar{\xv}\end{bmatrix}\right)\\
	= (\bar{\Gm}(\yv + \vv) + \bar{\Hm}\bar{\xv})\tran\bar{\Qm}(\bar{\Gm}(\yv + \vv) + \bar{\Hm}\bar{\xv})\\
	= (\bar{\Fm}\xv + \bar{\Gm}\vv)\tran\bar{\Qm}(\bar{\Fm}\xv + \bar{\Gm}\vv)\geq \epsilon^2 \geq 0.
	\end{multline*}

	\changed{Showing \eqref{eq:newjumpcond} for $i=2$ is slightly more involved.} First, notice the following fact:
	\begin{equation}\label{eq:FQFneg}
	\lambda_{\min}(\bar{\Fm}\tran\bar{\Qm}\bar{\Fm}) < 0.
	\end{equation}
	This is true because $\xv\tran\bar{\Fm}\tran\Qbar\bar{\Fm}\xv$ is just another representation of the triggering function \eqref{eq:lintrig}; thus, it can be expressed as $|\zv-\hat{\zv}|^2-\sigma^2|\zv|^2$ for some $\zv, \hat{\zv} \in \R^{\ny+\nup}$. This expression is negative if, e.g., $\zv = \hat{\zv} \neq \O$.%
	
	Again, let us construct one $\xv''$ that satisfies \eqref{eq:newjumpcond} for every $\xv, \vv, \epsilon.$ This is ${\xv''}\tran = \begin{bmatrix}-\vv\tran & \O\end{bmatrix} + \qv\tran$, where $\qv$ is the vector along the eigendirection corresponding to $\lambda_{\min}(\bar{\Fm}\tran\bar{\Qm}\bar{\Fm})$, \changed{i.e.~, $\bar{\Fm}\tran\bar{\Qm}\bar{\Fm}\qv = \lambda_{\min}(\bar{\Fm}\tran\bar{\Qm}\bar{\Fm})\qv$, satisfying}
	\begin{gather}
	(\bar{\Gm}(\yv + \vv) + \bar{\Hm}\bar{\xv})\tran\bar{\Qm}\bar{\Fm}\qv \geq 0, \label{eq:qsign}\\
	\norm[\qv]^2 = \norm[\lambda_{\min}(\bar{\Fm}\tran\bar{\Qm}\bar{\Fm})]^{-1}\epsilon^2. \label{eq:qnorm}
	\end{gather}
	One can always find such $\qv$: \eqref{eq:qnorm} determines its norm; and, if \eqref{eq:qsign} is not satisfied, $-\qv$ satisfies it. This gives
	\begin{equation}\label{eq:qFQFq}
	\qv\tran\bar{\Fm}\tran\bar{\Qm}\bar{\Fm}\qv = \frac{\lambda_{\min}(\bar{\Fm}\tran\bar{\Qm}\bar{\Fm})}{\norm[\lambda_{\min}(\bar{\Fm}\tran\bar{\Qm}\bar{\Fm})]}\epsilon^2 = -\epsilon^2,
	\end{equation}
	\changed{where the negative sign comes from Eq.~\eqref{eq:FQFneg}}. Therefore, the \changed{second inequality in \eqref{eq:newjumpcond}} satisfies
	\begin{multline*}%
	(\xv-\xv'')\tran\bar{\Fm}\tran\bar{\Qm}\bar{\Fm}(\xv-\xv'')\\ =(\bar{\Gm}(\yv + \vv) + \bar{\Hm}\bar{\xv})\tran\bar{\Qm}(\bar{\Gm}(\yv + \vv) + \bar{\Hm}\bar{\xv}) \\ -2(\bar{\Gm}(\yv + \vv) + \bar{\Hm}\bar{\xv})\tran\bar{\Qm}\bar{\Fm}\qv + \qv\tran\changed{\bar{\Fm}}\tran\bar{\Qm}\bar{\Fm}\qv \\
	\stackrel{\eqref{eq:d2}}{\leq} \epsilon^2 -  2(\bar{\Gm}(\yv + \vv) + \bar{\Hm}\bar{\xv})\tran\bar{\Qm}\bar{\Fm}\qv + \qv\tran\bar{\Fm}\tran\bar{\Qm}\bar{\Fm}\qv\\
	\stackrel{\eqref{eq:qsign},\eqref{eq:qFQFq}}{\leq} \epsilon^2 - \epsilon^2 = 0.
	\end{multline*}
	Additionally, the norm of $\xv''$ satisfies
    $$
    \norm[\xv''] \leq \norm[\vv]+\norm[\qv] = \norm[\vv]+\norm[\lambda_{\min}(\bar{\Fm}\tran\bar{\Qm}\bar{\Fm})\changed{^{-\frac{1}{2}}}]\norm[\epsilon] \leq L(\norm[\vv] + \norm[\epsilon]),
    $$
    for $L \coloneqq \max\left(1,\norm[\lambda_{\min}(\bar{\Fm}\tran\bar{\Qm}\bar{\Fm})]^{-\frac{1}{2}}\right)$. Now, it is easy to see that
    $$ (\norm[\vv] + \norm[\epsilon])^2 \leq 2\vv\tran\vv + 2\epsilon^2. $$ 
    Hence, ${\xv''}\tran\xv''\leq \LD^2(\vv\tran\vv+\epsilon^2)$ holds with $L_D=\sqrt{2}L.$
\section{Proof of Theorem \ref{thm:bigbound}}\label{ap:proof_theorem_bound}
First, we introduce the following Lemma:
\begin{lem}\label{lem:xdecomp}
Let $\Mm \in \S^n_{+}$. Then, for any $\xv\in\R^n$ such that $\xv\in\Es(\O,\Mm)$, there exist a vector $\sv$ with $\norm[\sv] \leq 1$ and a matrix $\Sm$ such that $\xv = \Sm\sv$ and $\Sm\Sm\tran=\Mm$.
\end{lem}
\begin{proof} 
	Since $\Mm$ is symmetric, it admits the singular value decomposition
	$$ \Mm = \Um\tran\begin{bmatrix}\Dm & \O \\ \O & \O\end{bmatrix}\Um,$$
	with $\Um$ invertible and $\Dm\in\S_{++}$ diagonal. From Definition \ref{def:ell}, it must hold that, for all $\lv\in\R^n$,
	\begin{equation}\label{eq:boundpf1}
	\lv\tran\xv \leq (\lv\tran\Mm\lv)^{1/2} = \left(\lv\tran\Um\tran\begin{bmatrix}\Dm & \O \\ \O & \O\end{bmatrix}\Um\lv\right)^{1/2}.
	\end{equation}
	Take $\lv' \coloneqq \Um\lv$ and $\sv' \coloneqq \Um\traninv\xv.$ Then, \eqref{eq:boundpf1} becomes
	\begin{equation}\label{eq:boundpf2}
	{\lv'}\tran\sv' \leq \left({\lv'}\tran\begin{bmatrix}\Dm & \O \\ \O & \O\end{bmatrix}\lv'\right)^{\!\!1/2} \!\!\!= ({\lv'_1}\tran\Dm{\lv'_1})^{1/2},
	\end{equation}
	where $\lv'$ is partitioned into $\begin{bmatrix}{\lv'_1}\tran & {\lv'_2}\tran\end{bmatrix}\tran$ according to $\begin{bsmallmatrix}
	\Dm & \O \\ \O & \O
	\end{bsmallmatrix}.$ Likewise, partition $\sv'$ into $\begin{bmatrix}{\sv'_1}\tran & {\sv'_2}\tran\end{bmatrix}\tran.$ Then, \eqref{eq:boundpf2} becomes
	$$ {\lv_1'}\tran\sv_1' + {\lv_2'}\tran\sv_2' \leq ({\lv'_1}\tran\Dm{\lv'_1})^{1/2},$$
	which, to hold for all $\lv'_1$ and$\lv'_2$, requires that $\sv_2'=0$. As ${\lv_1'}\tran\sv_1' \leq ({\lv'_1}\tran\Dm{\lv'_1})^{1/2}$ is the definition of the ellipsoid $\Es(\O,\Dm)$, we also conclude that ${\sv'_1}\tran\Dm^{-1}\sv'_1 \leq 1.$ Finally, the choice $\sv = \Dm^{-1/2}\sv'_1$ satisfies ${\sv}\tran\sv \leq 1$. Moreover,
	$$ \Um\traninv\xv = \begin{bmatrix}\sv_1' \\ \sv_2'\end{bmatrix} = \begin{bmatrix}\Dm^{1/2} \\ \O\end{bmatrix}\sv \iff \xv = \Um\tran\begin{bmatrix}\Dm^{1/2} \\ \O\end{bmatrix}\sv, $$
	so, $\Sm = \Um\tran\begin{bmatrix}\Dm^{1/2} \\ \O\end{bmatrix}$ gives $\xv=\Sm\sv$ and
	$ \Sm\Sm\tran = \Mm. $
\end{proof}
With the result above, the following Lemma introduces some useful bounds:
\begin{lem}\label{lem:bounds}
	Let $\Mm_i \in \S^n_{+}, i\in\{1,2\},$ $\pv \in \R^m,\Fm\in\R^{n\times m},$ and $\Qm\in\S^{n}$. For any $\xv_i\in\R^n$ such that $\changed{\xv_i}\in\Es(\O,\Mm_i)$, the following inequalities hold:
	\begin{subequations}\label{eq:bounds}
		\begin{align}
		\pv\tran\Fm\xv_i &\leq \sqrt{\pv\tran\Fm\Mm_i\Fm\tran\pv}, \label{eq:pFx}\\
		\xv_i\tran\Qm\xv_i &\leq \lambda_{\max}(\Mm_i\Qm), \label{eq:xQx}\\
		\xv_1\tran\Fm\xv_2 &\leq \sqrt{\lambda_{\max}(\Fm\Mm_2\Fm\tran\Mm_1)}. \label{eq:x1Fx2}
		\end{align}  
	\end{subequations}
\end{lem}
\begin{proof}	
	Using Lemma \ref{lem:xdecomp}, take $\sv_i,\Sm_i$ satisfying $\Sm_i\Sm_i\tran = \Mm_i$ and $\xv_i = \Sm_i\sv_i$ such that $\norm[\sv_i] \leq 1$. Thus, $\pv\tran\Fm\xv_i = \pv\tran\Fm\Sm_i\sv_i \leq \norm[\pv\tran\Fm\Sm_i]$; $\xv_i\tran\Qm\xv_i = \sv_i\tran\Sm_i\tran\Qm\Sm_i\sv_i \leq \lambda_{\max}(\Sm_i\tran\Qm\Sm_i)$; and $\xv_1\tran\Fm\xv_2 = \sv_1\tran\Sm_1\tran\Fm\Sm_2\sv_2 \leq |\Sm_1\tran\Fm\Sm_2| = \sqrt{\lambda_{\max}(\Sm_1\tran\Fm\Sm_2\Sm_2\tran\Fm\tran\Sm_1)}$. Using the fact that $\lambda(\Am\Bm) = \lambda(\Bm\Am)$ for any $\Am,\Bm\in\R^{n\times n}$ and replacing $\Sm_i\Sm_i\tran$ with $\Mm_i$ provides (\ref{eq:bounds}).
\end{proof}	

Now we can proceed to the proof of Theorem \ref{thm:bigbound}:

\begin{proof} Let $\ev \coloneqq \xp - \xptilde$. Hence,
\begin{align}
&\pv = \ptilde + [\ev\tran \ \ \O \ \ \O]\tran\,\label{eq:ptilde}
\end{align}
and, from \eqref{eq:optquadx}, $\ev\tran\changed{\Xm}^{-1}\ev \leq 1.$ Rewrite Eq.~(\ref{eq:optquadmin}) as a function of $\ptilde, \ev, \dv,$ and $\vv'$ by replacing $\zv',\zv$ and $\pv$ from Eqs.~(\ref{eq:optquadzeta}), (\ref{eq:optquadz}) and (\ref{eq:ptilde}):
\begin{multline*}
\eta(\zv',\zv)=\eta'(\kappa,\ptilde,\ev,\vv',\dv) = \\
\ppluset\Qk\ppluse + \\
2\ppluset\Fv(\dk)\vv' + 2\ppluset\Fw(\dk)\dv + \\
2{\vv'}\tran\Cv\tran\Qbar\Cw\dv + {\vv'}\tran\Qv\vv' + \dv\tran\Qw\dv,
\end{multline*}
Which results in
\begin{multline}\label{eq:etafull}
\eta'(\kappa,\ptilde,\ev,\vv',\dv) = \ptilde\tran\Qk\ptilde + 2\ptilde\tran\Qk|_{\bullet,\Ns}\ev \\ + \ev\tran\Qk|_{\Ns,\Ns}\ev +
2\ptilde\tran\Fv(\dk)\vv' + 2\ev\tran\Fv(\dk)|_{\Ns,\bullet}\vv' \\ + 2\ptilde\tran\Fw(\dk)\dv +  2\ev\tran\Fw(\dk)|_{\Ns,\bullet}\dv \\ +
2{\vv'}\tran\Cv\tran\Qbar\Cw\dv + {\vv'}\tran\Qv\vv' + \dv\tran\Qw\dv.
\end{multline}
Now Lemma~\ref{lem:bounds} is used. The only known term in Eq.~(\ref{eq:etafull}) is the first. Eq.~(\ref{eq:pFx}) is used for second, fourth and sixth terms; Eq.~(\ref{eq:xQx}) for the third, ninth and tenth; and Eq.~(\ref{eq:x1Fx2}) for the fifth, seventh and eighth terms. Mere replacement provides $\bar{\eta}(\kappa,\ptilde,\Xm)$.
\end{proof}

\section{Observer Initialization}\label{ap:init}    
\changed{For Assumption \ref{assum:sets} to hold, we need to construct a bounded set $\Xstilde$ containing the initial state. Fortunately, this can be achieved for our class of systems in a finite number of steps, as detailed in this Section. During these first few steps, the PSTC must trigger periodically with $\kappa^* = 1$. The construction of $\Xstilde$ requires the following:}
\begin{assum}\label{assum:obsdiscrete} The matrix $\Phip(1)$ is invertible and the pair $(\Phip(1),\Cp)$ is observable. \end{assum}
This is not a limiting assumption: one can always find $h$ such that $\Phip(1) = \e^{\Ap h}$ is invertible.\footnote{For $h=0$, $\e^{\Ap h}=\I$; from continuity, $\e^{\Ap h} \approx \I$ for small enough values of $h$, hence it is invertible.} Likewise, since the pair $(\Ap,\Cp)$ is observable, so is $(\Phip(1),\Cp)$ with the proper selection of $h$.\footnote{See \cite[Sec.~6.8]{gopal1993modern} for the pathological selections of $h$ for which it does not hold.} For compactness of expressions, denote $\Phip(1)$ as $\Phip$ \changed{and $\Gammap(1)$ as $\Gammap$} throughout the rest of this Appendix.

Instead of following the standard recursive GSE, which would require Minkowski sums of unbounded sets,\footnote{There are tools for that, but it is both unnecessary and computationally inefficient to do so. During the initialization, the STC has to trigger periodically, hence there is no advantage in keeping track of the best state estimate.} we collect sets relating the current state to each specific measurement up to a certain instant, then compute an intersection outer-approximation. Let $\Om(k)$ be the observability matrix for $k+1$ instants:
$$ \Om(k) \coloneqq \begin{bmatrix}\Cp \\ \Cp\Phip \\ \vdots \\ \Cp\Phip^k\end{bmatrix}. $$
Denote $\bar{k} \coloneqq \inf_{k\in\No}\rank(\Om(k)) = \np$. This is the number of steps needed to reconstruct the initial state on linear systems. We will see that it is also the minimum number of steps for getting a bounded set estimate from measurements with bounded noise. For now, denote $\deltav(k_1,k_2) \coloneqq \int_{hk_1}^{hk_2}{\e^{\Ap (hk_2-s)}\Em\omegav(s)}\d s$ as the contribution of disturbances to state from $k_1$ to $k_2$, and let $\tilde{\psiv}(k,\bar{k}) \coloneqq \psiv(hk) + \Cp\sum_{j=k}^{\bar{k}-1}\Phip^{k-1-j}\Gammap\changed{\hat{\upsilonv}}(hj)$ \changed{be the prediction of the output at time $\bar{k}$ from the output at $k$ and inputs from $k$ to $\bar{k}-1$.} The following holds:
\begin{lem}\label{lem:cylinders}
	Consider system \eqref{eq:linplant},\eqref{eq:lincont} with $b=k$ (periodic triggering), and let Assumption \ref{assum:sets} hold. Then, for all $k\leq\bar{k},$ 
	$$\Cp\Phip^{k-\bar{k}}\xip(h\bar{k}) \in \Es(\tilde{\psiv}(k,\bar{k}),\!\Vm) + \Cp\Phip^{k-\bar{k}}\Xwtilde(\bar{k}-k).$$
\end{lem}
\begin{proof} We can assess the contribution of the information $\psiv(hk), k \leq \bar{k}$ to the instant $\bar{k}$ in a similar manner to Eq.(\ref{eq:linreach}):
\begin{equation*}
\xip(h\bar{k}) = \Phip^{\bar{k}-k}\xip(hk) +\sum_{j=k}^{\bar{k}-1}\Phip^{\bar{k}-1-j}\Gammap\hat{\upsilonv}(hj) + \deltav(k,\bar{k}),
\end{equation*}
which implies, if $\Phip$ is invertible,
\begin{multline}\label{eq:initupdate}
\Cp\Phip^{k-\bar{k}}\xip(h\bar{k}) = \Cp\xip(hk)\\ +\Cp\sum_{j=k}^{\bar{k}-1}\Phip^{k-1-j}\Gammap\changed{\hat{\upsilonv}}(hj) + \Cp\Phip^{k-\bar{k}}\deltav(k,\bar{k}).
\end{multline}
Since $\Cp\xip(hk) = \psiv(hk) - \nuv(hk),$ it belongs to the input uncertainty set $\Es(\psiv(kh),\!\Vm)$, which after summing with the contribution from inputs $\Cp\sum_{j=k}^{\bar{k}-1}\Phip^{k-1-j}\Gammap\hat{\upsilonv}(hj)$ yields $\Es(\tilde{\psiv}(k,\bar{k}),\!\Vm)$. The remaining term is the contribution from disturbances after $\bar{k}-k$ steps, which belongs to $\Xwtilde(\bar{k}-k)$, followed by the linear transformation through $\Cp\Phip^{k-\bar{k}}$.
\end{proof}
Denote the outer-approximation (Eq.~\ref{eq:ellipsoidmink}) of the Minkowski sum in Lemma \ref{lem:cylinders} as $\Es(\tilde{\psiv}(k,\bar{k}),\!\tilde{\Vm}(k))$. From Definition \ref{def:cyl}, if $\Cp\Phip^{k-\bar{k}}\xip(h\bar{k}) \in \Es(\tilde{\psiv}(k,\bar{k}),\!\tilde{\Vm}(k))$, then $\xip(h\bar{k}) \in \Ks(\tilde{\psiv}(k,\bar{k}),\!\tilde{\Vm}(k),\Cp\Phip^{k-\bar{k}}) \changed{= \Xstilde(\bar{k}|k).}$ \changed{That is, we have found the elliptical cylinder that contains $\xip(h\bar{k})$ given information at $k$. Since this is true for all $k \in \{0,1,..\bar{k}\}$, we have that}
\begin{equation}\label{eq:cylintersection}
\xip(h\bar{k}) \in \bigcap_{k=0}^{k\leq\bar{k}}\Xstilde(\bar{k}|k).
\end{equation}
\changed{An ellipsoidal outer-approximation of this intersection of elliptical cylinders can be derived with the following:}
\begin{lem}\label{lem:ellipfromcyls} Let $\Cm_i \in \R^{m\times n}, \Mm_i \in \S_{++}^n, \yv_i \in \R^m, i\in\{1,...,q\}$. Denote $\bar{\Cm} \coloneqq [\Cm_1\tran \quad \Cm_2\tran \quad \cdots \quad \Cm_q\tran\,]$ and assume $\rank(\bar{\Cm}) = n$. Denote $\bar{\yv}\tran \coloneqq [\yv_1\tran \quad \yv_2\tran \quad \cdots \quad \yv_q\tran\,]$ and 
	$$ \bar{\Mm} \coloneqq \begin{bmatrix}\frac{1}{\mu_1}\Mm_1\! & \O & \cdots & \O \\
	\O & \!\frac{1}{\mu_2}\Mm_2\! & \cdots & \O \\
	\vdots &  \vdots & \ddots & \vdots \\
	\O & \O & \cdots & \!\frac{1}{\mu_q}\Mm_q\end{bmatrix}, $$
with $\sum_{i=1}^q\mu_i = 1$.
Then, %
\changed{$$\cap_i\Ks(\yv_i,\Mm_i,\Cm_i) \subseteq \Es(\bar{\Cm}^\dagger\bar{\yv},{\bar{\Cm}}^\dagger\bar{\Mm}{\bar{\Cm}^\dagger}{}\tran).$$}
\end{lem}

\begin{proof} \changed{The intersection means that $(\Cm_i\xv - \yv_i)\tran\changed{\Mm_i}^{-1}(\Cm_i\xv - \yv_i) \leq 1$ for all $i$; thus,} it holds that $\sum_{i=1}^q\lambda_i(\Cm_i\xv - \yv_i)\tran\changed{\Mm_i}^{-1}(\Cm_i\xv - \yv_i) \leq \sum_{i=1}^q\lambda_i$ for any $\lambda_i > 0$. Divide both sides by $\sum_{i=1}^q\lambda_i$ and denote $\mu_i = \lambda_i/(\sum_{i=1}^q\lambda_i)$. Putting in matrix form,
$$ (\bar{\Cm}\xv-\bar{\yv})\tran 
	\begin{bmatrix}\mu_1\Mm_1^{-1}\! & \O & \cdots & \O \\
				\O & \!\mu_2\Mm_2^{-1}\! & \cdots & \O \\
				\vdots &  \vdots & \ddots & \vdots \\
				\O & \O & \cdots & \!\mu_q\Mm_q^{-1}\end{bmatrix}(\bar{\Cm}\xv-\bar{\yv}) \leq 1. $$
The middle matrix is $\bar{\Mm}^{-1}$. Hence, $\bar{\Cm}\xv \in \Es(\bar{\yv},\bar{\Mm})$.
Since $\bar{\Cm}$ is full rank, then $mq \geq n$, which implies that $\bar{\Cm}^\dagger\bar{\Cm} = \I$. Therefore, \changed{$\xv = \bar{\Cm}^\dagger\bar{\Cm}\xv \in \bar{\Cm}^\dagger\Es(\bar{\yv},\bar{\Mm}).$ Finally, applying the linear transformation on the latter ellipsoid gives $\xv \in \Es(\bar{\Cm}^\dagger\bar{\yv},{\bar{\Cm}}^\dagger\bar{\Mm}{\bar{\Cm}^\dagger}{}\tran)$.}
\end{proof}
\changed{Finally, using the fact that  $\Om(\bar{k})$ is full-rank, we apply Lemma \ref{lem:ellipfromcyls} with $\mu_i = \bar{k}+1$ to Eq.~(\ref{eq:cylintersection}), obtaining the main initialization step:}
\begin{thm} Let $\bar{\Om}(\bar{k}) \coloneqq \Om(\bar{k})\Phip^{-\bar{k}}$ and
\begin{align*}
\bar{\psiv}(\bar{k})\tran &\coloneqq \begin{bmatrix}\tilde{\psiv}(0,\bar{k})\tran & \tilde{\psiv}(1,\bar{k})\tran & \cdots & \psiv(\bar{k})\tran\,\end{bmatrix},\\
\bar{\Vm}(\bar{k}) &\coloneqq \begin{bmatrix}(\bar{k}+1)\tilde{\Vm}(0)\! & \O & \cdots & \O \\
				\O & \!(\bar{k}+1)\tilde{\Vm}(1)\! & \cdots & \O \\
				\vdots &  \vdots & \ddots & \vdots \\
				\O & \O & \cdots & \!(\bar{k}+1)\tilde{\Vm}(\bar{k})\end{bmatrix}.
\end{align*}
Then $\xip(h\bar{k}) \in \Es\left(\bar{\Om}(\bar{k})^\dagger\bar{\psiv}(\bar{k}),\bar{\Om}(\bar{k})^\dagger\bar{\Vm}(\bar{k})\bar{\Om}(\bar{k})^\dagger{}\tran\,\right)$.
\end{thm}
Matrices $\bar{\Om}(\bar{k})^\dagger, \bar{\Vm}(\bar{k}), \bar{\Om}(\bar{k})^\dagger\bar{\Vm}(\bar{k})\bar{\Om}(\bar{k})^\dagger{}\tran$ and $\Phip^{-k}, k \in \{1,...,\bar{k}\}$ can be computed off-line. On-line, $\tilde{\psi}(k,\bar{k})$ are calculated and, at $k = \bar{k}$, the center of the state estimate $\Xstilde$, $\bar{\Om}(\bar{k})^\dagger\bar{\psiv}(\bar{k})$ is computed. The main loop with Algorithm \ref{alg} then follows.

\end{document}

%% file: tikzsettings.tex
\usepackage{pgfplots}
\usepackage{grffile}
\pgfplotsset{compat=newest}
\usetikzlibrary{spy, automata}
\usetikzlibrary{arrows, arrows.meta, decorations.markings, patterns}
\usepgfplotslibrary{patchplots}
\usetikzlibrary{shapes, positioning, fit, backgrounds}
\usepgfplotslibrary{fillbetween}

\usetikzlibrary{overlay-beamer-styles}
\usetikzlibrary{calc}

\usepackage{xcolor}
\usepackage{calc}

\definecolor{tudcyan}{RGB}{0,166,214}
\definecolor{tudmagenta}{RGB}{109,23,127}
\definecolor{tudpurple}{RGB}{29,28,115}
\definecolor{tudgraygreen}{RGB}{107,134,137}

\colorlet{lighttudcyan}{tudcyan!20}
\colorlet{lighttudmagenta}{tudmagenta!20}

\newlength{\hatchspread}
\newlength{\hatchthickness}
\newlength{\hatchshift}
\newcommand{\hatchcolor}{}
\tikzset{hatchspread/.code={\setlength{\hatchspread}{#1}},
	hatchthickness/.code={\setlength{\hatchthickness}{#1}},
	hatchshift/.code={\setlength{\hatchshift}{#1}},
	hatchcolor/.code={\renewcommand{\hatchcolor}{#1}}}
\tikzset{hatchspread=7pt,
	hatchthickness=0.5pt,
	hatchshift=0pt,
	hatchcolor=black}
\pgfdeclarepatternformonly[\hatchspread,\hatchthickness,\hatchshift,\hatchcolor]
{custom north west lines}
{\pgfqpoint{\dimexpr-2\hatchthickness}{\dimexpr-2\hatchthickness}}
{\pgfqpoint{\dimexpr\hatchspread+2\hatchthickness}{\dimexpr\hatchspread+2\hatchthickness}}
{\pgfqpoint{\dimexpr\hatchspread}{\dimexpr\hatchspread}}
{
	\pgfsetlinewidth{\hatchthickness}
	\pgfpathmoveto{\pgfqpoint{0pt}{\dimexpr\hatchspread+\hatchshift}}
	\pgfpathlineto{\pgfqpoint{\dimexpr\hatchspread+0.15pt+\hatchshift}{-0.15pt}}
	\ifdim \hatchshift > 0pt
	\pgfpathmoveto{\pgfqpoint{0pt}{\hatchshift}}
	\pgfpathlineto{\pgfqpoint{\dimexpr0.15pt+\hatchshift}{-0.15pt}}
	\fi
	\pgfsetstrokecolor{\hatchcolor}
	\pgfusepath{stroke}
}

\def\centerarc[#1](#2)(#3:#4:#5)
{ \draw[#1] ($(#2)+({#5*cos(#3)},{#5*sin(#3)})$) arc (#3:#4:#5); }

%% file: reach_ellipsoids.tex
\begin{tikzpicture}[x=2em,y=2em,decoration={markings,
	mark=at position 0.70 with {\coordinate (A);}
}]

\draw[-Latex, thick] (-5,0) -- (5,0) node[anchor=south] {$x_1$};
\draw[-Latex, thick] (0,-1.5) -- (0,2) node[anchor=east] {$x_2$};

\begin{scope}[rotate=15]
\filldraw[fill=lighttudmagenta, thick, dashed] (0,0) ellipse (4.5 and 0.9);
\node[pin={[pin edge={black,thin}]150:{$\Xwtilde(t)$}}, inner sep = 0] at (90:4.5 and 0.9) {};
\end{scope}
\filldraw[thick, fill=lighttudcyan, postaction={decorate}] (-4,-1.4) to [out=80,in=185, looseness=0.4] (4,1.4) to [out=-80,in=5, looseness=0.4] (-4,-1.4);
\node[pin={[pin edge={black,thin}]-45:{$\Xw(t)$}}, inner sep = 0] at (A) {};


\end{tikzpicture}

%% file: block_diagram_stc.tex
\tikzstyle{block} 		= [draw, rectangle, minimum height=2em, minimum width=4em]
\tikzstyle{sum} 		= [draw, circle, inner sep=0, minimum size=0.2cm]
\tikzstyle{input} 		= [coordinate]
\tikzstyle{output} 		= [coordinate]
\tikzstyle{split} 		= [coordinate]
\tikzstyle{pinstyle} 	= [pin edge={to-,thin,black}]
\tikzstyle{branch}=[fill,shape=circle,minimum size=3pt,inner sep=0pt]

\begin{tikzpicture}[auto, node distance=2em,>=latex']

	\node [block, align=center, minimum height=2.0em] (P) {Plant \eqref{eq:linplant}};	
	\node [block, below =1.5 em of P] (STC) {STC Algorithm};
	\node [block, below = 2.5 em of STC] (C) {Controller \eqref{eq:lincont}};
	\node [sum, right=of P] (sum) {};
	\node [above=1.5em of sum] (noise) {$\nuv(t)$};
	\node [left=of P.167] (dist) {$\omegav(t)$};
	\node[coordinate, right = 5 em of C.east](br){};
	\node[coordinate, output, left = 5 em of C.west](bl){};
	\draw (P.east -| br) node[coordinate] (ur){};
	\draw (P.193 -| bl) node[coordinate] (ul){};
	\path (ur) |- coordinate[midway](mra) (STC);
	\node[coordinate, below = 0.75 em of mra] (mr) {};
	\path (ul) |- coordinate(ml) (STC);
	\node[rectangle, draw, below = 0.00 em of ul, anchor=center] (ZOH) {ZOH};
	
	\draw [->] (dist) -- (P.167);
	\draw [->] (P) --node[above, pos=0.8] {$+$} (sum);
	\draw [->] (noise) -- (sum);
	\draw [-] (sum) --node[below]{$\psiv(t)$} (ur);
	\draw [->] (br) --node[below]{$\hat{\psiv}(k_b)$} node[midway, branch, anchor=center, name=yhat]{} (C.east);
	\draw [-] (C.west) --node[below]{$\upsilonv(k)$} node[midway, branch, anchor=center, name=u]{} (bl);
	\draw [->] (ZOH) --node[below]{$\hat{\upsilonv}(t)$} (P.193);
	\draw [->] (C) --node[left]{$\xic(k)$} (STC);
	\draw [->] (ml) -- (ZOH);
	\draw [->] (yhat) -- +(0,0.6) -- (STC.south east);
	\draw [->] (u) -- +(0,0.6) -- (STC.south west);
	
	\node[coordinate, above = 0.75 em of mr] (rbegs){};
	\node[coordinate, below = 1.5 em of rbegs](rends){};
	\node[coordinate, below left = 1.5 em of rbegs](rups){};
	\draw [-] (ur) -- (rbegs);
	\draw [-] (rbegs) -- (rups);
	\draw [-] (rends) -- (br);
	\node[coordinate, left = 0.75 em of rbegs](rarrs){};
	\node[coordinate, below = 0.00 em of mr] (rarre){};
	\draw [-{Latex[length=0.25em,width=0.25em]}] (rarrs) to [out=-90, in=180, looseness=1] (rarre);
	
	\node[coordinate, above = 0.00 em of ml] (lbegs){};
	\node[coordinate, below = 1.5 em of lbegs](lends){};
	\node[coordinate, below right = 1.5 em of lbegs](lups){};
	\draw [-] (ZOH) --node[left]{$\upsilonv(k_b)$} (lbegs);
	\draw [-] (lbegs) -- (lups);
	\draw [-] (lends) -- (bl);
	\node[coordinate, right = 0.75 em of lbegs](larrs){};
	\node[coordinate, below = 0.75 em of ml] (larre){};
	\draw [-{Latex[length=0.25em,width=0.25em]}] (larrs) to [out=-90, in=0, looseness=1] (larre);

	\node[coordinate, above = 0.00 em of ml] (larre){};
	
	\node[coordinate, left = 1. em of mr] (sr) {};
	\node[coordinate, right = 1. em of ml] (sl) {};
	\draw[dash pattern={on 0.2em off 0.2em}] (STC) -- (rarrs);
	\draw[dash pattern={on 0.2em off 0.2em}] (STC) -- (larrs);
%
\end{tikzpicture}

%% file: gse.tex
\begin{tikzpicture}[x=1em,y=1em,decoration={markings,
	mark=at position 0.90 with {\coordinate (A);}
}]

\draw[-Latex, thick] (-5,-3) -- (5,-3) node[anchor=south] {$x_1$};
\draw[-Latex, thick] (-5,-3) -- (-5,3) node[anchor=east] {$x_2$};

\draw[thick, rotate=0, fill=red, fill opacity = 0.2] (-1,-1) ellipse (1 and 1.5);
\node[pin={[pin edge={black,thin}]90:{$\Xstilde(t_b|t_b)$}}, inner sep = 0] at ($(-1,-1)+(90:1 and 1.5)$) {};

\begin{scope}[shift={(12,0)}]
	\draw[-Latex, thick] (-5,-3) -- (5,-3) node[anchor=south] {$x_1$};
	\draw[-Latex, thick] (-5,-3) -- (-5,3) node[anchor=east] {$x_2$};
	
	\draw[rotate=0, dashed, gray, thick] (-1,-1) ellipse (1 and 1.5);
	\begin{scope}[rotate=30]
		\draw[thick, fill=tudgraygreen, fill opacity = 0.2] (0,0) ellipse (1.5 and 2);
		\node[pin={[pin edge={black,thin}, pin distance=1.6em]90:{$\Phip(\kappa^*)\Es(\xptilde,\Xm) + \Gammap(\kappa^*)\uv$}}, inner sep = 0] at (80:1.5 and 2) {};
	\end{scope}
\end{scope}

\begin{scope}[shift={(0,-9)}]
	\draw[-Latex, thick] (-5,-3) -- (5,-3) node[anchor=south] {$x_1$};
	\draw[-Latex, thick] (-5,-3) -- (-5,3) node[anchor=east] {$x_2$};
	\draw[thick, gray, dashed, rotate=30] (0,0) ellipse (1.5 and 2);
	\begin{scope}[rotate=30]
		\foreach \x in {0,45,...,330}
		\draw[tudcyan, fill=tudcyan, fill opacity = 0.2, rotate=-15] (\x:1.5 and 2) ellipse (1.8 and 0.4);
		\node[pin={[pin edge={black,thin}, pin distance=2em]90:{\small $\Xwtilde(h\kappa^*)$}}, inner sep = 0] at ($(0:1.5 and 2) + (1,-0.35)$) {};
	\end{scope}
	\begin{scope}[rotate=15]
		\draw[thick, fill=orange, fill opacity = 0.2] (0,0) ellipse (3.4 and 2.75);
		\node[pin={[pin edge={black,thin}, pin distance=1em]120:{$\Xstilde(t_{b+1}|t_b)$}}, inner sep = 0] at (100:3.4 and 2.75) {};
	\end{scope}
\end{scope}

\begin{scope}[shift={(12,-9)}]
    \draw[-Latex, thick] (-5,-3) -- (5,-3) node[anchor=south] {$x_1$};
    \draw[-Latex, thick] (-5,-3) -- (-5,3) node[anchor=east] {$x_2$};
    \draw[thick, dashed, gray, rotate=15] (0,0) ellipse (3.4 and 2.75);
	\fill[fill=tudmagenta, fill opacity=0.2] (-4.5,0)--(4,2)--(4,3)--(-4.5,1)--cycle;
	\draw[tudmagenta, dashdotted] (-4.5,0)--(4,2);
	\draw[tudmagenta, dashdotted, postaction={decorate}] (-4.5,1)--(4,3);
	\node[pin={[pin edge={black,thin}, pin distance=0.7em]135:{\small $\Xsy(t_{b+1})$}}, inner sep = 0] at (A) {};
	\begin{scope}[rotate=13]
		\draw[thick, fill=green, fill opacity = 0.3] (0.0,1.45) ellipse (4 and 0.7);
		\node[pin={[pin edge={black,thin}, pin distance=1em]-90:{$\Xstilde(t_{b+1}|t_{b+1})$}}, inner sep = 0] at ($(0.0,1.45) + (-90:4 and 0.7)$) {};
	\end{scope}
\end{scope}


\end{tikzpicture}

%% file: automatica_vs_necsys.tex
%
\definecolor{mycolor1}{rgb}{0.00000,0.44700,0.74100}%
\definecolor{mycolor2}{rgb}{0.85000,0.32500,0.09800}%
\definecolor{mycolor3}{rgb}{0.92900,0.69400,0.12500}%
\begin{tikzpicture}

\begin{axis}[%
width=70mm,
height=20.93mm,
at={(0mm,27mm)},
scale only axis,
xmin=0,
xmax=10,
ymin=0,
ymax=20,
ylabel style={font=\color{white!15!black}},
ylabel={State norm $|\xiv(t)|$},
axis background/.style={fill=white},
xmajorgrids,
ymajorgrids,
legend style={legend cell align=left, align=left, draw=white!15!black}
]
\addplot [color=mycolor1, line width=1.0pt]
table[row sep=crcr]{%
0	20\\
0.0100000000000016	17.2428177818324\\
0.0300000000000011	13.351602632782\\
0.0500000000000007	11.2956643209226\\
0.0700000000000003	10.2886669278217\\
0.0899999999999999	9.80045595177121\\
0.109999999999999	9.53675696451746\\
0.129999999999999	9.35813285617197\\
0.210000000000001	8.82490343035522\\
0.27	8.34275541875097\\
0.329999999999998	7.84072181896223\\
0.399999999999999	7.27361599741633\\
0.48	6.66638920450217\\
0.57	6.04677827265195\\
0.670000000000002	5.44365491864984\\
0.800000000000001	4.77794527936135\\
0.98	4.06882923570683\\
1.12	3.69437211430827\\
1.24	3.36032642999466\\
1.31	3.21582703167296\\
1.39	3.03677973661205\\
1.53	2.75091769412351\\
1.66	2.50449412412836\\
1.78	2.29067418287607\\
1.9	2.08951407001454\\
2.02	1.90054203948393\\
2.14	1.72348240954255\\
2.26	1.55814861467626\\
2.38	1.40434612346108\\
2.45	1.31955882297401\\
2.52	1.23888759845678\\
2.59	1.16189730223545\\
2.66	1.08852639432618\\
2.79	0.961786183235997\\
2.91	0.854901360712251\\
3.03	0.7573769953697\\
3.15	0.668687248870235\\
3.27	0.588266579367264\\
3.39	0.515544061479794\\
3.51	0.44995464973746\\
3.62	0.395721352191288\\
3.72	0.350806672702603\\
3.82	0.309809119861857\\
3.92	0.27247374412844\\
4.02	0.238536741586024\\
4.17	0.193460652169293\\
4.29	0.162038400151175\\
4.41	0.134238178242583\\
4.57	0.102346630426617\\
4.73	0.0758070285080876\\
4.89	0.0542243702342091\\
5.09	0.0327070337733915\\
5.2	0.026437243589168\\
5.36	0.0228559043384813\\
6.48	0.00870890199320939\\
8.27	0.00681091231964714\\
10	0.00214940438958777\\
};
\addlegendentry{PSTC}

\addplot [color=mycolor2, dashed, line width=1.0pt]
  table[row sep=crcr]{%
0	20\\
0.0199999999999996	14.9367798044312\\
0.0399999999999991	12.1470760844376\\
0.0599999999999987	10.7368766484167\\
0.0700000000000003	10.3305946678222\\
0.0799999999999983	10.0355496618482\\
0.0899999999999999	9.81756190485391\\
0.109999999999999	9.520302126269\\
0.129999999999999	9.31596657700351\\
0.16	9.06809414801933\\
0.280000000000001	8.11564189331379\\
0.379999999999999	7.3270774575987\\
0.449999999999999	6.80713755484056\\
0.510000000000002	6.38904399846749\\
0.559999999999999	6.06186218340922\\
0.620000000000001	5.69329787693519\\
0.68	5.35488934322808\\
0.739999999999998	5.0453918513868\\
0.800000000000001	4.76325309716764\\
0.859999999999999	4.50652453101285\\
0.920000000000002	4.27295508473205\\
0.98	4.05841347461145\\
1.04	3.86348783262495\\
1.1	3.68494392907279\\
1.17	3.4929354696325\\
1.25	3.29330107192329\\
1.33	3.11049639058274\\
1.42	2.9194079489005\\
1.52	2.72302371943472\\
1.62	2.53924001448821\\
1.72	2.36552228657364\\
1.82	2.2004864456055\\
1.92	2.04335947362068\\
2.02	1.89373984742091\\
2.12	1.75143733855165\\
2.22	1.61636666334076\\
2.32	1.4884801258139\\
2.42	1.36772738157031\\
2.52	1.25403336353284\\
2.62	1.14728797862385\\
2.72	1.04734319376059\\
2.82	0.954014587163098\\
2.94	0.850633435273082\\
3.02	0.786503773837616\\
3.1	0.726144102223667\\
3.18	0.669417981022942\\
3.3	0.590819125172313\\
3.42	0.519555144786541\\
3.52	0.46549741375982\\
3.64	0.406382063272876\\
3.76	0.35314807400507\\
3.88	0.305351137144278\\
4	0.2625433433677\\
4.12	0.224300932217297\\
4.26	0.18498051863083\\
4.4	0.150724153946527\\
4.54	0.121044938369952\\
4.68	0.0955070402668561\\
4.84	0.0709302288101128\\
5.02	0.0482615844218479\\
5.11	0.0393029681412393\\
5.22	0.032960335025308\\
5.41	0.0272091124162444\\
6.29	0.00810508963347445\\
6.62	0.0068244083990372\\
8.21	0.00657877569552667\\
10	0.00216481938736734\\
};
\addlegendentry{GM18-STC}

\addplot [color=mycolor3, dotted, line width=1.0pt]
table[row sep=crcr]{%
	0	20\\
	0.0100000000000016	17.2428179662696\\
	0.0199999999999996	15.0965785847668\\
	0.0300000000000011	13.454546222489\\
	0.0399999999999991	12.2208796711499\\
	0.0500000000000007	11.3166159498195\\
	0.0599999999999987	10.6637107776044\\
	0.0700000000000003	10.1992279006078\\
	0.0799999999999983	9.86924697981979\\
	0.0899999999999999	9.63404095093406\\
	0.100000000000001	9.46250366699428\\
	0.120000000000001	9.23143889186197\\
	0.140000000000001	9.07106275005012\\
	0.199999999999999	8.66733928277265\\
	0.260000000000002	8.24305749671173\\
	0.460000000000001	6.80053106993208\\
	0.530000000000001	6.33559022188656\\
	0.600000000000001	5.90265891640519\\
	0.66	5.55969430545123\\
	0.710000000000001	5.29358708962175\\
	0.77	4.99847220403142\\
	0.82	4.77129598817633\\
	0.879999999999999	4.52173196313041\\
	0.940000000000001	4.29207337196783\\
	0.989999999999998	4.11757423359697\\
	1.05	3.92555145712403\\
	1.11	3.74797056129528\\
	1.17	3.58352970555785\\
	1.23	3.43375524031975\\
	1.3	3.27159016673881\\
	1.37	3.12169190550865\\
	1.45	2.96175347225465\\
	1.54	2.79255238238958\\
	1.64	2.61439584690753\\
	1.75	2.42771812327089\\
	1.86	2.24927715113099\\
	1.97	2.07851188440118\\
	2.08	1.91534522295681\\
	2.19	1.759941623927\\
	2.29	1.62552461326112\\
	2.39	1.49776055933654\\
	2.49	1.3767801496216\\
	2.59	1.26261132232446\\
	2.69	1.15523491090668\\
	2.79	1.05454762967301\\
	2.89	0.960430754047284\\
	2.99	0.872690135134981\\
	3.09	0.79113285459443\\
	3.19	0.715500561703735\\
	3.29	0.645539338339866\\
	3.39	0.580960396461609\\
	3.49	0.521473587784225\\
	3.59	0.466789933790537\\
	3.69	0.416605584540989\\
	3.8	0.366260149378235\\
	3.91	0.320648087835973\\
	4.02	0.279412343802761\\
	4.14	0.239018158473222\\
	4.26	0.203011183231979\\
	4.39	0.168508805715053\\
	4.52	0.138259690858757\\
	4.66	0.109989104166779\\
	4.81	0.0841877126411354\\
	4.96	0.0626521830462892\\
	5.11	0.0451050944954794\\
	5.22	0.0377875811794937\\
	5.4	0.0308367759230634\\
	6.05	0.0126419663377213\\
	6.36	0.00780792904379624\\
	6.69	0.00727099411819054\\
	7.81	0.00807235088600322\\
	9.99	0.00216480668560237\\
};
\addlegendentry{PETC}

\end{axis}

\begin{axis}[%
width=70mm,
height=20.93mm,
at={(0mm,0mm)},
scale only axis,
xmin=0,
xmax=10,
ymin=0,
ymax=25,
ylabel style={font=\color{white!15!black}},
ylabel={$\dk^*$ (samples)},
xlabel={$t$},
axis background/.style={fill=white},
axis x line*=bottom,
axis y line*=left,
xmajorgrids,
ymajorgrids,
legend style={legend cell align=left, align=left, draw=white!15!black, at={(0.5, 0.99)}, anchor=north},
]
\addplot[only marks, mark=o, mark options={}, mark size=3.0000pt, draw=mycolor1] table[row sep=crcr]{%
	x	y\\
	0	1\\
	0.01	2\\
	0.03	2\\
	0.05	2\\
	0.07	2\\
	0.09	2\\
	0.11	2\\
	0.13	3\\
	0.16	5\\
	0.21	6\\
	0.27	6\\
	0.33	7\\
	0.4	8\\
	0.48	9\\
	0.57	10\\
	0.67	13\\
	0.8	18\\
	0.98	14\\
	1.12	12\\
	1.24	7\\
	1.31	8\\
	1.39	7\\
	1.46	7\\
	1.53	7\\
	1.6	6\\
	1.66	6\\
	1.72	6\\
	1.78	6\\
	1.84	6\\
	1.9	6\\
	1.96	6\\
	2.02	6\\
	2.08	6\\
	2.14	6\\
	2.2	6\\
	2.26	6\\
	2.32	6\\
	2.38	7\\
	2.45	7\\
	2.52	7\\
	2.59	7\\
	2.66	7\\
	2.73	6\\
	2.79	6\\
	2.85	6\\
	2.91	6\\
	2.97	6\\
	3.03	6\\
	3.09	6\\
	3.15	6\\
	3.21	6\\
	3.27	6\\
	3.33	6\\
	3.39	6\\
	3.45	6\\
	3.51	6\\
	3.57	5\\
	3.62	5\\
	3.67	5\\
	3.72	5\\
	3.77	5\\
	3.82	5\\
	3.87	5\\
	3.92	5\\
	3.97	5\\
	4.02	5\\
	4.07	5\\
	4.12	5\\
	4.17	4\\
	4.21	4\\
	4.25	4\\
	4.29	4\\
	4.33	4\\
	4.37	4\\
	4.41	4\\
	4.45	4\\
	4.49	4\\
	4.53	4\\
	4.57	4\\
	4.61	4\\
	4.65	4\\
	4.69	4\\
	4.73	4\\
	4.77	3\\
	4.8	3\\
	4.83	3\\
	4.86	3\\
	4.89	3\\
	4.92	3\\
	4.95	3\\
	4.98	3\\
	5.01	3\\
	5.04	3\\
	5.07	2\\
	5.09	2\\
	5.11	2\\
	5.13	2\\
	5.15	2\\
	5.17	2\\
	5.19	1\\
	5.2	1\\
	5.21	1\\
	5.22	1\\
	5.23	1\\
	5.24	1\\
	5.25	1\\
	5.26	1\\
	5.27	1\\
	5.28	1\\
	5.29	1\\
	5.3	1\\
	5.31	1\\
	5.32	1\\
	5.33	1\\
	5.34	1\\
	5.35	1\\
	5.36	1\\
	5.37	1\\
	5.38	1\\
	5.39	1\\
	5.4	1\\
	5.41	1\\
	5.42	1\\
	5.43	2\\
	5.45	2\\
	5.47	2\\
	5.49	2\\
	5.51	2\\
	5.53	2\\
	5.55	2\\
	5.57	2\\
	5.59	2\\
	5.61	2\\
	5.63	2\\
	5.65	2\\
	5.67	2\\
	5.69	2\\
	5.71	2\\
	5.73	2\\
	5.75	2\\
	5.77	2\\
	5.79	2\\
	5.81	2\\
	5.83	2\\
	5.85	2\\
	5.87	2\\
	5.89	2\\
	5.91	2\\
	5.93	1\\
	5.94	1\\
	5.95	1\\
	5.96	1\\
	5.97	1\\
	5.98	1\\
	5.99	1\\
	6	1\\
	6.01	1\\
	6.02	1\\
	6.03	1\\
	6.04	1\\
	6.05	1\\
	6.06	1\\
	6.07	1\\
	6.08	1\\
	6.09	1\\
	6.1	1\\
	6.11	1\\
	6.12	1\\
	6.13	1\\
	6.14	1\\
	6.15	1\\
	6.16	1\\
	6.17	1\\
	6.18	1\\
	6.19	1\\
	6.2	1\\
	6.21	1\\
	6.22	1\\
	6.23	1\\
	6.24	1\\
	6.25	1\\
	6.26	1\\
	6.27	1\\
	6.28	1\\
	6.29	1\\
	6.3	1\\
	6.31	1\\
	6.32	1\\
	6.33	1\\
	6.34	1\\
	6.35	1\\
	6.36	1\\
	6.37	1\\
	6.38	1\\
	6.39	1\\
	6.4	1\\
	6.41	1\\
	6.42	1\\
	6.43	1\\
	6.44	1\\
	6.45	1\\
	6.46	1\\
	6.47	1\\
	6.48	1\\
	6.49	1\\
	6.5	1\\
	6.51	1\\
	6.52	1\\
	6.53	1\\
	6.54	1\\
	6.55	1\\
	6.56	1\\
	6.57	1\\
	6.58	1\\
	6.59	1\\
	6.6	1\\
	6.61	1\\
	6.62	1\\
	6.63	1\\
	6.64	1\\
	6.65	1\\
	6.66	1\\
	6.67	1\\
	6.68	1\\
	6.69	1\\
	6.7	1\\
	6.71	1\\
	6.72	1\\
	6.73	1\\
	6.74	1\\
	6.75	1\\
	6.76	1\\
	6.77	1\\
	6.78	1\\
	6.79	1\\
	6.8	1\\
	6.81	1\\
	6.82	1\\
	6.83	1\\
	6.84	1\\
	6.85	1\\
	6.86	1\\
	6.87	1\\
	6.88	1\\
	6.89	1\\
	6.9	1\\
	6.91	1\\
	6.92	1\\
	6.93	1\\
	6.94	1\\
	6.95	1\\
	6.96	1\\
	6.97	1\\
	6.98	1\\
	6.99	1\\
	7	1\\
	7.01	1\\
	7.02	1\\
	7.03	1\\
	7.04	1\\
	7.05	1\\
	7.06	1\\
	7.07	1\\
	7.08	1\\
	7.09	1\\
	7.1	1\\
	7.11	1\\
	7.12	1\\
	7.13	1\\
	7.14	1\\
	7.15	1\\
	7.16	1\\
	7.17	1\\
	7.18	1\\
	7.19	1\\
	7.2	1\\
	7.21	1\\
	7.22	1\\
	7.23	1\\
	7.24	1\\
	7.25	1\\
	7.26	1\\
	7.27	1\\
	7.28	1\\
	7.29	1\\
	7.3	1\\
	7.31	1\\
	7.32	1\\
	7.33	1\\
	7.34	1\\
	7.35	1\\
	7.36	1\\
	7.37	1\\
	7.38	1\\
	7.39	1\\
	7.4	1\\
	7.41	1\\
	7.42	1\\
	7.43	1\\
	7.44	1\\
	7.45	1\\
	7.46	1\\
	7.47	1\\
	7.48	1\\
	7.49	1\\
	7.5	1\\
	7.51	1\\
	7.52	1\\
	7.53	1\\
	7.54	1\\
	7.55	1\\
	7.56	1\\
	7.57	1\\
	7.58	1\\
	7.59	1\\
	7.6	1\\
	7.61	1\\
	7.62	1\\
	7.63	1\\
	7.64	1\\
	7.65	1\\
	7.66	1\\
	7.67	1\\
	7.68	1\\
	7.69	1\\
	7.7	1\\
	7.71	1\\
	7.72	1\\
	7.73	1\\
	7.74	1\\
	7.75	1\\
	7.76	1\\
	7.77	1\\
	7.78	1\\
	7.79	1\\
	7.8	1\\
	7.81	1\\
	7.82	1\\
	7.83	1\\
	7.84	1\\
	7.85	1\\
	7.86	1\\
	7.87	1\\
	7.88	1\\
	7.89	1\\
	7.9	1\\
	7.91	1\\
	7.92	1\\
	7.93	1\\
	7.94	1\\
	7.95	1\\
	7.96	1\\
	7.97	1\\
	7.98	1\\
	7.99	1\\
	8	1\\
	8.01	1\\
	8.02	1\\
	8.03	1\\
	8.04	1\\
	8.05	1\\
	8.06	1\\
	8.07	1\\
	8.08	1\\
	8.09	1\\
	8.1	1\\
	8.11	1\\
	8.12	1\\
	8.13	1\\
	8.14	1\\
	8.15	1\\
	8.16	1\\
	8.17	1\\
	8.18	1\\
	8.19	1\\
	8.2	1\\
	8.21	1\\
	8.22	1\\
	8.23	1\\
	8.24	1\\
	8.25	1\\
	8.26	1\\
	8.27	1\\
	8.28	1\\
	8.29	1\\
	8.3	1\\
	8.31	1\\
	8.32	1\\
	8.33	1\\
	8.34	1\\
	8.35	1\\
	8.36	1\\
	8.37	1\\
	8.38	1\\
	8.39	1\\
	8.4	1\\
	8.41	1\\
	8.42	1\\
	8.43	1\\
	8.44	1\\
	8.45	1\\
	8.46	1\\
	8.47	1\\
	8.48	1\\
	8.49	1\\
	8.5	1\\
	8.51	1\\
	8.52	1\\
	8.53	1\\
	8.54	1\\
	8.55	1\\
	8.56	1\\
	8.57	1\\
	8.58	1\\
	8.59	1\\
	8.6	1\\
	8.61	1\\
	8.62	1\\
	8.63	1\\
	8.64	1\\
	8.65	1\\
	8.66	1\\
	8.67	1\\
	8.68	1\\
	8.69	1\\
	8.7	1\\
	8.71	1\\
	8.72	1\\
	8.73	1\\
	8.74	1\\
	8.75	1\\
	8.76	1\\
	8.77	1\\
	8.78	1\\
	8.79	1\\
	8.8	1\\
	8.81	1\\
	8.82	1\\
	8.83	1\\
	8.84	1\\
	8.85	1\\
	8.86	1\\
	8.87	1\\
	8.88	1\\
	8.89	1\\
	8.9	1\\
	8.91	1\\
	8.92	1\\
	8.93	1\\
	8.94	1\\
	8.95	1\\
	8.96	1\\
	8.97	1\\
	8.98	1\\
	8.99	1\\
	9	1\\
	9.01	1\\
	9.02	1\\
	9.03	1\\
	9.04	1\\
	9.05	1\\
	9.06	1\\
	9.07	1\\
	9.08	1\\
	9.09	1\\
	9.1	1\\
	9.11	1\\
	9.12	1\\
	9.13	1\\
	9.14	1\\
	9.15	1\\
	9.16	1\\
	9.17	1\\
	9.18	1\\
	9.19	1\\
	9.2	1\\
	9.21	1\\
	9.22	1\\
	9.23	1\\
	9.24	1\\
	9.25	1\\
	9.26	1\\
	9.27	1\\
	9.28	1\\
	9.29	1\\
	9.3	1\\
	9.31	1\\
	9.32	1\\
	9.33	1\\
	9.34	1\\
	9.35	1\\
	9.36	1\\
	9.37	1\\
	9.38	1\\
	9.39	1\\
	9.4	1\\
	9.41	1\\
	9.42	1\\
	9.43	1\\
	9.44	1\\
	9.45	1\\
	9.46	1\\
	9.47	1\\
	9.48	1\\
	9.49	1\\
	9.5	1\\
	9.51	1\\
	9.52	1\\
	9.53	1\\
	9.54	1\\
	9.55	1\\
	9.56	1\\
	9.57	1\\
	9.58	1\\
	9.59	1\\
	9.6	1\\
	9.61	1\\
	9.62	1\\
	9.63	1\\
	9.64	1\\
	9.65	1\\
	9.66	1\\
	9.67	1\\
	9.68	1\\
	9.69	1\\
	9.7	1\\
	9.71	1\\
	9.72	1\\
	9.73	1\\
	9.74	1\\
	9.75	1\\
	9.76	1\\
	9.77	1\\
	9.78	1\\
	9.79	1\\
	9.8	1\\
	9.81	1\\
	9.82	1\\
	9.83	1\\
	9.84	1\\
	9.85	1\\
	9.86	1\\
	9.87	1\\
	9.88	1\\
	9.89	1\\
	9.9	1\\
	9.91	1\\
	9.92	1\\
	9.93	1\\
	9.94	1\\
	9.95	1\\
	9.96	1\\
	9.97	1\\
	9.98	1\\
	9.99	1\\
	10	1\\
};
\addlegendentry{PSTC}

\addplot[only marks, mark=*, mark options={}, mark size=1.0000pt, draw=mycolor2, fill=mycolor2] table[row sep=crcr]{%
x	y\\
0	2\\
0.02	2\\
0.04	2\\
0.06	1\\
0.07	1\\
0.08	1\\
0.09	1\\
0.1	1\\
0.11	1\\
0.12	1\\
0.13	1\\
0.14	1\\
0.15	1\\
0.16	1\\
0.17	1\\
0.18	1\\
0.19	1\\
0.2	1\\
0.21	1\\
0.22	1\\
0.23	1\\
0.24	1\\
0.25	1\\
0.26	1\\
0.27	1\\
0.28	1\\
0.29	1\\
0.3	1\\
0.31	1\\
0.32	1\\
0.33	1\\
0.34	1\\
0.35	1\\
0.36	1\\
0.37	1\\
0.38	1\\
0.39	1\\
0.4	1\\
0.41	1\\
0.42	1\\
0.43	1\\
0.44	1\\
0.45	1\\
0.46	1\\
0.47	1\\
0.48	1\\
0.49	1\\
0.5	1\\
0.51	1\\
0.52	1\\
0.53	1\\
0.54	1\\
0.55	1\\
0.56	2\\
0.58	2\\
0.6	2\\
0.62	2\\
0.64	2\\
0.66	2\\
0.68	2\\
0.7	2\\
0.72	2\\
0.74	2\\
0.76	2\\
0.78	2\\
0.8	2\\
0.82	2\\
0.84	2\\
0.86	2\\
0.88	2\\
0.9	2\\
0.92	3\\
0.95	3\\
0.98	3\\
1.01	3\\
1.04	3\\
1.07	3\\
1.1	3\\
1.13	4\\
1.17	4\\
1.21	4\\
1.25	4\\
1.29	4\\
1.33	4\\
1.37	5\\
1.42	5\\
1.47	5\\
1.52	5\\
1.57	5\\
1.62	5\\
1.67	5\\
1.72	5\\
1.77	5\\
1.82	5\\
1.87	5\\
1.92	5\\
1.97	5\\
2.02	5\\
2.07	5\\
2.12	5\\
2.17	5\\
2.22	5\\
2.27	5\\
2.32	5\\
2.37	5\\
2.42	5\\
2.47	5\\
2.52	5\\
2.57	5\\
2.62	5\\
2.67	5\\
2.72	5\\
2.77	5\\
2.82	4\\
2.86	4\\
2.9	4\\
2.94	4\\
2.98	4\\
3.02	4\\
3.06	4\\
3.1	4\\
3.14	4\\
3.18	4\\
3.22	4\\
3.26	4\\
3.3	4\\
3.34	4\\
3.38	4\\
3.42	4\\
3.46	3\\
3.49	3\\
3.52	3\\
3.55	3\\
3.58	3\\
3.61	3\\
3.64	3\\
3.67	3\\
3.7	3\\
3.73	3\\
3.76	3\\
3.79	3\\
3.82	3\\
3.85	3\\
3.88	3\\
3.91	3\\
3.94	3\\
3.97	3\\
4	3\\
4.03	3\\
4.06	3\\
4.09	3\\
4.12	3\\
4.15	3\\
4.18	2\\
4.2	2\\
4.22	2\\
4.24	2\\
4.26	2\\
4.28	2\\
4.3	2\\
4.32	2\\
4.34	2\\
4.36	2\\
4.38	2\\
4.4	2\\
4.42	2\\
4.44	2\\
4.46	2\\
4.48	2\\
4.5	2\\
4.52	2\\
4.54	2\\
4.56	2\\
4.58	2\\
4.6	2\\
4.62	2\\
4.64	2\\
4.66	2\\
4.68	2\\
4.7	2\\
4.72	2\\
4.74	2\\
4.76	2\\
4.78	2\\
4.8	2\\
4.82	2\\
4.84	2\\
4.86	2\\
4.88	2\\
4.9	2\\
4.92	2\\
4.94	2\\
4.96	2\\
4.98	2\\
5	2\\
5.02	2\\
5.04	2\\
5.06	1\\
5.07	1\\
5.08	1\\
5.09	1\\
5.1	1\\
5.11	1\\
5.12	1\\
5.13	1\\
5.14	1\\
5.15	1\\
5.16	1\\
5.17	1\\
5.18	1\\
5.19	1\\
5.2	1\\
5.21	1\\
5.22	1\\
5.23	1\\
5.24	1\\
5.25	1\\
5.26	1\\
5.27	1\\
5.28	1\\
5.29	1\\
5.3	1\\
5.31	1\\
5.32	1\\
5.33	1\\
5.34	1\\
5.35	1\\
5.36	1\\
5.37	1\\
5.38	1\\
5.39	1\\
5.4	1\\
5.41	1\\
5.42	1\\
5.43	1\\
5.44	1\\
5.45	1\\
5.46	1\\
5.47	1\\
5.48	1\\
5.49	1\\
5.5	1\\
5.51	1\\
5.52	1\\
5.53	1\\
5.54	1\\
5.55	1\\
5.56	1\\
5.57	1\\
5.58	1\\
5.59	1\\
5.6	1\\
5.61	1\\
5.62	1\\
5.63	1\\
5.64	1\\
5.65	1\\
5.66	1\\
5.67	1\\
5.68	1\\
5.69	1\\
5.7	1\\
5.71	1\\
5.72	1\\
5.73	1\\
5.74	1\\
5.75	1\\
5.76	1\\
5.77	1\\
5.78	1\\
5.79	1\\
5.8	1\\
5.81	1\\
5.82	1\\
5.83	1\\
5.84	1\\
5.85	1\\
5.86	1\\
5.87	1\\
5.88	1\\
5.89	1\\
5.9	1\\
5.91	1\\
5.92	1\\
5.93	1\\
5.94	1\\
5.95	1\\
5.96	1\\
5.97	1\\
5.98	1\\
5.99	1\\
6	1\\
6.01	1\\
6.02	1\\
6.03	1\\
6.04	1\\
6.05	1\\
6.06	1\\
6.07	1\\
6.08	1\\
6.09	1\\
6.1	1\\
6.11	1\\
6.12	1\\
6.13	1\\
6.14	1\\
6.15	1\\
6.16	1\\
6.17	1\\
6.18	1\\
6.19	1\\
6.2	1\\
6.21	1\\
6.22	1\\
6.23	1\\
6.24	1\\
6.25	1\\
6.26	1\\
6.27	1\\
6.28	1\\
6.29	1\\
6.3	1\\
6.31	1\\
6.32	1\\
6.33	1\\
6.34	1\\
6.35	1\\
6.36	1\\
6.37	1\\
6.38	1\\
6.39	1\\
6.4	1\\
6.41	1\\
6.42	1\\
6.43	1\\
6.44	1\\
6.45	1\\
6.46	1\\
6.47	1\\
6.48	1\\
6.49	1\\
6.5	1\\
6.51	1\\
6.52	1\\
6.53	1\\
6.54	1\\
6.55	1\\
6.56	1\\
6.57	1\\
6.58	1\\
6.59	1\\
6.6	1\\
6.61	1\\
6.62	1\\
6.63	1\\
6.64	1\\
6.65	1\\
6.66	1\\
6.67	1\\
6.68	1\\
6.69	1\\
6.7	1\\
6.71	1\\
6.72	1\\
6.73	1\\
6.74	1\\
6.75	1\\
6.76	1\\
6.77	1\\
6.78	1\\
6.79	1\\
6.8	1\\
6.81	1\\
6.82	1\\
6.83	1\\
6.84	1\\
6.85	1\\
6.86	1\\
6.87	1\\
6.88	1\\
6.89	1\\
6.9	1\\
6.91	1\\
6.92	1\\
6.93	1\\
6.94	1\\
6.95	1\\
6.96	1\\
6.97	1\\
6.98	1\\
6.99	1\\
7	1\\
7.01	1\\
7.02	1\\
7.03	1\\
7.04	1\\
7.05	1\\
7.06	1\\
7.07	1\\
7.08	1\\
7.09	1\\
7.1	1\\
7.11	1\\
7.12	1\\
7.13	1\\
7.14	1\\
7.15	1\\
7.16	1\\
7.17	1\\
7.18	1\\
7.19	1\\
7.2	1\\
7.21	1\\
7.22	1\\
7.23	1\\
7.24	1\\
7.25	1\\
7.26	1\\
7.27	1\\
7.28	1\\
7.29	1\\
7.3	1\\
7.31	1\\
7.32	1\\
7.33	1\\
7.34	1\\
7.35	1\\
7.36	1\\
7.37	1\\
7.38	1\\
7.39	1\\
7.4	1\\
7.41	1\\
7.42	1\\
7.43	1\\
7.44	1\\
7.45	1\\
7.46	1\\
7.47	1\\
7.48	1\\
7.49	1\\
7.5	1\\
7.51	1\\
7.52	1\\
7.53	1\\
7.54	1\\
7.55	1\\
7.56	1\\
7.57	1\\
7.58	1\\
7.59	1\\
7.6	1\\
7.61	1\\
7.62	1\\
7.63	1\\
7.64	1\\
7.65	1\\
7.66	1\\
7.67	1\\
7.68	1\\
7.69	1\\
7.7	1\\
7.71	1\\
7.72	1\\
7.73	1\\
7.74	1\\
7.75	1\\
7.76	1\\
7.77	1\\
7.78	1\\
7.79	1\\
7.8	1\\
7.81	1\\
7.82	1\\
7.83	1\\
7.84	1\\
7.85	1\\
7.86	1\\
7.87	1\\
7.88	1\\
7.89	1\\
7.9	1\\
7.91	1\\
7.92	1\\
7.93	1\\
7.94	1\\
7.95	1\\
7.96	1\\
7.97	1\\
7.98	1\\
7.99	1\\
8	1\\
8.01	1\\
8.02	1\\
8.03	1\\
8.04	1\\
8.05	1\\
8.06	1\\
8.07	1\\
8.08	1\\
8.09	1\\
8.1	1\\
8.11	1\\
8.12	1\\
8.13	1\\
8.14	1\\
8.15	1\\
8.16	1\\
8.17	1\\
8.18	1\\
8.19	1\\
8.2	1\\
8.21	1\\
8.22	1\\
8.23	1\\
8.24	1\\
8.25	1\\
8.26	1\\
8.27	1\\
8.28	1\\
8.29	1\\
8.3	1\\
8.31	1\\
8.32	1\\
8.33	1\\
8.34	1\\
8.35	1\\
8.36	1\\
8.37	1\\
8.38	1\\
8.39	1\\
8.4	1\\
8.41	1\\
8.42	1\\
8.43	1\\
8.44	1\\
8.45	1\\
8.46	1\\
8.47	1\\
8.48	1\\
8.49	1\\
8.5	1\\
8.51	1\\
8.52	1\\
8.53	1\\
8.54	1\\
8.55	1\\
8.56	1\\
8.57	1\\
8.58	1\\
8.59	1\\
8.6	1\\
8.61	1\\
8.62	1\\
8.63	1\\
8.64	1\\
8.65	1\\
8.66	1\\
8.67	1\\
8.68	1\\
8.69	1\\
8.7	1\\
8.71	1\\
8.72	1\\
8.73	1\\
8.74	1\\
8.75	1\\
8.76	1\\
8.77	1\\
8.78	1\\
8.79	1\\
8.8	1\\
8.81	1\\
8.82	1\\
8.83	1\\
8.84	1\\
8.85	1\\
8.86	1\\
8.87	1\\
8.88	1\\
8.89	1\\
8.9	1\\
8.91	1\\
8.92	1\\
8.93	1\\
8.94	1\\
8.95	1\\
8.96	1\\
8.97	1\\
8.98	1\\
8.99	1\\
9	1\\
9.01	1\\
9.02	1\\
9.03	1\\
9.04	1\\
9.05	1\\
9.06	1\\
9.07	1\\
9.08	1\\
9.09	1\\
9.1	1\\
9.11	1\\
9.12	1\\
9.13	1\\
9.14	1\\
9.15	1\\
9.16	1\\
9.17	1\\
9.18	1\\
9.19	1\\
9.2	1\\
9.21	1\\
9.22	1\\
9.23	1\\
9.24	1\\
9.25	1\\
9.26	1\\
9.27	1\\
9.28	1\\
9.29	1\\
9.3	1\\
9.31	1\\
9.32	1\\
9.33	1\\
9.34	1\\
9.35	1\\
9.36	1\\
9.37	1\\
9.38	1\\
9.39	1\\
9.4	1\\
9.41	1\\
9.42	1\\
9.43	1\\
9.44	1\\
9.45	1\\
9.46	1\\
9.47	1\\
9.48	1\\
9.49	1\\
9.5	1\\
9.51	1\\
9.52	1\\
9.53	1\\
9.54	1\\
9.55	1\\
9.56	1\\
9.57	1\\
9.58	1\\
9.59	1\\
9.6	1\\
9.61	1\\
9.62	1\\
9.63	1\\
9.64	1\\
9.65	1\\
9.66	1\\
9.67	1\\
9.68	1\\
9.69	1\\
9.7	1\\
9.71	1\\
9.72	1\\
9.73	1\\
9.74	1\\
9.75	1\\
9.76	1\\
9.77	1\\
9.78	1\\
9.79	1\\
9.8	1\\
9.81	1\\
9.82	1\\
9.83	1\\
9.84	1\\
9.85	1\\
9.86	1\\
9.87	1\\
9.88	1\\
9.89	1\\
9.9	1\\
9.91	1\\
9.92	1\\
9.93	1\\
9.94	1\\
9.95	1\\
9.96	1\\
9.97	1\\
9.98	1\\
9.99	1\\
10	1\\
};
\addlegendentry{GM18-STC}

\addplot[only marks, mark=x, mark options={}, mark size=3.0000pt, draw=mycolor3] table[row sep=crcr]{%
	x	y\\
	0	3\\
	0.03	2\\
	0.05	2\\
	0.07	2\\
	0.09	2\\
	0.11	2\\
	0.13	2\\
	0.15	3\\
	0.18	5\\
	0.23	6\\
	0.29	6\\
	0.35	7\\
	0.42	7\\
	0.49	8\\
	0.57	9\\
	0.66	11\\
	0.77	15\\
	0.92	20\\
	1.12	15\\
	1.27	9\\
	1.36	7\\
	1.43	7\\
	1.5	6\\
	1.56	6\\
	1.62	6\\
	1.68	5\\
	1.73	5\\
	1.78	5\\
	1.83	5\\
	1.88	5\\
	1.93	5\\
	1.98	5\\
	2.03	6\\
	2.09	6\\
	2.15	6\\
	2.21	6\\
	2.27	6\\
	2.33	6\\
	2.39	6\\
	2.45	6\\
	2.51	7\\
	2.58	7\\
	2.65	7\\
	2.72	7\\
	2.79	7\\
	2.86	7\\
	2.93	7\\
	3	7\\
	3.07	7\\
	3.14	7\\
	3.21	7\\
	3.28	7\\
	3.35	7\\
	3.42	7\\
	3.49	7\\
	3.56	7\\
	3.63	6\\
	3.69	6\\
	3.75	6\\
	3.81	6\\
	3.87	6\\
	3.93	6\\
	3.99	6\\
	4.05	6\\
	4.11	6\\
	4.17	6\\
	4.23	6\\
	4.29	6\\
	4.35	6\\
	4.41	6\\
	4.47	6\\
	4.53	6\\
	4.59	6\\
	4.65	6\\
	4.71	6\\
	4.77	6\\
	4.83	6\\
	4.89	6\\
	4.95	6\\
	5.01	5\\
	5.06	3\\
	5.09	3\\
	5.12	3\\
	5.15	2\\
	5.17	2\\
	5.19	2\\
	5.21	2\\
	5.23	2\\
	5.25	2\\
	5.27	2\\
	5.29	2\\
	5.31	2\\
	5.33	2\\
	5.35	2\\
	5.37	2\\
	5.39	2\\
	5.41	2\\
	5.43	2\\
	5.45	2\\
	5.47	2\\
	5.49	2\\
	5.51	2\\
	5.53	2\\
	5.55	2\\
	5.57	2\\
	5.59	2\\
	5.61	3\\
	5.64	3\\
	5.67	3\\
	5.7	3\\
	5.73	3\\
	5.76	3\\
	5.79	3\\
	5.82	3\\
	5.85	3\\
	5.88	3\\
	5.91	3\\
	5.94	3\\
	5.97	3\\
	6	3\\
	6.03	3\\
	6.06	4\\
	6.1	4\\
	6.14	4\\
	6.18	4\\
	6.22	4\\
	6.26	4\\
	6.3	4\\
	6.34	4\\
	6.38	5\\
	6.43	5\\
	6.48	5\\
	6.53	5\\
	6.58	5\\
	6.63	6\\
	6.69	6\\
	6.75	7\\
	6.82	7\\
	6.89	8\\
	6.97	8\\
	7.05	9\\
	7.14	10\\
	7.24	11\\
	7.35	12\\
	7.47	13\\
	7.6	15\\
	7.75	16\\
	7.91	17\\
	8.08	17\\
	8.25	17\\
	8.42	16\\
	8.58	15\\
	8.73	14\\
	8.87	13\\
	9	12\\
	9.12	11\\
	9.23	11\\
	9.34	10\\
	9.44	10\\
	9.54	9\\
	9.63	9\\
	9.72	9\\
	9.81	8\\
	9.89	8\\
};
\addlegendentry{PETC}

\end{axis}
\end{tikzpicture}%

%% file: stc_vs_etc_automatica.tex
%
\definecolor{mycolor1}{rgb}{0.00000,0.44700,0.74100}%
\definecolor{mycolor2}{rgb}{0.85000,0.32500,0.09800}%
\begin{tikzpicture}

\begin{axis}[%
width=70mm,
height=20.93mm,
at={(0mm,54mm)},
scale only axis,
xmin=0,
xmax=10,
ymin=0,
ymax=20,
ylabel style={font=\color{white!15!black}},
ylabel={State norm $\|\xiv\|$},
axis background/.style={fill=white},
xmajorgrids,
ymajorgrids,
legend style={legend cell align=left, align=left, draw=white!15!black}
]
\input{with_noise_data.tex}
\end{axis}

\begin{axis}[%
width=35mm,
height=10mm,
at={(35mm,54mm)},
scale only axis,
domain={5,10},
restrict y to domain=0:0.05,
xmin=5,
xmax=10,
ymin=0,
ymax=0.05,
axis background/.style={fill=white},
xmajorgrids,
ymajorgrids,
xticklabels={},
scaled ticks=false,
ytick={0.02, 0.04}
]
\input{with_noise_data.tex}
\legend{}
\end{axis}

\begin{axis}[%
clip mode=individual,
width=70mm,
height=20.93mm,
at={(0mm,27mm)},
scale only axis,
xmin=0,
xmax=10,
ymin=0,
ymax=25.5,
ylabel style={font=\color{white!15!black}},
ylabel={$\dk^*$ (samples)},
axis background/.style={fill=white},
axis x line*=bottom,
axis y line*=left,
xmajorgrids,
ymajorgrids,
legend style={legend cell align=left, align=left, draw=white!15!black}
]
\addplot[only marks, mark=o, mark options={}, mark size=3.0000pt, draw=mycolor1] table[row sep=crcr]{%
	x	y\\
	0	1\\
	0.01	2\\
	0.03	2\\
	0.05	2\\
	0.07	2\\
	0.09	1\\
	0.1	1\\
	0.11	2\\
	0.13	2\\
	0.15	4\\
	0.19	5\\
	0.24	6\\
	0.3	6\\
	0.36	7\\
	0.43	8\\
	0.51	9\\
	0.6	10\\
	0.7	12\\
	0.82	15\\
	0.97	14\\
	1.11	9\\
	1.2	9\\
	1.29	6\\
	1.35	7\\
	1.42	6\\
	1.48	6\\
	1.54	6\\
	1.6	5\\
	1.65	5\\
	1.7	5\\
	1.75	5\\
	1.8	5\\
	1.85	5\\
	1.9	4\\
	1.94	4\\
	1.98	5\\
	2.03	3\\
	2.06	4\\
	2.1	4\\
	2.14	3\\
	2.17	4\\
	2.21	2\\
	2.23	2\\
	2.25	3\\
	2.28	3\\
	2.31	2\\
	2.33	3\\
	2.36	2\\
	2.38	2\\
	2.4	3\\
	2.43	2\\
	2.45	1\\
	2.46	2\\
	2.48	1\\
	2.49	1\\
	2.5	1\\
	2.51	1\\
	2.52	1\\
	2.53	1\\
	2.54	1\\
	2.55	1\\
	2.56	1\\
	2.57	1\\
	2.58	1\\
	2.59	1\\
	2.6	1\\
	2.61	1\\
	2.62	1\\
	2.63	1\\
	2.64	1\\
	2.65	1\\
	2.66	1\\
	2.67	1\\
	2.68	1\\
	2.69	1\\
	2.7	1\\
	2.71	1\\
	2.72	1\\
	2.73	1\\
	2.74	1\\
	2.75	1\\
	2.76	1\\
	2.77	1\\
	2.78	1\\
	2.79	1\\
	2.8	1\\
	2.81	1\\
	2.82	1\\
	2.83	1\\
	2.84	1\\
	2.85	1\\
	2.86	1\\
	2.87	1\\
	2.88	1\\
	2.89	1\\
	2.9	1\\
	2.91	1\\
	2.92	1\\
	2.93	1\\
	2.94	1\\
	2.95	1\\
	2.96	1\\
	2.97	1\\
	2.98	1\\
	2.99	1\\
	3	1\\
	3.01	1\\
	3.02	1\\
	3.03	1\\
	3.04	1\\
	3.05	1\\
	3.06	1\\
	3.07	1\\
	3.08	1\\
	3.09	1\\
	3.1	1\\
	3.11	1\\
	3.12	1\\
	3.13	1\\
	3.14	1\\
	3.15	1\\
	3.16	1\\
	3.17	1\\
	3.18	1\\
	3.19	1\\
	3.2	1\\
	3.21	1\\
	3.22	1\\
	3.23	1\\
	3.24	1\\
	3.25	1\\
	3.26	1\\
	3.27	1\\
	3.28	1\\
	3.29	1\\
	3.3	1\\
	3.31	1\\
	3.32	1\\
	3.33	1\\
	3.34	1\\
	3.35	1\\
	3.36	1\\
	3.37	1\\
	3.38	1\\
	3.39	1\\
	3.4	1\\
	3.41	1\\
	3.42	1\\
	3.43	1\\
	3.44	1\\
	3.45	1\\
	3.46	1\\
	3.47	1\\
	3.48	1\\
	3.49	1\\
	3.5	1\\
	3.51	1\\
	3.52	1\\
	3.53	1\\
	3.54	1\\
	3.55	1\\
	3.56	1\\
	3.57	1\\
	3.58	1\\
	3.59	1\\
	3.6	1\\
	3.61	1\\
	3.62	1\\
	3.63	1\\
	3.64	1\\
	3.65	1\\
	3.66	1\\
	3.67	1\\
	3.68	1\\
	3.69	1\\
	3.7	1\\
	3.71	1\\
	3.72	1\\
	3.73	1\\
	3.74	1\\
	3.75	1\\
	3.76	1\\
	3.77	1\\
	3.78	1\\
	3.79	1\\
	3.8	1\\
	3.81	1\\
	3.82	1\\
	3.83	1\\
	3.84	1\\
	3.85	1\\
	3.86	1\\
	3.87	1\\
	3.88	1\\
	3.89	1\\
	3.9	1\\
	3.91	1\\
	3.92	1\\
	3.93	1\\
	3.94	1\\
	3.95	1\\
	3.96	1\\
	3.97	1\\
	3.98	1\\
	3.99	1\\
	4	1\\
	4.01	1\\
	4.02	1\\
	4.03	1\\
	4.04	1\\
	4.05	1\\
	4.06	1\\
	4.07	1\\
	4.08	1\\
	4.09	1\\
	4.1	1\\
	4.11	1\\
	4.12	1\\
	4.13	1\\
	4.14	1\\
	4.15	1\\
	4.16	1\\
	4.17	1\\
	4.18	1\\
	4.19	1\\
	4.2	1\\
	4.21	1\\
	4.22	1\\
	4.23	1\\
	4.24	1\\
	4.25	1\\
	4.26	1\\
	4.27	1\\
	4.28	1\\
	4.29	1\\
	4.3	1\\
	4.31	1\\
	4.32	1\\
	4.33	1\\
	4.34	1\\
	4.35	1\\
	4.36	1\\
	4.37	1\\
	4.38	1\\
	4.39	1\\
	4.4	1\\
	4.41	1\\
	4.42	1\\
	4.43	1\\
	4.44	1\\
	4.45	1\\
	4.46	1\\
	4.47	1\\
	4.48	1\\
	4.49	1\\
	4.5	1\\
	4.51	1\\
	4.52	1\\
	4.53	1\\
	4.54	1\\
	4.55	1\\
	4.56	1\\
	4.57	1\\
	4.58	1\\
	4.59	1\\
	4.6	1\\
	4.61	1\\
	4.62	1\\
	4.63	1\\
	4.64	1\\
	4.65	1\\
	4.66	1\\
	4.67	1\\
	4.68	1\\
	4.69	1\\
	4.7	1\\
	4.71	1\\
	4.72	1\\
	4.73	1\\
	4.74	1\\
	4.75	1\\
	4.76	1\\
	4.77	1\\
	4.78	1\\
	4.79	1\\
	4.8	1\\
	4.81	1\\
	4.82	1\\
	4.83	1\\
	4.84	1\\
	4.85	1\\
	4.86	1\\
	4.87	1\\
	4.88	1\\
	4.89	1\\
	4.9	1\\
	4.91	1\\
	4.92	1\\
	4.93	1\\
	4.94	1\\
	4.95	1\\
	4.96	1\\
	4.97	1\\
	4.98	1\\
	4.99	1\\
	5	1\\
	5.01	1\\
	5.02	1\\
	5.03	1\\
	5.04	1\\
	5.05	1\\
	5.06	1\\
	5.07	1\\
	5.08	1\\
	5.09	1\\
	5.1	1\\
	5.11	1\\
	5.12	1\\
	5.13	1\\
	5.14	1\\
	5.15	1\\
	5.16	1\\
	5.17	1\\
	5.18	1\\
	5.19	1\\
	5.2	1\\
	5.21	1\\
	5.22	1\\
	5.23	1\\
	5.24	1\\
	5.25	1\\
	5.26	1\\
	5.27	1\\
	5.28	1\\
	5.29	1\\
	5.3	1\\
	5.31	1\\
	5.32	1\\
	5.33	1\\
	5.34	1\\
	5.35	1\\
	5.36	1\\
	5.37	1\\
	5.38	1\\
	5.39	1\\
	5.4	1\\
	5.41	1\\
	5.42	1\\
	5.43	1\\
	5.44	1\\
	5.45	1\\
	5.46	1\\
	5.47	1\\
	5.48	1\\
	5.49	1\\
	5.5	1\\
	5.51	1\\
	5.52	1\\
	5.53	1\\
	5.54	1\\
	5.55	1\\
	5.56	1\\
	5.57	1\\
	5.58	1\\
	5.59	1\\
	5.6	1\\
	5.61	1\\
	5.62	1\\
	5.63	1\\
	5.64	1\\
	5.65	1\\
	5.66	1\\
	5.67	1\\
	5.68	1\\
	5.69	1\\
	5.7	1\\
	5.71	1\\
	5.72	1\\
	5.73	1\\
	5.74	1\\
	5.75	1\\
	5.76	1\\
	5.77	1\\
	5.78	1\\
	5.79	1\\
	5.8	1\\
	5.81	1\\
	5.82	1\\
	5.83	1\\
	5.84	1\\
	5.85	1\\
	5.86	1\\
	5.87	1\\
	5.88	1\\
	5.89	1\\
	5.9	1\\
	5.91	1\\
	5.92	1\\
	5.93	1\\
	5.94	1\\
	5.95	1\\
	5.96	1\\
	5.97	1\\
	5.98	1\\
	5.99	1\\
	6	1\\
	6.01	1\\
	6.02	1\\
	6.03	1\\
	6.04	1\\
	6.05	1\\
	6.06	1\\
	6.07	1\\
	6.08	1\\
	6.09	1\\
	6.1	1\\
	6.11	1\\
	6.12	1\\
	6.13	1\\
	6.14	1\\
	6.15	1\\
	6.16	1\\
	6.17	1\\
	6.18	1\\
	6.19	1\\
	6.2	1\\
	6.21	1\\
	6.22	1\\
	6.23	1\\
	6.24	1\\
	6.25	1\\
	6.26	1\\
	6.27	1\\
	6.28	1\\
	6.29	1\\
	6.3	1\\
	6.31	1\\
	6.32	1\\
	6.33	1\\
	6.34	1\\
	6.35	1\\
	6.36	1\\
	6.37	1\\
	6.38	1\\
	6.39	1\\
	6.4	1\\
	6.41	1\\
	6.42	1\\
	6.43	1\\
	6.44	1\\
	6.45	1\\
	6.46	1\\
	6.47	1\\
	6.48	1\\
	6.49	1\\
	6.5	1\\
	6.51	1\\
	6.52	1\\
	6.53	1\\
	6.54	1\\
	6.55	1\\
	6.56	1\\
	6.57	1\\
	6.58	1\\
	6.59	1\\
	6.6	1\\
	6.61	1\\
	6.62	1\\
	6.63	1\\
	6.64	1\\
	6.65	1\\
	6.66	1\\
	6.67	1\\
	6.68	1\\
	6.69	1\\
	6.7	1\\
	6.71	1\\
	6.72	1\\
	6.73	1\\
	6.74	1\\
	6.75	1\\
	6.76	1\\
	6.77	1\\
	6.78	1\\
	6.79	1\\
	6.8	1\\
	6.81	1\\
	6.82	1\\
	6.83	1\\
	6.84	1\\
	6.85	1\\
	6.86	1\\
	6.87	1\\
	6.88	1\\
	6.89	1\\
	6.9	1\\
	6.91	1\\
	6.92	1\\
	6.93	1\\
	6.94	1\\
	6.95	1\\
	6.96	1\\
	6.97	1\\
	6.98	1\\
	6.99	1\\
	7	1\\
	7.01	1\\
	7.02	1\\
	7.03	1\\
	7.04	1\\
	7.05	1\\
	7.06	1\\
	7.07	1\\
	7.08	1\\
	7.09	1\\
	7.1	1\\
	7.11	1\\
	7.12	1\\
	7.13	1\\
	7.14	1\\
	7.15	1\\
	7.16	1\\
	7.17	1\\
	7.18	1\\
	7.19	1\\
	7.2	1\\
	7.21	1\\
	7.22	1\\
	7.23	1\\
	7.24	1\\
	7.25	1\\
	7.26	1\\
	7.27	1\\
	7.28	1\\
	7.29	1\\
	7.3	1\\
	7.31	1\\
	7.32	1\\
	7.33	1\\
	7.34	1\\
	7.35	1\\
	7.36	1\\
	7.37	1\\
	7.38	1\\
	7.39	1\\
	7.4	1\\
	7.41	1\\
	7.42	1\\
	7.43	1\\
	7.44	1\\
	7.45	1\\
	7.46	1\\
	7.47	1\\
	7.48	1\\
	7.49	1\\
	7.5	1\\
	7.51	1\\
	7.52	1\\
	7.53	1\\
	7.54	1\\
	7.55	1\\
	7.56	1\\
	7.57	1\\
	7.58	1\\
	7.59	1\\
	7.6	1\\
	7.61	1\\
	7.62	1\\
	7.63	1\\
	7.64	1\\
	7.65	1\\
	7.66	1\\
	7.67	1\\
	7.68	1\\
	7.69	1\\
	7.7	1\\
	7.71	1\\
	7.72	1\\
	7.73	1\\
	7.74	1\\
	7.75	1\\
	7.76	1\\
	7.77	1\\
	7.78	1\\
	7.79	1\\
	7.8	1\\
	7.81	1\\
	7.82	1\\
	7.83	1\\
	7.84	1\\
	7.85	1\\
	7.86	1\\
	7.87	1\\
	7.88	1\\
	7.89	1\\
	7.9	1\\
	7.91	1\\
	7.92	1\\
	7.93	1\\
	7.94	1\\
	7.95	1\\
	7.96	1\\
	7.97	1\\
	7.98	1\\
	7.99	1\\
	8	1\\
	8.01	1\\
	8.02	1\\
	8.03	1\\
	8.04	1\\
	8.05	1\\
	8.06	1\\
	8.07	1\\
	8.08	1\\
	8.09	1\\
	8.1	1\\
	8.11	1\\
	8.12	1\\
	8.13	1\\
	8.14	1\\
	8.15	1\\
	8.16	1\\
	8.17	1\\
	8.18	1\\
	8.19	1\\
	8.2	1\\
	8.21	1\\
	8.22	1\\
	8.23	1\\
	8.24	1\\
	8.25	1\\
	8.26	1\\
	8.27	1\\
	8.28	1\\
	8.29	1\\
	8.3	1\\
	8.31	1\\
	8.32	1\\
	8.33	1\\
	8.34	1\\
	8.35	1\\
	8.36	1\\
	8.37	1\\
	8.38	1\\
	8.39	1\\
	8.4	1\\
	8.41	1\\
	8.42	1\\
	8.43	1\\
	8.44	1\\
	8.45	1\\
	8.46	1\\
	8.47	1\\
	8.48	1\\
	8.49	1\\
	8.5	1\\
	8.51	1\\
	8.52	1\\
	8.53	1\\
	8.54	1\\
	8.55	1\\
	8.56	1\\
	8.57	1\\
	8.58	1\\
	8.59	1\\
	8.6	1\\
	8.61	1\\
	8.62	1\\
	8.63	1\\
	8.64	1\\
	8.65	1\\
	8.66	1\\
	8.67	1\\
	8.68	1\\
	8.69	1\\
	8.7	1\\
	8.71	1\\
	8.72	1\\
	8.73	1\\
	8.74	1\\
	8.75	1\\
	8.76	1\\
	8.77	1\\
	8.78	1\\
	8.79	1\\
	8.8	1\\
	8.81	1\\
	8.82	1\\
	8.83	1\\
	8.84	1\\
	8.85	1\\
	8.86	1\\
	8.87	1\\
	8.88	1\\
	8.89	1\\
	8.9	1\\
	8.91	1\\
	8.92	1\\
	8.93	1\\
	8.94	1\\
	8.95	1\\
	8.96	1\\
	8.97	1\\
	8.98	1\\
	8.99	1\\
	9	1\\
	9.01	1\\
	9.02	1\\
	9.03	1\\
	9.04	1\\
	9.05	1\\
	9.06	1\\
	9.07	1\\
	9.08	1\\
	9.09	1\\
	9.1	1\\
	9.11	1\\
	9.12	1\\
	9.13	1\\
	9.14	1\\
	9.15	1\\
	9.16	1\\
	9.17	1\\
	9.18	1\\
	9.19	1\\
	9.2	1\\
	9.21	1\\
	9.22	1\\
	9.23	1\\
	9.24	1\\
	9.25	1\\
	9.26	1\\
	9.27	1\\
	9.28	1\\
	9.29	1\\
	9.3	1\\
	9.31	1\\
	9.32	1\\
	9.33	1\\
	9.34	1\\
	9.35	1\\
	9.36	1\\
	9.37	1\\
	9.38	1\\
	9.39	1\\
	9.4	1\\
	9.41	1\\
	9.42	1\\
	9.43	1\\
	9.44	1\\
	9.45	1\\
	9.46	1\\
	9.47	1\\
	9.48	1\\
	9.49	1\\
	9.5	1\\
	9.51	1\\
	9.52	1\\
	9.53	1\\
	9.54	1\\
	9.55	1\\
	9.56	1\\
	9.57	1\\
	9.58	1\\
	9.59	1\\
	9.6	1\\
	9.61	1\\
	9.62	1\\
	9.63	1\\
	9.64	1\\
	9.65	1\\
	9.66	1\\
	9.67	1\\
	9.68	1\\
	9.69	1\\
	9.7	1\\
	9.71	1\\
	9.72	1\\
	9.73	1\\
	9.74	1\\
	9.75	1\\
	9.76	1\\
	9.77	1\\
	9.78	1\\
	9.79	1\\
	9.8	1\\
	9.81	1\\
	9.82	1\\
	9.83	1\\
	9.84	1\\
	9.85	1\\
	9.86	1\\
	9.87	1\\
	9.88	1\\
	9.89	1\\
	9.9	1\\
	9.91	1\\
	9.92	1\\
	9.93	1\\
	9.94	1\\
	9.95	1\\
	9.96	1\\
	9.97	1\\
	9.98	1\\
	9.99	1\\
	10	1\\
};
\addlegendentry{PSTC}

\addplot [color=mycolor2, line width=1.0pt]
  table[row sep=crcr]{%
  	0	3\\
  	0.01	3\\
  	0.03	2\\
  	0.05	2\\
  	0.07	2\\
  	0.09	2\\
  	0.1	2\\
  	0.11	2\\
  	0.13	3\\
  	0.15	4\\
  	0.19	6\\
  	0.24	6\\
  	0.3	7\\
  	0.36	7\\
  	0.43	8\\
  	0.51	9\\
  	0.6	11\\
  	0.7	13\\
  	0.82	20\\
  	0.97	17\\
  	1.11	10\\
  	1.2	11\\
  	1.29	7\\
  	1.35	8\\
  	1.42	7\\
  	1.48	7\\
  	1.54	7\\
  	1.6	6\\
  	1.65	6\\
  	1.7	6\\
  	1.75	6\\
  	1.8	6\\
  	1.85	6\\
  	1.9	6\\
  	1.94	6\\
  	1.98	6\\
  	2.03	5\\
  	2.06	6\\
  	2.1	7\\
  	2.14	5\\
  	2.17	7\\
  	2.21	5\\
  	2.23	5\\
  	2.25	7\\
  	2.28	6\\
  	2.31	6\\
  	2.33	7\\
  	2.36	6\\
  	2.38	6\\
  	2.4	8\\
  	2.43	7\\
  	2.45	4\\
  	2.46	8\\
  	2.48	7\\
  	2.49	7\\
  	2.5	5\\
  	2.51	6\\
  	2.52	7\\
  	2.53	7\\
  	2.54	5\\
  	2.55	7\\
  	2.56	6\\
  	2.57	7\\
  	2.58	5\\
  	2.59	5\\
  	2.6	8\\
  	2.61	7\\
  	2.62	5\\
  	2.63	7\\
  	2.64	6\\
  	2.65	8\\
  	2.66	6\\
  	2.67	7\\
  	2.68	6\\
  	2.69	6\\
  	2.7	6\\
  	2.71	6\\
  	2.72	7\\
  	2.73	5\\
  	2.74	6\\
  	2.75	6\\
  	2.76	8\\
  	2.77	7\\
  	2.78	8\\
  	2.79	7\\
  	2.8	6\\
  	2.81	7\\
  	2.82	7\\
  	2.83	7\\
  	2.84	6\\
  	2.85	7\\
  	2.86	6\\
  	2.87	7\\
  	2.88	7\\
  	2.89	6\\
  	2.9	4\\
  	2.91	7\\
  	2.92	5\\
  	2.93	5\\
  	2.94	6\\
  	2.95	6\\
  	2.96	4\\
  	2.97	6\\
  	2.98	6\\
  	2.99	6\\
  	3	8\\
  	3.01	7\\
  	3.02	4\\
  	3.03	5\\
  	3.04	7\\
  	3.05	6\\
  	3.06	6\\
  	3.07	7\\
  	3.08	5\\
  	3.09	5\\
  	3.1	7\\
  	3.11	4\\
  	3.12	3\\
  	3.13	7\\
  	3.14	6\\
  	3.15	6\\
  	3.16	3\\
  	3.17	7\\
  	3.18	5\\
  	3.19	5\\
  	3.2	7\\
  	3.21	3\\
  	3.22	7\\
  	3.23	3\\
  	3.24	7\\
  	3.25	2\\
  	3.26	6\\
  	3.27	6\\
  	3.28	7\\
  	3.29	4\\
  	3.3	4\\
  	3.31	3\\
  	3.32	4\\
  	3.33	4\\
  	3.34	7\\
  	3.35	4\\
  	3.36	7\\
  	3.37	7\\
  	3.38	7\\
  	3.39	6\\
  	3.4	7\\
  	3.41	6\\
  	3.42	5\\
  	3.43	5\\
  	3.44	7\\
  	3.45	6\\
  	3.46	6\\
  	3.47	4\\
  	3.48	7\\
  	3.49	6\\
  	3.5	5\\
  	3.51	4\\
  	3.52	2\\
  	3.53	3\\
  	3.54	1\\
  	3.55	5\\
  	3.56	3\\
  	3.57	6\\
  	3.58	3\\
  	3.59	3\\
  	3.6	6\\
  	3.61	6\\
  	3.62	7\\
  	3.63	5\\
  	3.64	4\\
  	3.65	1\\
  	3.66	4\\
  	3.67	4\\
  	3.68	2\\
  	3.69	2\\
  	3.7	1\\
  	3.71	5\\
  	3.72	2\\
  	3.73	4\\
  	3.74	4\\
  	3.75	6\\
  	3.76	3\\
  	3.77	5\\
  	3.78	7\\
  	3.79	1\\
  	3.8	7\\
  	3.81	1\\
  	3.82	2\\
  	3.83	5\\
  	3.84	5\\
  	3.85	3\\
  	3.86	2\\
  	3.87	5\\
  	3.88	1\\
  	3.89	5\\
  	3.9	4\\
  	3.91	3\\
  	3.92	7\\
  	3.93	7\\
  	3.94	1\\
  	3.95	2\\
  	3.96	4\\
  	3.97	5\\
  	3.98	6\\
  	3.99	5\\
  	4	2\\
  	4.01	1\\
  	4.02	2\\
  	4.03	2\\
  	4.04	1\\
  	4.05	3\\
  	4.06	3\\
  	4.07	3\\
  	4.08	2\\
  	4.09	1\\
  	4.1	4\\
  	4.11	5\\
  	4.12	2\\
  	4.13	1\\
  	4.14	1\\
  	4.15	1\\
  	4.16	2\\
  	4.17	4\\
  	4.18	1\\
  	4.19	3\\
  	4.2	2\\
  	4.21	1\\
  	4.22	1\\
  	4.23	3\\
  	4.24	3\\
  	4.25	1\\
  	4.26	1\\
  	4.27	3\\
  	4.28	2\\
  	4.29	1\\
  	4.3	3\\
  	4.31	2\\
  	4.32	2\\
  	4.33	1\\
  	4.34	1\\
  	4.35	1\\
  	4.36	1\\
  	4.37	3\\
  	4.38	2\\
  	4.39	1\\
  	4.4	1\\
  	4.41	1\\
  	4.42	1\\
  	4.43	1\\
  	4.44	2\\
  	4.45	1\\
  	4.46	2\\
  	4.47	2\\
  	4.48	1\\
  	4.49	2\\
  	4.5	1\\
  	4.51	1\\
  	4.52	1\\
  	4.53	1\\
  	4.54	2\\
  	4.55	2\\
  	4.56	1\\
  	4.57	1\\
  	4.58	4\\
  	4.59	1\\
  	4.6	2\\
  	4.61	1\\
  	4.62	3\\
  	4.63	1\\
  	4.64	1\\
  	4.65	4\\
  	4.66	1\\
  	4.67	3\\
  	4.68	1\\
  	4.69	1\\
  	4.7	1\\
  	4.71	1\\
  	4.72	1\\
  	4.73	2\\
  	4.74	1\\
  	4.75	1\\
  	4.76	1\\
  	4.77	3\\
  	4.78	2\\
  	4.79	2\\
  	4.8	1\\
  	4.81	1\\
  	4.82	2\\
  	4.83	1\\
  	4.84	1\\
  	4.85	1\\
  	4.86	1\\
  	4.87	1\\
  	4.88	3\\
  	4.89	2\\
  	4.9	1\\
  	4.91	1\\
  	4.92	1\\
  	4.93	1\\
  	4.94	1\\
  	4.95	1\\
  	4.96	1\\
  	4.97	1\\
  	4.98	1\\
  	4.99	1\\
  	5	1\\
  	5.01	1\\
  	5.02	1\\
  	5.03	1\\
  	5.04	1\\
  	5.05	1\\
  	5.06	2\\
  	5.07	1\\
  	5.08	1\\
  	5.09	1\\
  	5.1	1\\
  	5.11	1\\
  	5.12	1\\
  	5.13	1\\
  	5.14	1\\
  	5.15	1\\
  	5.16	1\\
  	5.17	1\\
  	5.18	1\\
  	5.19	1\\
  	5.2	1\\
  	5.21	1\\
  	5.22	1\\
  	5.23	1\\
  	5.24	1\\
  	5.25	1\\
  	5.26	2\\
  	5.27	1\\
  	5.28	1\\
  	5.29	1\\
  	5.3	1\\
  	5.31	1\\
  	5.32	2\\
  	5.33	1\\
  	5.34	1\\
  	5.35	1\\
  	5.36	1\\
  	5.37	1\\
  	5.38	1\\
  	5.39	1\\
  	5.4	1\\
  	5.41	1\\
  	5.42	2\\
  	5.43	1\\
  	5.44	1\\
  	5.45	1\\
  	5.46	1\\
  	5.47	1\\
  	5.48	1\\
  	5.49	1\\
  	5.5	1\\
  	5.51	1\\
  	5.52	1\\
  	5.53	2\\
  	5.54	1\\
  	5.55	1\\
  	5.56	1\\
  	5.57	1\\
  	5.58	1\\
  	5.59	1\\
  	5.6	1\\
  	5.61	1\\
  	5.62	1\\
  	5.63	1\\
  	5.64	2\\
  	5.65	1\\
  	5.66	1\\
  	5.67	1\\
  	5.68	1\\
  	5.69	1\\
  	5.7	1\\
  	5.71	1\\
  	5.72	1\\
  	5.73	1\\
  	5.74	1\\
  	5.75	1\\
  	5.76	2\\
  	5.77	1\\
  	5.78	1\\
  	5.79	2\\
  	5.8	1\\
  	5.81	1\\
  	5.82	1\\
  	5.83	1\\
  	5.84	1\\
  	5.85	1\\
  	5.86	1\\
  	5.87	1\\
  	5.88	1\\
  	5.89	1\\
  	5.9	1\\
  	5.91	1\\
  	5.92	1\\
  	5.93	1\\
  	5.94	1\\
  	5.95	1\\
  	5.96	1\\
  	5.97	1\\
  	5.98	1\\
  	5.99	1\\
  	6	1\\
  	6.01	1\\
  	6.02	1\\
  	6.03	1\\
  	6.04	1\\
  	6.05	1\\
  	6.06	1\\
  	6.07	1\\
  	6.08	1\\
  	6.09	1\\
  	6.1	1\\
  	6.11	1\\
  	6.12	1\\
  	6.13	1\\
  	6.14	1\\
  	6.15	2\\
  	6.16	1\\
  	6.17	1\\
  	6.18	2\\
  	6.19	1\\
  	6.2	1\\
  	6.21	1\\
  	6.22	1\\
  	6.23	1\\
  	6.24	1\\
  	6.25	1\\
  	6.26	1\\
  	6.27	1\\
  	6.28	1\\
  	6.29	1\\
  	6.3	1\\
  	6.31	1\\
  	6.32	1\\
  	6.33	1\\
  	6.34	1\\
  	6.35	1\\
  	6.36	1\\
  	6.37	1\\
  	6.38	1\\
  	6.39	1\\
  	6.4	1\\
  	6.41	1\\
  	6.42	1\\
  	6.43	1\\
  	6.44	1\\
  	6.45	1\\
  	6.46	1\\
  	6.47	1\\
  	6.48	1\\
  	6.49	1\\
  	6.5	1\\
  	6.51	1\\
  	6.52	1\\
  	6.53	1\\
  	6.54	1\\
  	6.55	1\\
  	6.56	1\\
  	6.57	1\\
  	6.58	1\\
  	6.59	1\\
  	6.6	1\\
  	6.61	1\\
  	6.62	2\\
  	6.63	1\\
  	6.64	2\\
  	6.65	1\\
  	6.66	1\\
  	6.67	1\\
  	6.68	1\\
  	6.69	1\\
  	6.7	1\\
  	6.71	1\\
  	6.72	1\\
  	6.73	1\\
  	6.74	1\\
  	6.75	2\\
  	6.76	1\\
  	6.77	1\\
  	6.78	1\\
  	6.79	1\\
  	6.8	1\\
  	6.81	1\\
  	6.82	1\\
  	6.83	1\\
  	6.84	1\\
  	6.85	1\\
  	6.86	1\\
  	6.87	1\\
  	6.88	1\\
  	6.89	3\\
  	6.9	2\\
  	6.91	1\\
  	6.92	2\\
  	6.93	1\\
  	6.94	1\\
  	6.95	1\\
  	6.96	1\\
  	6.97	1\\
  	6.98	1\\
  	6.99	1\\
  	7	1\\
  	7.01	2\\
  	7.02	1\\
  	7.03	1\\
  	7.04	1\\
  	7.05	1\\
  	7.06	1\\
  	7.07	1\\
  	7.08	1\\
  	7.09	1\\
  	7.1	1\\
  	7.11	1\\
  	7.12	1\\
  	7.13	1\\
  	7.14	1\\
  	7.15	1\\
  	7.16	1\\
  	7.17	1\\
  	7.18	1\\
  	7.19	1\\
  	7.2	1\\
  	7.21	1\\
  	7.22	1\\
  	7.23	2\\
  	7.24	1\\
  	7.25	1\\
  	7.26	2\\
  	7.27	1\\
  	7.28	1\\
  	7.29	1\\
  	7.3	1\\
  	7.31	1\\
  	7.32	1\\
  	7.33	1\\
  	7.34	1\\
  	7.35	1\\
  	7.36	2\\
  	7.37	1\\
  	7.38	1\\
  	7.39	2\\
  	7.4	1\\
  	7.41	1\\
  	7.42	1\\
  	7.43	1\\
  	7.44	1\\
  	7.45	1\\
  	7.46	1\\
  	7.47	2\\
  	7.48	1\\
  	7.49	1\\
  	7.5	1\\
  	7.51	1\\
  	7.52	1\\
  	7.53	1\\
  	7.54	1\\
  	7.55	1\\
  	7.56	1\\
  	7.57	1\\
  	7.58	1\\
  	7.59	1\\
  	7.6	2\\
  	7.61	1\\
  	7.62	1\\
  	7.63	1\\
  	7.64	1\\
  	7.65	1\\
  	7.66	1\\
  	7.67	1\\
  	7.68	1\\
  	7.69	1\\
  	7.7	1\\
  	7.71	1\\
  	7.72	1\\
  	7.73	2\\
  	7.74	1\\
  	7.75	1\\
  	7.76	1\\
  	7.77	2\\
  	7.78	1\\
  	7.79	1\\
  	7.8	1\\
  	7.81	1\\
  	7.82	1\\
  	7.83	1\\
  	7.84	1\\
  	7.85	1\\
  	7.86	1\\
  	7.87	1\\
  	7.88	1\\
  	7.89	1\\
  	7.9	1\\
  	7.91	1\\
  	7.92	4\\
  	7.93	3\\
  	7.94	2\\
  	7.95	1\\
  	7.96	1\\
  	7.97	1\\
  	7.98	1\\
  	7.99	1\\
  	8	1\\
  	8.01	1\\
  	8.02	1\\
  	8.03	1\\
  	8.04	1\\
  	8.05	1\\
  	8.06	1\\
  	8.07	1\\
  	8.08	1\\
  	8.09	1\\
  	8.1	1\\
  	8.11	1\\
  	8.12	1\\
  	8.13	1\\
  	8.14	1\\
  	8.15	1\\
  	8.16	1\\
  	8.17	1\\
  	8.18	1\\
  	8.19	1\\
  	8.2	1\\
  	8.21	1\\
  	8.22	1\\
  	8.23	1\\
  	8.24	1\\
  	8.25	2\\
  	8.26	1\\
  	8.27	1\\
  	8.28	1\\
  	8.29	1\\
  	8.3	1\\
  	8.31	1\\
  	8.32	1\\
  	8.33	1\\
  	8.34	1\\
  	8.35	1\\
  	8.36	1\\
  	8.37	1\\
  	8.38	1\\
  	8.39	1\\
  	8.4	1\\
  	8.41	1\\
  	8.42	1\\
  	8.43	1\\
  	8.44	1\\
  	8.45	1\\
  	8.46	1\\
  	8.47	1\\
  	8.48	1\\
  	8.49	1\\
  	8.5	1\\
  	8.51	1\\
  	8.52	1\\
  	8.53	2\\
  	8.54	1\\
  	8.55	1\\
  	8.56	1\\
  	8.57	1\\
  	8.58	1\\
  	8.59	1\\
  	8.6	1\\
  	8.61	1\\
  	8.62	1\\
  	8.63	1\\
  	8.64	1\\
  	8.65	1\\
  	8.66	1\\
  	8.67	1\\
  	8.68	1\\
  	8.69	1\\
  	8.7	1\\
  	8.71	1\\
  	8.72	1\\
  	8.73	1\\
  	8.74	1\\
  	8.75	1\\
  	8.76	1\\
  	8.77	1\\
  	8.78	1\\
  	8.79	1\\
  	8.8	1\\
  	8.81	1\\
  	8.82	1\\
  	8.83	1\\
  	8.84	1\\
  	8.85	1\\
  	8.86	1\\
  	8.87	1\\
  	8.88	1\\
  	8.89	1\\
  	8.9	1\\
  	8.91	1\\
  	8.92	1\\
  	8.93	2\\
  	8.94	1\\
  	8.95	1\\
  	8.96	1\\
  	8.97	1\\
  	8.98	1\\
  	8.99	1\\
  	9	1\\
  	9.01	1\\
  	9.02	1\\
  	9.03	1\\
  	9.04	1\\
  	9.05	1\\
  	9.06	1\\
  	9.07	1\\
  	9.08	2\\
  	9.09	1\\
  	9.1	1\\
  	9.11	1\\
  	9.12	2\\
  	9.13	1\\
  	9.14	2\\
  	9.15	1\\
  	9.16	1\\
  	9.17	1\\
  	9.18	1\\
  	9.19	1\\
  	9.2	1\\
  	9.21	1\\
  	9.22	1\\
  	9.23	1\\
  	9.24	1\\
  	9.25	2\\
  	9.26	1\\
  	9.27	1\\
  	9.28	1\\
  	9.29	1\\
  	9.3	1\\
  	9.31	1\\
  	9.32	1\\
  	9.33	1\\
  	9.34	1\\
  	9.35	1\\
  	9.36	1\\
  	9.37	1\\
  	9.38	1\\
  	9.39	1\\
  	9.4	2\\
  	9.41	1\\
  	9.42	1\\
  	9.43	1\\
  	9.44	1\\
  	9.45	1\\
  	9.46	1\\
  	9.47	1\\
  	9.48	1\\
  	9.49	1\\
  	9.5	1\\
  	9.51	1\\
  	9.52	1\\
  	9.53	1\\
  	9.54	1\\
  	9.55	1\\
  	9.56	1\\
  	9.57	1\\
  	9.58	1\\
  	9.59	2\\
  	9.6	1\\
  	9.61	1\\
  	9.62	1\\
  	9.63	1\\
  	9.64	2\\
  	9.65	1\\
  	9.66	1\\
  	9.67	1\\
  	9.68	1\\
  	9.69	1\\
  	9.7	1\\
  	9.71	1\\
  	9.72	1\\
  	9.73	1\\
  	9.74	1\\
  	9.75	1\\
  	9.76	1\\
  	9.77	1\\
  	9.78	1\\
  	9.79	1\\
  	9.8	1\\
  	9.81	2\\
  	9.82	1\\
  	9.83	1\\
  	9.84	1\\
  	9.85	1\\
  	9.86	1\\
  	9.87	1\\
  	9.88	1\\
  	9.89	1\\
  	9.9	1\\
  	9.91	1\\
  	9.92	1\\
  	9.93	1\\
  	9.94	1\\
  	9.95	1\\
  	9.96	2\\
  	9.97	1\\
  	9.98	1\\
  	9.99	1\\
  	10	1\\
  };
\addlegendentry{PETC}

\node[draw=none, fill=white] (title) at (5,20) {$\epsilon = 0$};
\end{axis}

\begin{axis}[%
clip mode=individual,
width=70.00mm,
height=20.93mm,
at={(0mm,0mm)},
scale only axis,
xmin=0,
xmax=10,
ymin=0,
ymax=25.5,
ylabel style={font=\color{white!15!black}},
ylabel={$\dk^*$ (samples)},
xlabel={$t$},
axis background/.style={fill=white},
axis x line*=bottom,
axis y line*=left,
xmajorgrids,
ymajorgrids,
]
\addplot[only marks, mark=o, mark options={}, mark size=3.0000pt, draw=mycolor1] table[row sep=crcr]{%
	x	y\\
	0	1\\
	0.01	2\\
	0.03	2\\
	0.05	2\\
	0.07	2\\
	0.09	1\\
	0.1	1\\
	0.11	2\\
	0.13	2\\
	0.15	4\\
	0.19	6\\
	0.25	6\\
	0.31	7\\
	0.38	7\\
	0.45	8\\
	0.53	9\\
	0.62	11\\
	0.73	13\\
	0.86	16\\
	1.02	14\\
	1.16	11\\
	1.27	9\\
	1.36	9\\
	1.45	8\\
	1.53	7\\
	1.6	8\\
	1.68	8\\
	1.76	7\\
	1.83	9\\
	1.92	7\\
	1.99	9\\
	2.08	10\\
	2.18	9\\
	2.27	8\\
	2.35	11\\
	2.46	11\\
	2.57	12\\
	2.69	10\\
	2.79	11\\
	2.9	10\\
	3	11\\
	3.11	8\\
	3.19	12\\
	3.31	12\\
	3.43	11\\
	3.54	11\\
	3.65	12\\
	3.77	12\\
	3.89	13\\
	4.02	15\\
	4.17	15\\
	4.32	13\\
	4.45	13\\
	4.58	10\\
	4.68	15\\
	4.83	12\\
	4.95	14\\
	5.09	12\\
	5.21	15\\
	5.36	15\\
	5.51	15\\
	5.66	12\\
	5.78	15\\
	5.93	15\\
	6.08	14\\
	6.22	15\\
	6.37	15\\
	6.52	15\\
	6.67	15\\
	6.82	13\\
	6.95	15\\
	7.1	15\\
	7.25	15\\
	7.4	15\\
	7.55	15\\
	7.7	11\\
	7.81	12\\
	7.93	12\\
	8.05	15\\
	8.2	13\\
	8.33	12\\
	8.45	11\\
	8.56	9\\
	8.65	10\\
	8.75	12\\
	8.87	15\\
	9.02	15\\
	9.17	15\\
	9.32	14\\
	9.46	12\\
	9.58	12\\
	9.7	12\\
	9.82	14\\
	9.96	15\\
};

\addplot [color=mycolor2, line width=1.0pt]
  table[row sep=crcr]{%
  	0	3\\
  	0.01	3\\
  	0.03	2\\
  	0.05	2\\
  	0.07	2\\
  	0.09	2\\
  	0.1	2\\
  	0.11	2\\
  	0.13	3\\
  	0.15	4\\
  	0.19	6\\
  	0.25	6\\
  	0.31	7\\
  	0.38	7\\
  	0.45	8\\
  	0.53	10\\
  	0.62	12\\
  	0.73	16\\
  	0.86	21\\
  	1.02	16\\
  	1.16	13\\
  	1.27	10\\
  	1.36	10\\
  	1.45	9\\
  	1.53	8\\
  	1.6	10\\
  	1.68	9\\
  	1.76	8\\
  	1.83	11\\
  	1.92	9\\
  	1.99	11\\
  	2.08	12\\
  	2.18	12\\
  	2.27	11\\
  	2.35	15\\
  	2.46	15\\
  	2.57	15\\
  	2.69	12\\
  	2.79	22\\
  	2.9	14\\
  	3	23\\
  	3.11	13\\
  	3.19	25\\
  	3.31	17\\
  	3.43	25\\
  	3.54	17\\
  	3.65	25\\
  	3.77	25\\
  	3.89	25\\
  	4.02	25\\
  	4.17	25\\
  	4.32	25\\
  	4.45	25\\
  	4.58	25\\
  	4.68	25\\
  	4.83	25\\
  	4.95	25\\
  	5.09	25\\
  	5.21	25\\
  	5.36	25\\
  	5.51	25\\
  	5.66	25\\
  	5.78	25\\
  	5.93	25\\
  	6.08	25\\
  	6.22	25\\
  	6.37	25\\
  	6.52	25\\
  	6.67	25\\
  	6.82	25\\
  	6.95	25\\
  	7.1	25\\
  	7.25	25\\
  	7.4	25\\
  	7.55	25\\
  	7.7	22\\
  	7.81	25\\
  	7.93	25\\
  	8.05	25\\
  	8.2	25\\
  	8.33	25\\
  	8.45	21\\
  	8.56	19\\
  	8.65	22\\
  	8.75	25\\
  	8.87	25\\
  	9.02	25\\
  	9.17	25\\
  	9.32	25\\
  	9.46	25\\
  	9.58	25\\
  	9.7	25\\
  	9.82	25\\
  	9.96	25\\
  };

\node[draw=none, fill=white] (title) at (5,20) {$\epsilon = 0.1$};
\end{axis}
\end{tikzpicture}%

%% file: arxiv.bbl
\begin{thebibliography}{10}

\bibitem{almeida2014self}
Jo{\~a}o Almeida, Carlos Silvestre, and Ant{\'o}nio~M. Pascoal.
\newblock Self-triggered output feedback control of linear plants in the
  presence of unknown disturbances.
\newblock {\em IEEE Transactions on Automatic Control}, 59(11):3040--3045,
  2014.

\bibitem{anta2008self}
Adolfo Anta and Paulo Tabuada.
\newblock Self-triggered stabilization of homogeneous control systems.
\newblock In {\em American Control Conference, 2008}, pages 4129--4134. IEEE,
  2008.

\bibitem{aastrom2002comparison}
Karl~Johan {\AA}str{\"o}m and Bo~Bernhardsson.
\newblock Comparison of riemann and lebesque sampling for first order
  stochastic systems.
\newblock In {\em Proceedings of the 41st IEEE Conference on Decision and
  Control, 2002}, volume~2, pages 2011--2016. IEEE, 2002.

\bibitem{blanchini2008set}
Franco Blanchini and Stefano Miani.
\newblock {\em Set-theoretic methods in control}.
\newblock Springer, 2008.

\bibitem{borgers2014event}
DPN Borgers and WPMH Heemels.
\newblock Event-separation properties of event-triggered control systems.
\newblock {\em IEEE Transactions on Automatic Control}, 59(10):2644--2656,
  2014.

\bibitem{boyd2004convex}
Stephen Boyd and Lieven Vandenberghe.
\newblock {\em Convex optimization}.
\newblock Cambridge university press, 2004.

\bibitem{brunner2019event}
Florian~David Brunner, W~P M~H Heemels, and Frank Allg{\"o}wer.
\newblock Event-triggered and self-triggered control for linear systems based
  on reachable sets.
\newblock {\em Automatica}, 101:15--26, 2019.

\bibitem{cai2009characterizations}
Chaohong Cai and Andrew~R Teel.
\newblock Characterizations of input-to-state stability for hybrid systems.
\newblock {\em Systems \& Control Letters}, 58(1):47--53, 2009.

\bibitem{demmel2007fast}
James Demmel, Ioana Dumitriu, and Olga Holtz.
\newblock Fast linear algebra is stable.
\newblock {\em Numerische Mathematik}, 108(1):59--91, 2007.

\bibitem{el2001robust}
Laurent El~Ghaoui and Giuseppe Calafiore.
\newblock Robust filtering for discrete-time systems with bounded noise and
  parametric uncertainty.
\newblock {\em IEEE Transactions on Automatic Control}, 46(7):1084--1089, 2001.

\bibitem{gleizer2018selftriggered}
Gabriel~A Gleizer and Manuel Mazo~Jr.
\newblock Self-triggered output feedback control for perturbed linear systems.
\newblock {\em IFAC-PapersOnLine}, 51(23):248--253, 2018.

\bibitem{gopal1993modern}
Madan Gopal.
\newblock {\em Modern control system theory}.
\newblock New Age International, 1993.

\bibitem{heemels2013periodic}
WPMH Heemels, MCF Donkers, and AR~Teel.
\newblock Periodic event-triggered control for linear systems.
\newblock {\em IEEE Transactions on Automatic Control}, 58(4):847--861, 2013.

\bibitem{kolarijani2016formal}
Arman~Sharifi Kolarijani and Manuel Mazo~Jr.
\newblock A formal traffic characterization of {LTI} event-triggered control
  systems.
\newblock {\em IEEE Transactions on Control of Network Systems}, 2016.

\bibitem{kurzhanski1997ellipsoidal}
Alexander~B. Kurzhanski{\u\i} and Istv{\'a}n V{\'a}lyi.
\newblock {\em Ellipsoidal calculus for estimation and control}.
\newblock Birkh{\"a}user, 1997.

\bibitem{kurzhanski2006ellipsoidal}
Alexander~B. Kurzhanski{\u\i} and Istv{\'a}n V{\'a}lyi.
\newblock Ellipsoidal toolbox, 2006.
\newblock Technical Report.

\bibitem{mazo2008event}
Manuel Mazo and Paulo Tabuada.
\newblock On event-triggered and self-triggered control over sensor/actuator
  networks.
\newblock In {\em Decision and Control, 2008. CDC 2008. 47th IEEE Conference
  on}, pages 435--440. IEEE, 2008.

\bibitem{mazo2010iss}
Manuel Mazo~Jr., Adolfo Anta, and Paulo Tabuada.
\newblock An {ISS} self-triggered implementation of linear controllers.
\newblock {\em Automatica}, 46(8):1310--1314, 2010.

\bibitem{moreira2019observer}
L~G Moreira, S~Tarbouriech, A~Seuret, and J~M~Gomes da~Silva~Jr.
\newblock Observer-based event-triggered control in the presence of
  cone-bounded nonlinear inputs.
\newblock {\em Nonlinear Analysis: Hybrid Systems}, 33:17--32, 2019.

\bibitem{nesic2013finite}
Dragan Ne{\v{s}}i{\'c}, Andrew~R Teel, Giorgio Valmorbida, and Luca Zaccarian.
\newblock Finite-gain lp stability for hybrid dynamical systems.
\newblock {\em Automatica}, 49(8):2384--2396, 2013.

\bibitem{ros2002ellipsoidal}
Llu{\'\i}s Ros, Assumpta Sabater, and Federico Thomas.
\newblock An ellipsoidal calculus based on propagation and fusion.
\newblock {\em IEEE Transactions on Systems, Man, and Cybernetics, Part B
  (Cybernetics)}, 32(4):430--442, 2002.

\bibitem{schweppe1968recursive}
Fred Schweppe.
\newblock Recursive state estimation: Unknown but bounded errors and system
  inputs.
\newblock {\em IEEE Transactions on Automatic Control}, 13(1):22--28, 1968.

\bibitem{scott2016constrained}
Joseph~K Scott, Davide~M Raimondo, Giuseppe~Roberto Marseglia, and Richard~D
  Braatz.
\newblock Constrained zonotopes: A new tool for set-based estimation and fault
  detection.
\newblock {\em Automatica}, 69:126--136, 2016.

\bibitem{tabuada2007event}
Paulo Tabuada.
\newblock Event-triggered real-time scheduling of stabilizing control tasks.
\newblock {\em IEEE Transactions on Automatic Control}, 52(9):1680--1685, 2007.

\bibitem{velasco2003self}
Manel Velasco, Josep Fuertes, and Pau Marti.
\newblock The self triggered task model for real-time control systems.
\newblock In {\em Work-in-Progress Session of the 24th IEEE Real-Time Systems
  Symposium (RTSS03)}, volume 384, 2003.

\bibitem{walsh2001scheduling}
Gregory~C Walsh and Hong Ye.
\newblock Scheduling of networked control systems.
\newblock {\em Control Systems, IEEE}, 21(1):57--65, 2001.

\end{thebibliography}
